\documentclass[onecolumn,showpacs,11pt,nofootinbib]{revtex4-2}
\usepackage{axodraw}
\usepackage{graphicx}
\usepackage{dcolumn}
\usepackage{bm}
\usepackage{color}
\usepackage{eurosym}
\usepackage{multirow}
\usepackage{amsfonts}
\usepackage{amsmath}
\usepackage{hyperref}
\usepackage{amssymb}
\usepackage[english]{babel}
\usepackage{graphicx}
\usepackage{epsfig}
\usepackage{subfigure}
\usepackage{unicode}
\begin{document}
%%%%%%%%%%%%%%%%%%%%%%%%
\newcommand{\hs}{\hspace*{0.2cm}}
\newcommand{\hsp}{\hspace*{0.5cm}}
\newcommand{\vs}{\vspace*{0.5cm}}
\newcommand{\be}{\begin{equation}}
\newcommand{\ee}{\end{equation}}
\newcommand{\bea}{\begin{eqnarray}}
\newcommand{\eea}{\end{eqnarray}}
\newcommand{\ben}{\begin{enumerate}}
\newcommand{\een}{\end{enumerate}}
\newcommand{\bde}{\begin{widetext}}
\newcommand{\ede}{\end{widetext}}
\newcommand{\nn}{\nonumber}
\newcommand{\crn}{\nonumber \\}
\newcommand{\Tr}{\mathrm{Tr}}
\newcommand{\non}{\nonumber}
\newcommand{\noi}{\noindent}
\newcommand{\al}{\alpha}
\newcommand{\la}{\lambda}
\newcommand{\bet}{\beta}
\newcommand{\ga}{\gamma}
\newcommand{\va}{\varphi}
\newcommand{\om}{\omega}
\newcommand{\pa}{\partial}
\newcommand{\+}{\dagger}
\newcommand{\fr}{\frac}
\newcommand{\sq}{\sqrt}
\newcommand{\bc}{\begin{center}}
\newcommand{\ec}{\end{center}}
\newcommand{\Ga}{\Gamma}
\newcommand{\de}{\delta}
\newcommand{\De}{\Delta}
\newcommand{\ep}{\epsilon}
\newcommand{\varep}{\varepsilon}
\newcommand{\ka}{\kappa}
\newcommand{\La}{\Lambda}
\newcommand{\si}{\sigma}
\newcommand{\Si}{\Sigma}
\newcommand{\ta}{\tau}
\newcommand{\up}{\upsilon}
\newcommand{\Up}{\Upsilon}
\newcommand{\ze}{\zeta}
\newcommand{\ps}{\psi}
\newcommand{\Ps}{\Psi}
\newcommand{\ph}{\phi}
\newcommand{\vph}{\varphi}
\newcommand{\Ph}{\Phi}
\newcommand{\Om}{\Omega}
%%%%%%%%%%%%%%%%%%%%%%%%
\newcommand{\Revised}[1]{{\color{red}#1}}
\newcommand{\Red}[1]{{\color{red}#1}}
\newcommand{\Blue}[1]{{\color{blue}#1}}
\newcommand{\Black}[1]{{\color{black}#1}}
\title{$B-L$ model with $A_4\times Z_3\times Z_4$ symmetry for $3+1$ active$-$sterile neutrino mixing}%neutrino mixing}
\author{V. V. Vien$^{a,b}$}
%\author{V. V. Vien}
\email{vovanvien@tdmu.edu.vn}
\affiliation{$^{a)}$Institute of Applied Technology, Thu Dau Mot University, Binh Duong Province, Vietnam, }
\affiliation{$^{b)}$Department of Physics, Tay Nguyen University, %567 Le Duan street, Buon Ma Thuot city,
Daklak province, Vietnam.}
%\email{vvvien@ttn.edu.vn}
%\affiliation{$^{a}$Institute of Research and Development, Duy Tan University, 182 Nguyen Van Linh, Da Nang City, Vietnam, and\\
%$^{b}$Department of Physics, Tay Nguyen University, 567 Le Duan, Buon Ma Thuot, DakLak, Vietnam.}

\begin{abstract}
We construct a $B-L$ model with $A_4\times Z_3\times Z_4$ flavor symmetry which can accounts for the recent $3+1$ active-sterile neutrino data. The tiny neutrino mass and the mass hierarchy are obtained by the type-I seesaw mechanism. The hierarchy of the lepton masses is satisfied by a factor of $v_H \left(\frac{v_l}{\Lambda}\right)^2 \sim 10^{-4}\, \mathrm{GeV}$ of the electron mass compared to the muon and tau masses of the order of $\frac{v_H v_l}{\Lambda} \sim 10^{-1}\, \mathrm{GeV}$. The $3+1$ active-sterile neutrino mixings are predicted to be $0.015 \leq|U_{1 4}|^2\leq 0.045$, $0.004 \leq|U_{2 4}|^2\leq 0.012$ and $0.004 \leq|U_{3 4}|^2\leq 0.014$ for normal hierarchy while $0.020\leq|U_{1 4}|^2\leq 0.045$, $0.008 \leq|U_{2 4}|^2\leq 0.018$ and $0.008\leq|U_{3 4}|^2\leq 0.022$ for inverted hierarchy. Sterile neutrino masses are predicted to be $0.7 \lesssim  m_s \, (\mathrm{eV}) \lesssim 3.16$ for normal hierarchy and $2.6 \lesssim  m_s \, (\mathrm{eV}) \lesssim 7.1$  for inverted hierarchy. For three neutrino scheme the model predicts $0.3401 \leq \sin^2\theta_{12}\leq 0.3415, \, 0.460 \leq \sin^2\theta_{23}\leq 0.540,\, -0.60 \leq \sin\delta_{CP}\leq -0.20$ for normal hierarchy and $0.3402 \leq \sin^2\theta_{12}\leq 0.3416,\, 0.434\leq\sin^2\theta_{23}\leq 0.610,\, -0.95 \leq \sin\delta_{CP}\leq -0.60$ for inverted hierarchy. %The effective neutrino masses are predicted to be $35.70 \leq \langle m_{ee}\rangle [\mbox{meV}] \leq 36.50$ in 3+1 scheme and $3.65 \leq \langle m^{(3)}_{ee}\rangle [\mbox{meV}] \leq 4.10$ in three neutrino scheme for NH while $160.0 \leq \langle m_{ee}\rangle [\mbox{meV}] \leq 168.0$ in 3+1 scheme and $47.80 \leq \langle m^{(3)}_{ee}\rangle [\mbox{meV}] \leq 48.70$ in three neutrino scheme for for IH which are all in agreement with the recent experimental data.
\end{abstract}
\date{\today}
\keywords{Neutrino mass and mixing;
Extensions of electroweak Higgs sector; Non-standard-model neutrinos, right-handed neutrinos, discrete
symmetry.}
\pacs{12.15 Ff; 12.60.Fr; 14.60.St.}

\maketitle
\section{\label{intro} Introduction}
Apart from the current unconfirmed parameters, % in the neutrino sector,
including the mass hierarchy, the octant of %the atmospheric mixing angle
$\theta_{23}$ and the Dirac CP phase \cite{Salas2020}, there exist some experimental results in the neutrino sector that cannot be explained within the three neutrino scheme \cite{Aguilar2001,Arevalo2013, Maltoni2003,Acero2008, Arevalo2010, Mention2011, An2014,%Abe2015,
Arevalo2018,Adamson2019, Adamson2020, Aartsen2020, Beheraa2019,BeheraPRD2020,Gariazzo2016}. However, these parameters %observations
could be interpreted by adding at least
an additional %fourth
neutrino %(called sterile neutrinos)
%having non-trivial mixingwith active neutrinos
with mass in
the eV range ($\Delta m^2_{41}\gg |\Delta m^2_{31}|$). The mentioned neutrinos are $SU(2)_L$ singlets that do not contribute to the weak interactions but mix with the active neutrinos which can be probed by experiments. %in the oscillation experiments.

There have been some schemes with sterile neutrinos proposed in the literature such as (3+1) scheme \cite{Kang2013,Girardi2014,Agudelo2015,Gariazzo17,Coloma2018,Liu2018,Gariazzo2018plb,Dentler2018ea,
Gupta2018ea,Thakore2018ea,Dev2019,Miranda2019,Giunti2019,Boser2020,Giunti2020cz,Behera2020,Diaz2020}, (3+1+1) scheme \cite{Nelson2011,Fan2012,Kuflik2012,Huang2013,Giunti2013}, and (1+3+1) scheme \cite{Kopp2011, Kopp2013}, (3+2) scheme \cite{Sorel2004,Karagiorgi2007,Maltoni2007,Goswami2007,Karagiorgi2009,Giunti2011,Donini2012,Archidiacono2012}, neutrino non-standard interactions \cite{Akhmedov2010,Liao2016,Babu2016,Blennow2017,Esmaili2019} and radiative sterile neutrino decays \cite{Gninenko2012, Gninenko2012plb}. % of which the one with one additional sterile neutrino with mass in the eV range (called four neutrino scheme) is the simplest extension of standard three neutrino mixing.
Among four neutrino schemes, the one with three active neutrinos and one sterile neutrino (called 3+1 scheme) is preferred because the 1+3 scenario %in which the three active neutrinos are in eV scale and the sterile neutrino is lighter than the active neutrinos
is disfavored by cosmology and the 2+2 scheme is not compatible with the solar and the atmospheric data \cite{Maltoni2003, Krishnan2020}.

At present, the neutrino parameters %mass squared differences and the mixing angles
in the three neutrino framework \cite{Salas2020} and  (3+1) neutrino framework \cite{Gariazzo2016, Krishnan2020}, for the best-fit values and $3\sigma$ range, are given in Table \ref{experconstrain}.
\begin{table}[ht]
\begin{center}
\caption{Neutrino parameters in the three neutrino framework taken from Ref. \cite{Salas2020} and  (3+1) neutrino framework taken from \cite{Gariazzo2016, Krishnan2020}} \label{experconstrain}
\vspace{0.15 cm}
\begin{tabular}{|c|c|c|}
    \hline
    \multicolumn{1}{|c|}{\multirow{2}{*}{Parameters}}&\multicolumn{2}{c|}{3$\sigma$ range (best fit)}\\
    \cline{2-3} %\hline
    \multicolumn{1}{|c|}{}&\hspace{0.5 cm} Normal hierarchy&\hspace{0.5 cm} Inverted hierarchy\\    \hline\hline
    $\Delta m^{2}_{21}(10^{-5}\, \mathrm{eV}^2)$&$6.94\rightarrow 8.14\, (7.50)$&$6.94\rightarrow8.14\, (7.50)$\\ \hline
$| \Delta m^{2}_{31}|(10^{-3}\, \mathrm{eV}^2)$&$2.47\rightarrow2.63\, (2.55)$&$2.37\rightarrow2.53\, (2.45)$\\ \hline
$\sin^{2}\theta_{12}/10^{-1}$&$2.71\rightarrow3.69 \,(3.18)$&$2.71\rightarrow3.69\, (3.18)$ \\ \hline
$\sin^{2}\theta_{23}/10^{-1}$&$4.34\rightarrow6.10 \,(5.74)$&$4.33\rightarrow6.08\,(5.78)$\\ \hline
$\sin^{2}\theta_{13}/10^{-2}$&$2.000\rightarrow2.405\, (2.200)$&$2.018\rightarrow2.424 \,(2.225)$\\ \hline
$\delta/\pi$&$0.71\rightarrow1.99 \,(1.08)$&$1.11\rightarrow1.96 \,(1.58)$\\ \hline\hline
$|U_{14}|^{2}$&$0.012\rightarrow0.047$&$0.012\rightarrow0.047$\\ \hline
$|U_{24}|^{2}$&$0.005\rightarrow0.03$&$0.005\rightarrow0.03$\\ \hline
$|U_{34}|^{2}$&$0 \rightarrow 0.16$&$0 \rightarrow 0.16$\\ \hline
%$|U_{14}|^{2}$&0.0098-0.031 (0.020)&0.0098-0.031 (0.020)\\ \hline
%$|U_{24}|^{2}$&0.006-0.026 (0.015)&0.006-0.026 (0.015)\\ \hline
%$|U_{34}|^{2}$&0 - 0.039 (-)&0 - 0.039 (-)\\ \hline
%$\delta_{14}$&0 - 2$\pi$ (-)&0 - 2$\pi$ (-)\\ \hline
%$\delta_{24}$&0 - 2$\pi$ (-)&0 - 2$\pi$ (-)\\ \hline
\end{tabular}
\vspace{-0.25 cm}
\end{center}
\end{table}

One outstanding feature of %non-Abelian
discrete symmetries is imposed to explain the neutrino oscillation data which have been widely used in the literature. Among several discrete symmetries, the $A_4$ has attracted a lot of attention because it is the smallest symmetry which possesses one three-dimensional representation and three inequivalent one-dimensional representations. Recently, there have been different suggestions for generating the active-sterile neutrino mixing within different works \cite{Suematsu2001,Mohapatra2001,Babu2004,Mohapatra2005,
Barry2011,Barry2012,Zhang2012,Ghosh2012,Machado2013S3,Zhang2013, Zhang2013v2,Frank2013,Borah2013,
Merle2014,Nath2016,Borah2016xkc,Borah2016lrl,Dev2017,Nath2017,Borah2017azf,
Borah2017fqj,Das2019ea,Sarma2019,%Giunti2019,
Bhat2020,Pires2020,Kumar2020,Pinheiro2020,Krishnan2020,VienS3EPJC21}. In $3+1$ framework, the $B-L$ model based on $S_3$ symmetry was first proposed in Ref. \cite{Machado2013S3} without considering the active-sterile neutrino mixing. This problem has been improved in Ref. \cite{VienS3EPJC21} in which the active-sterile neutrino mass and mixing are obtained at the first-order of the perturbation theory and only the normal ordering of the active neutrino masses is considered. However, the charged-lepton mass hierarchy is not yet solved in Ref. \cite{VienS3EPJC21}\footnote{As will be shown in section \ref{ASmixing}, the charged-lepton mass hierarchy is naturally explained in the current model.}. In Refs. \cite{Barry2011, Barry2012, Zhang2012, Borah2017fqj, Das2019ea, Sarma2019, Krishnan2020}, based on $A_4$ symmetry, the active-sterile neutrino mass and mixing have been considered with non minimal scalar sector and many $SU(2)_L$ Higgs doublets, thus there are substantial differences between previous works and our present work. For instance, in Refs. \cite{Barry2011, Barry2012} the symmetry of the SM is supplemented by the $A_4\times Z_3\times U(1)_R$ symmetry in which two doublets are introduced to get sterile neutrino mass with the help of the dimension-six terms \cite{Barry2011} or the perturbation theory\cite{Barry2012}. In order to generate the sterile-active neutrino masses and mixings, in Ref. \cite{Zhang2012} the symmetry of the SM is supplemented by the $A_4\times Z_4$ symmetry in which one doublet and up to twelve singlets are introduced but only $|U_{e4}|$ is predicted without $|U_{\mu 4}|$ and $|U_{\tau4}|$; in Ref. \cite{Borah2017fqj}, the symmetry of the SM is supplemented by the $A_4\times Z_3\times Z^{'}_3\times U(1)_R$ symmetry in which two doublets and up to seventeen singlets are introduced; in Ref. \cite{Das2019ea}, the symmetry of the SM is supplemented by the $A_4\times Z_3\times Z_4$ symmetry\footnote{In our present work, although the discrete symmetry structure is similar to that of Ref. \cite{Das2019ea}, however, the alignment of fields and the number of scalars are completely different from each other.} in which three doublets and up to twelve singlets are introduced; in Ref. \cite{Sarma2019} the symmetry of the SM is supplemented by the $A_4\times Z_4$ symmetry in which one doublets and up to twelve singlets are introduced, and in Ref. \cite{Krishnan2020} the symmetry of the SM is supplemented by the $A_4\times C_4\times C_6 \times C_2\times U(1)_s$ symmetry in which one doublets and up to twelve singlets are introduced whereby only the normal ordering of the light neutrino masses is predicted. Therefore, it would be necessary to construct an $A_{4}$ flavor model with less scalar content than our previous model which satisfies the minimization condition of the scalar potential.
To the best of our knowledge, $A_4$ symmetry has not yet been combined with the $B-L$ model in the 3+1 scheme.

The remainder of this work is structured as follows. The description of the model is presented in section \ref{model}. The scalar potential of the model is described in section \ref{minimumcondi}. The active$-$sterile neutrino mixing is presented in section \ref{ASmixing} and section \ref{NR} is devoted to the numerical analysis. Finally, some conclusions are drawn in section \ref{conclusion}.

\section{The model \label{model}}
The $B-L$ gauge model\footnote{In the gauge $B-L$ model, the anomalies can be cancelled in various scenarios\cite{U1X1, U1X2,U1X3,U1X4,U1X5,U1X6,U1X7,U1X8,U1X9,U1X10,U1X11,U1X12,U1X13, U1X14, U1X14, U1X16, U1X17} with different charge assignments of $B-L$. In this study, we use the model proposed in Refs. \cite{U1X3, U1X4}% whereby all of three right-handed neutrinos possesses $B-L=-1$ which is different from the work in Ref. \cite{Machado2013S3} in which two of right-handed neutrinos owns $B-L=-4$ and the other one has $B-L=5$.
} is extended by the discrete symmetry $A_4\times Z_3\times Z_4$. Besides, three right-handed neutrinos ($\nu_{R}$), one sterile neutrino ($\nu_s$) and three $SU(2)_L$ singlet scalars ($\phi_l, \phi_\nu, \phi_s$) are introduced in addition to the $B-L$ model. In present work, each three left-handed leptons and three right-handed neutrinos are put in one $A_4$ triplet while each the right-handed charged leptons $l_{1R}, l_{2R}$ and $l_{3R}$ are, respectively, put in $\underline{1}, \underline{1}^{''}$ and $\underline{1}^{'}$ under $A_4$, and the sterile neutrino $\nu_{s}$ is put in $\underline{1}$ under $A_4$ symmetry. The particle content of the considered model, under $\mathbf{\Gamma}\equiv SU(3)_C\times SU(2)_L\times U(1)_Y\times U(1)_{B-L} \times A_4\times Z_3\times Z_4$
symmetry, is shown in Table \ref{lepcont}\footnote{All leptons and scalar fields are aligned in singlet $\mathbf{1}$ of $SU(3)_L$ symmetry.}.
%%%%%%%%%%%%%%%%%%%%%%%%%%%
\begin{table}[h]
\caption{\label{lepcont} The particle and scalar contents of the model.}
\begin{center}
\begin{tabular}{|c|cccc|ccccc|cc|c}
\hline
Fields  &\hspace{0.25cm}$\psi_{L}$  &\hspace{0.25cm}$l_{1R}$ &\hspace{0.25cm}$l_{2R}$&\hspace{0.25cm}$l_{3R}$\,\,&\,\,$H$&\hspace{0.05cm}\,\,$\phi_l$\,\,&\,\hspace{0.1cm}$\phi_\nu$\,\,&$\eta\, (\eta_s)$&$\phi_s$&
\hspace{0.2cm}$\nu_{R}$&\hspace{0.15cm}$\nu_s$ \\ \hline
%$\mathrm{SU}(3)_C$ &  $\textbf{1}$ &$\textbf{1}$&$\textbf{1}$&   $\textbf{1}$  & $\textbf{1}$ &$\textbf{1}$&  $\textbf{1}$   \\
$\mathrm{SU}(2)_L$  & \hspace{0.2cm}$\textbf{2}$ &\hspace{0.25cm}$\textbf{1}$&\hspace{0.25cm}$\textbf{1}$&\hspace{0.25cm}$\textbf{1}$&   $\textbf{2}$  &    $\textbf{1}$ &$\textbf{1}$& $\textbf{1}$& $\textbf{1}$&\hspace{0.2cm}$\textbf{1}$ &\hspace{0.15cm} $\textbf{1}$  \\
%$\mathrm{U}(1)_Y$  & $-\frac{1}{2}$ &$-1$&$-1$&$-1$& $\frac{1}{2}$  & $0$    &$0$& $0$& $0$&\hspace{0.2cm}$0$  &\hspace{0.15cm} $0$ &\hspace{0.0cm} $0$ \\
$\mathrm{U}(1)_{B-L}$ & $-1$ &$-1$&$-1$&$-1$& $0$  & $0$ &$0$& $2$& $0$ &$-1$& $-1$ \\
$A_4$&\hspace{0.15cm}  $\mathbf{3}$  &\hspace{0.2cm}$\mathbf{1}$&\hspace{0.15cm}$\mathbf{1^{''}}$&\hspace{0.15cm}$\mathbf{1^'}$ &$\mathbf{1}$&  $\mathbf{3}$ &$\mathbf{3}$ & $\mathbf{1}$&$\mathbf{3}$ &\hspace{0.2cm}$\mathbf{3}$&\hspace{0.15cm} $\mathbf{1}$  \\
%old$Z_3$&\hspace{0.15cm}  $\om$  &\hspace{0.15cm}$\om$&\hspace{0.15cm}$\om^2$&\hspace{0.15cm}$\om$&$\om$&$\om$&$1$&\hspace{0.05cm} $\om^2$  &\hspace{0.1cm} $\om^2$ &\hspace{0.2cm}$\om^2$&\hspace{0.15cm} $1$ \\
$Z_3$&\hspace{0.15cm}  $\om$  &\hspace{0.15cm}$\om$&\hspace{0.15cm}$\om$&\hspace{0.15cm}$\om$&$1$&$1$&$1$&\hspace{0.05cm} $\om$  &\hspace{0.1cm} $\om$ &\hspace{0.2cm}$\om$&\hspace{0.15cm} $1$ \\
$Z_4$&\hspace{0.15cm}  $i$  &$-i$&\hspace{0.1cm}$1$&$-1$& $1$& $i$ &$1$&$-1$& $-1$&\hspace{0.15cm}$i$& $-i$ \\ \hline
%old$Z_4$&\hspace{0.15cm}  $i$  &$-i$&\hspace{0.1cm}$1$&$-1$& $1$& $i$ &$1$&$-1$& $-1$&\hspace{0.15cm}$i$& $-i$ \\ \hline
%$Z_2$&\hspace{0.15cm}  $\Red{+}$  &$\Red{+}$&\hspace{0.1cm}$\Red{+}$&$\Red{+}$& $\Red{+}$& $\Red{+}$ &$\Red{+}$&$\Red{+} \, \Red{(-)}$& $\Red{+}$&\hspace{0.15cm}$\Red{+}$& $\Red{-}$ \\  \hline
\end{tabular}
\end{center}
\vspace*{-0.3cm}
\end{table}
%%%%%%%%%%%%%%%%%%%%%%%%%%%%%%%%%%%%%%%%%%%%%%%%%%%%%%%%

From the field assignments given in Tab. \ref{lepcont}, the following Yukawa terms, up to five-dimension which invariant under $\mathbf{\Gamma}$ symmtry, arise\footnote{The Lagrangian in Eq. (\ref{Lylep0}) is performed with the additional symmetry $Z_2$, where $\eta_s$ and $\nu_s$ are even under $Z_2$ while all other fields are odd under $Z_2$. This $Z_2$ symmetry ensures that the $\mathbf{\Gamma}$ invariant Yukawa term $\overline{\psi}_{L} \nu_{s} \widetilde{H} \phi_s$ is absent from the
Lagrangian.}:
\bea -\mathcal{L}^{l}_{Y}&=&\frac{x_{1cl}}{\Lambda^2}(\bar{\psi}_{L}l_{1R} )_{\underline{3}} \left(H \phi^2_l \right)_{\underline{3}}+ \frac{x_{2cl}}{\Lambda}(\bar{\psi}_{L}l_{2R} )_{\underline{3}} \left(H \phi_l \right)_{\underline{3}}+ \frac{x_{3cl}}{\Lambda}(\bar{\psi}_{L}l_{3R} )_{\underline{3}} \left(H \phi^*_l\right)_{\underline{3}} \crn
 &+&x_{1\nu} (\bar{\psi}_{L} \nu_{R} )_{\underline{1}}\widetilde{H}+ \frac{x_{2\nu}}{\Lambda} (\bar{\psi}_{L} \nu_{R})_{\underline{3}_s}(\widetilde{H}\phi_\nu)_{\underline{3}}+ \frac{x_{3\nu}}{\Lambda} (\bar{\psi}_{L} \nu_{R})_{\underline{3}_a}(\widetilde{H}\phi_\nu)_{\underline{3}}\crn
 &+&\frac{y_\nu}{2} (\bar{\nu}^c_{R}\nu_{R})_{\underline{1}}\eta + \frac{y_s}{\Lambda} (\bar{\nu}^c_{s}\nu_{R})_{\underline{3}}(\eta_{s}\phi_s)_{\underline{3}} + \mathrm{H.c}, \label{Lylep0}\eea
 where $x_{1,2,3cl}, x_{1,2,3\nu}, y_\nu$ and $y_s$ are the Yukawa-like dimensionless coupling constants, and $\Lambda$ is the cut-off scale.
The non-Abelian discrete symmetry $A_4$ along with Abelian symmetries $Z_3$ and $Z_4$ play an important role to generate a desired neutrino mass matrix structure consistent with the sterile-active neutrino mixing data. For instance, the charged lepton masses arise from
$\bar{\psi}_{L} l_{1,2,3R}$ to scalars, where under $\mathbf{\Gamma}$ symmetry, $\bar{\psi}_{L} l_{1R}\sim (\mathbf{2}, -\frac{1}{2}, 0, \underline{3}, 1, -1)$, $\bar{\psi}_{L} l_{2R}\sim (\mathbf{2}, -\frac{1}{2}, 0, \underline{3}, \om, -i)$ and
$\bar{\psi}_{L} l_{1R}\sim (\mathbf{2}, -\frac{1}{2}, 0, \underline{3}, 1, i)$.
For the known scalars, $\bar{\psi}_{L} l_{1R} H$ is prevented by three symmetries $A_4, Z_3$ and $Z_4$; $\bar{\psi}_{L} l_{1R} H\phi_l$ is prevented by two symmetries $Z_3$ and $Z_4$; $\bar{\psi}_{L} l_{1R} H\phi^*_l$ is prevented by $Z_4$ symmetry and so on. $\bar{\psi}_{L} l_{2R} H$
is prevented by $A_4, Z_3$ and $Z_4$ symmetries; $\bar{\psi}_{L} l_{2R} H\phi_\nu$
is prevented by $Z_3$ and $Z_4$ symmetries and so on. $\bar{\psi}_{L} l_{3R} H$
is prevented by $A_4, Z_3$ and $Z_4$ symmetries; $\bar{\psi}_{L} l_{3R} H\phi_l$
is prevented by $Z_3$ and $Z_4$ symmetries; $\bar{\psi}_{L} l_{3R} H\eta$
is prevented by $A_4, Z_3$ and $Z_4$ symmetries; $\bar{\psi}_{L} l_{3R}\phi_s$
is prevented by $Z_3$ and $Z_4$ symmetries and so on. Thus, under $\mathbf{\Gamma}$ symmetry,  there is only three invariant terms $(\bar{\psi}_{L}l_{1R} )_{\underline{3}} \left(H \phi^2_l \right)_{\underline{3}}, (\bar{\psi}_{L}l_{2R} )_{\underline{3}} \left(H \phi_l \right)_{\underline{3}}$ and $(\bar{\psi}_{L}l_{3R} )_{\underline{3}} \left(H \phi^*_l\right)_{\underline{3}}$
which are responsible for generating masses for charged leptons as given in Eq. (\ref{memt}). The situation is similar for the remaining couplings that generate the mass matrices for neutrino sector. Besides, $A_4$, $Z_3$ and $Z_4$ symmetries also prevent some interaction terms in the Higgs potential as discussed in Appendix \ref{Higgspotential}. It is well-known that 3+1 scheme-the minimal extended seesaw-by adding only one sterile neutrino $\nu_s$ cannot be anomaly free \cite{Barry2011, Zhang2012}, however, this may be solved by introducing additional fields required for anomaly cancelation \cite{Borah2017fqj} which is left for our future study.

It is interesting to note that, with the particle and scalar contents of the model in Table \ref{lepcont}, all the other quadratic terms forming a triplet under $A_4$, including $(\phi_\nu \phi_\nu)_{3_s}$, $(\phi_\nu\phi_\nu)_{3_a}$, $(\phi_s \phi_s)_{3_s}$ and $(\phi_s\phi_s)_{3_a}$ can combine with $H$ to form $A_4$ singlets but one cannot construct an invariant under $Z_3$ and/or $Z_4$. Thus, the charged lepton mass expressions in our model (\ref{memt}) is much simpler compared to that of Ref. \cite{Krishnan2020}.
In fact, there exist the other contributions via higher dimensional Weinberg operators\footnote{Another operator $\frac{1}{\Lambda^{2(k+l)+3}}\left(\overline{\psi}^c_L \psi_L\right)_{\mathbf{3}_a}  H^2 \left(\eta \phi_\nu\right)_{\mathbf{3}}(H^\+H)^k (\eta^\+\eta)^l$ which is invariant under $\mathrm{\Gamma}$ but vanished due to the symmetric in $\overline{\psi}^c_{iL}$ and $\psi_{jL}\, (i, j=1,2,3; i\neq j)$.}
%\bea
%\frac{1}{\Lambda^{2(k+l+1)}}\left(\overline{\psi}^c_L \psi_L\right)_{\mathbf{1}} H^2 \eta (H^\+H)^k (\eta^\+\eta)^l, \eea
%and
%\bea
%\frac{1}{\Lambda^{2(k+l)+3}}\left(\overline{\psi}^c_L \psi_L\right)_{\mathbf{3}_s}  H^2 \left(\eta \phi_\nu\right)_{\mathbf{3}}(H^\+H)^k (\eta^\+\eta)^l, \eea
$\frac{1}{\Lambda^{2(k+l+1)}}\left(\overline{\psi}^c_L \psi_L\right)_{\mathbf{1}} H^2 \eta (H^\+H)^k (\eta^\+\eta)^l$ and
$\frac{1}{\Lambda^{2(k+l)+3}}\left(\overline{\psi}^c_L \psi_L\right)_{\mathbf{3}_s}  H^2 \left(\eta \phi_\nu\right)_{\mathbf{3}}(H^\+H)^k (\eta^\+\eta)^l,$ with $k, l = 0, 1, 2, ...$. However, because $v_H\ll v_\eta, v_\nu \ll \Lambda$, the left-handed neutrinos generated by the Type-II seesaw mechanism, $\sim v_H \left(\frac{ v_H }{\Lambda}\right)^{2k+1} \left(\frac{v_\eta}{\Lambda}\right)^{2l+1}$ and $v_H \left(\frac{ v_H }{\Lambda}\right)^{2k+1} \left(\frac{ v_\nu }{\Lambda}\right)\left(\frac{v_\eta}{\Lambda}\right)^{2l+1}$, are very small compared to the ones generated via the canonical type-I seesaw mechanism obtained in Eq. (\ref{Mnu44}) and thus would be neglected in the Lagrangian (\ref{Lylep0}).
%\Red{A part of the reason for assigning the various Abelian charges to the flavons, Table \ref{lepcont}, is to ensure that this suppression occurs %so that our model leads to the standard MES framework where these terms are assumed to vanish (the first block of zeros in Eq. %(\ref{Mnu77})).}

The VEVs of scalar fields are chosen as follows
 \bea
&&\langle H \rangle =\left(%
\begin{array}{ccc}
  0 \\
  v_H\\
\end{array}%
\right),\hs \langle \phi_\nu \rangle = (0\hspace{0.2cm} \langle \phi_{2\nu} \rangle \hspace{0.2cm} 0), \hs \langle \phi_{2\nu} \rangle = v_{\nu}, \hs \langle \eta \rangle = v_{\eta}, \crn
&& \langle \eta_s \rangle = v_{\eta_s}, \hs  \langle \phi_l \rangle = (\langle \phi_{1l} \rangle \hspace{0.3cm} \langle \phi_{2l} \rangle \hspace{0.3cm} \langle \phi_{3l} \rangle), \hs \langle \phi_{1l} \rangle= \langle \phi_{2l} \rangle= \langle \phi_{3l} \rangle =v_{l}, \crn
&& \langle \phi_s \rangle = (\langle \phi_{1s} \rangle \hspace{0.3cm} \langle \phi_{2s} \rangle \hspace{0.3cm} \langle \phi_{3s} \rangle), \hs \langle \phi_{1s} \rangle=  \langle \phi_{3s} \rangle =v_{s}, \hs \langle \phi_{2s} \rangle=0. \label{scalarvev}\eea
The fact that the electroweak symmetry breaking scale is \cite{EBS1,EBS2}
\bea
v \sim 10^{2}\, \mathrm{GeV}. \label{SMscale} \eea
The scale of $B-L$ symmetry breaking is undetermined, spanning from TeV to much
higher scales \cite{BLscales, buchmuller2014, Pallis2013, Moursy2021}. In this study, assume that the $B-L$ symmetry breaking is at TeV scale \cite{BLscales} while %the VEV of singlets and
the cut-off scale\footnote{In Ref. \cite{cutoffscal21} the cut-off scale in the QED case is $\Lambda_{\mathrm{EW}}\simeq 3.8 \times 10^{13}\, \mathrm{GeV}$; thus, we chose $\Lambda=10^{13}$ GeV for its scale.} is at a very high scale \cite{Krishnan2020},
\bea
&&v_{\eta_s} \sim \, v_\eta \sim 10^{3}\, \mathrm{GeV}, \hs\hs  %\label{BLscale} \\
%&&v_l \sim v_\nu  \sim v_s \sim 10^{10} \, \mathrm{GeV}, \hs
\Lambda \sim 10^{13}\, \mathrm{GeV}. \label{BLscale}
\eea
On the other hand, the model result  in Eq. (\ref{memt}) tells us that $\frac{m_e}{m_\mu}= \frac{2 v_l}{\Lambda} \frac{x_{1cl}}{ x_{3cl}}\sim \frac{v_l}{\Lambda}$, i.e., $\frac{m_e}{m_\mu}\sim \frac{v_l}{\Lambda}$ provided that $x_{1cl}\sim x_{3cl}$. Furthermore, the experimental data \cite{PDG2020} implies that $\frac{m_e}{m_\mu}\sim 10^{-3}$. %Thus, $v_l \sim 10^{10} \, \mathrm{GeV}$.
The VEV of singlets with $B-L=0$ are assumed to be in the same scale, i.e., $v_l\sim v_\nu\sim v_s$. In this sense, we get the scale of VEVs of the flavons as follows: \bea
&&v_l \sim v_\nu  \sim v_s \sim 10^{10} \, \mathrm{GeV}. \label{vevsingletscale}
\eea
From Eq. (\ref{Lylep0}), with the help of Eqs. (\ref{SMscale})-(\ref{vevsingletscale}), after spontaneous symmetry breaking one can estimate the scales of the
%our
mass terms as follows:
\bea
&&\bar{l}_{L}l_{1R}\sim \frac{v_H v^2_l}{\Lambda^2}\sim 10^{-4}\, \mathrm{GeV},\hs \bar{l}_{L}l_{2R}\sim \bar{l}_{L}l_{3R}\sim \frac{v_H v_l}{\Lambda}\sim 10^{-1}\, \mathrm{GeV},\crn
&&\bar{\nu}_{L} \nu_{R}\sim v_H \sim 10^2 \, \mathrm{GeV}, \hs
\bar{\nu}^c_{R}\nu_{R}\sim v_\eta \sim 10^{3} \, \mathrm{GeV}, \hs
\bar{\nu}^c_{s}\nu_{R}\sim \frac{v_{\eta_{\Blue{s}}} v_s}{\Lambda}\sim 1 \, \mathrm{GeV}. \label{massscale}
\eea
\section{\label{minimumcondi}Scalar potential minimum condition}
\begin{figure}[h]
\begin{center}
\vspace{-0.25 cm}
\hspace{-1.0 cm}\includegraphics[width=0.65\textwidth]{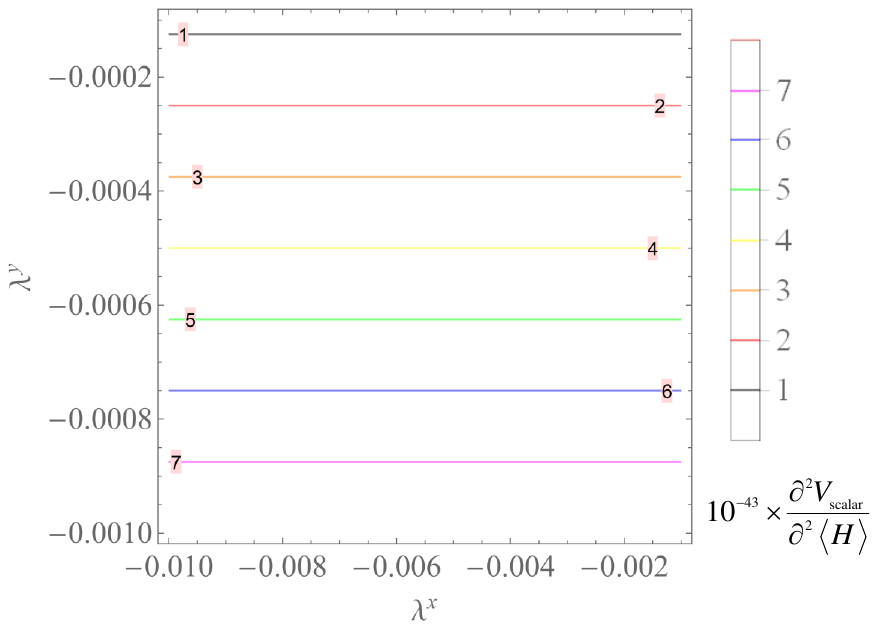}\hspace{-3.1 cm}
\includegraphics[width=0.65\textwidth]{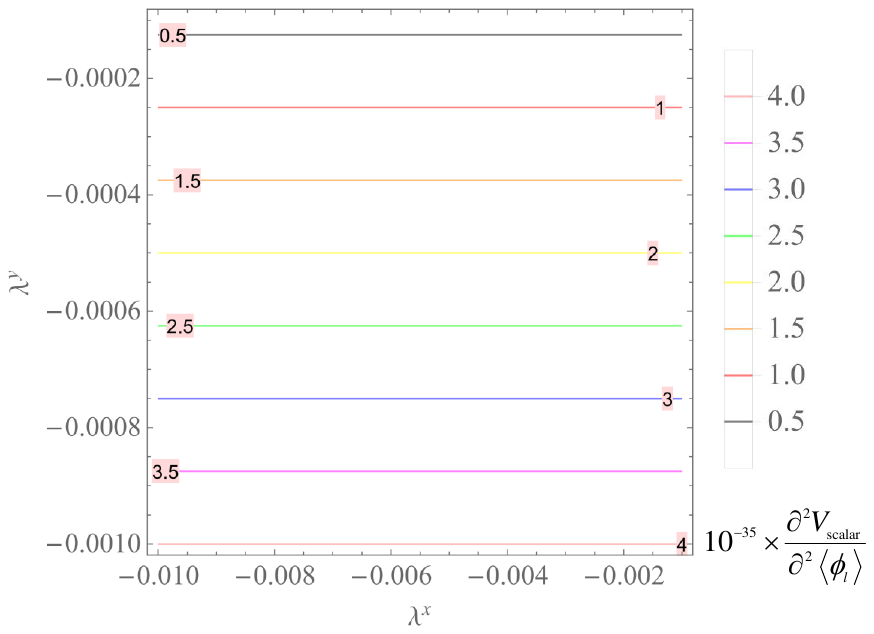}\hspace*{-1.25 cm}\hspace{-15.5 cm}\\
\vspace{-9.15 cm}
\hspace{-1.0 cm}\includegraphics[width=0.65\textwidth]{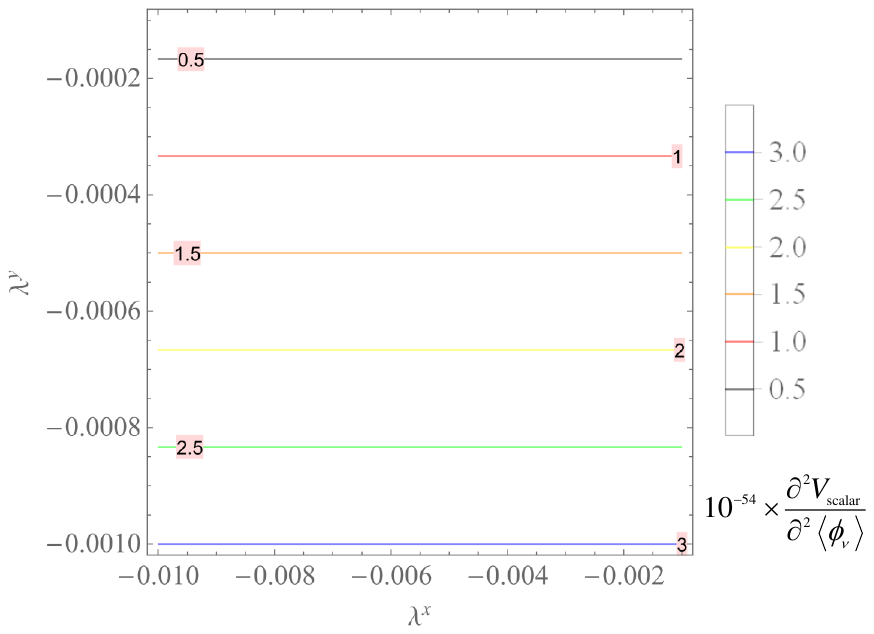}\hspace{-3.1 cm}
\includegraphics[width=0.65\textwidth]{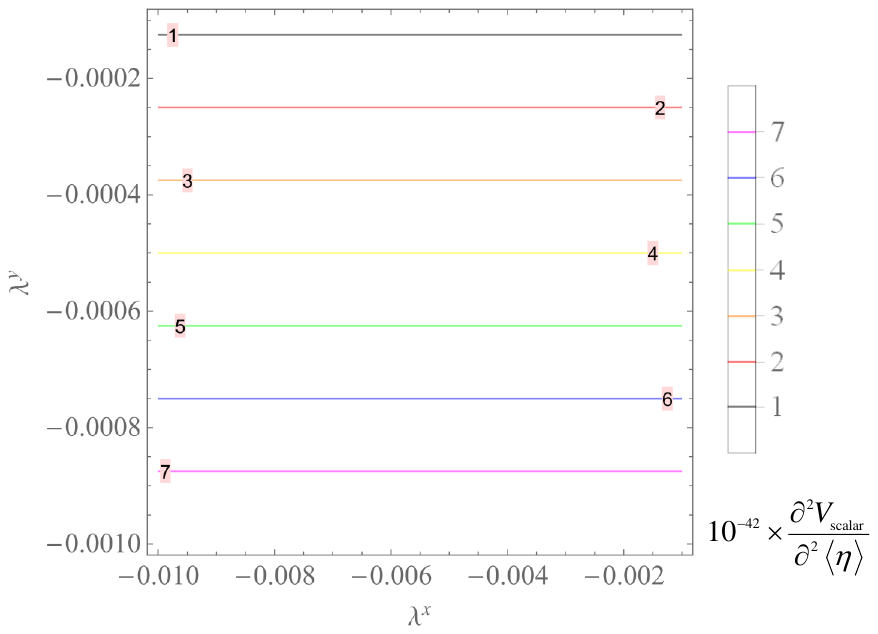}\hspace*{-1.25 cm}\hspace{-15.5 cm}\\
\vspace{-9.15 cm}
\hspace{-1.0 cm}\includegraphics[width=0.65\textwidth]{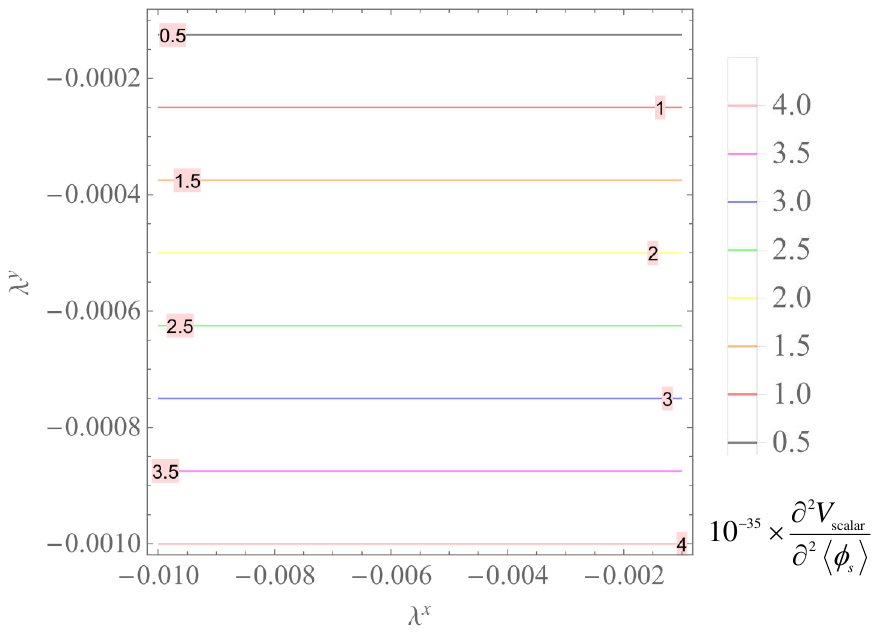}\hspace{-3.1 cm}
\includegraphics[width=0.65\textwidth]{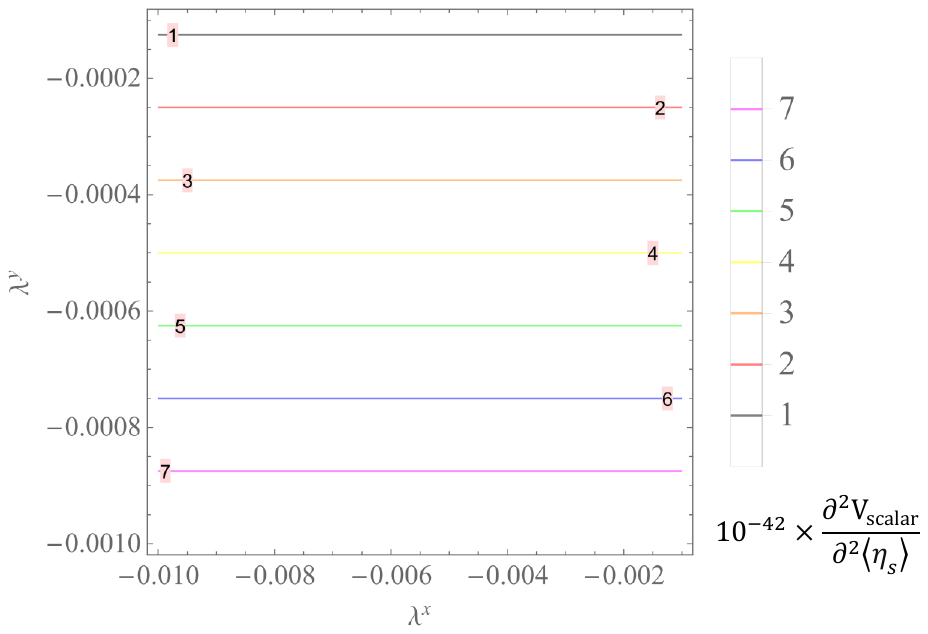}\hspace*{-1.25 cm}\hspace{-15.5 cm}
\end{center}
\vspace{-9.5 cm}
\caption{$\frac{\partial^2 V_{\mathrm{scalar}}}{\partial \langle \Phi \rangle^2}\, (\Phi =H, \phi_l, \phi_\nu, \eta, \phi_s, \eta_s)$ versus $\lambda^{x}$ and $\lambda^{y}$ with $\lambda^{x} \in (-10^{-2}, -10^{-3})$ and $\lambda^{y} \in (-10^{-3}, -10^{-4})$.}
\label{gilaxlay}
%\vspace{1.0 cm}
\end{figure}

In this section we will show that the VEV alignments in Eq. (\ref{scalarvev}) satisfies the
minimization condition of the scalar potential $V_{\mathrm{scalar}}$ in Appendix \ref{Higgspotential}. Indeed, in the
minimization condition of $V_{\mathrm{scalar}}$, we put $v_{\phi_{1l}}= v_{\phi_{2l}}=v_{\phi_{3l}}=v_{l}$,\, $v_{\phi_{1\nu}}= v_{\phi_{3\nu}}=0,\, v_{\phi_{2\nu}}=v_{\nu}$,\,  $v_{\phi_{2s}}=0,\,  v_{\phi_{1s}}=v_{\phi_{3s}}=v_{s}$ and $v^*_{H}=v_{H}, \, v^*_{l}=v_{l}, \, v^{*}_{\nu}=v_{\nu},\, v^*_{\eta}=v_{\eta},\, v^*_{\eta_s}=v_{\eta_s}$ and $ v^*_{s}=v_{s}$
 which leads to
\bea
\frac{\partial V_{\mathrm{scalar}}}{\partial v^*_k} &=&\frac{\partial V_{\mathrm{scalar}}}{\partial v_k},\hs \frac{\partial^2 V_{\mathrm{scalar}}}{\partial v^{*2}_k}=\frac{\partial^2 V_{\mathrm{scalar}}}{\partial v^2_k} \,\, (v_k=v_H, v_{l}, v_{\nu}, v_{\eta}, v_{\eta_s}, v_s), \label{eqreduced}
\eea
and the scalar potential minimum condition reduces to
\bea
\mu_H^2  + 2 \lambda^{H} v_H^2 + 3\lambda^{H\phi_l} v_l^2 + \lambda^{H\phi_\nu}v_\nu^2+ \lambda^{H\eta}v_\eta^2 +
 2\lambda^{H\phi_s}v_s^2 + 2 \Lambda^{H\phi_l \phi_\nu \phi_s} v_\nu +\lambda^{H\eta_s} v_{\eta_s}^2=0, && \label{eq1}\\
3 \mu_{\phi_l}^2 + 3\lambda^{H\phi_l}v_H^2 + 6\lambda^{\phi_l}v_l^2+ 3\lambda^{\phi_l\eta}v_\eta^2 + \lambda^{\phi_l\phi_\nu}v_\nu^2 +\lambda^{\phi_l\phi_s} v_s^2+ 2 \Lambda^{H\phi_l\phi_\nu \eta\phi_s}_1 v_\nu \hspace{0.65 cm}&&\crn
+\, (3 \lambda^{\phi_l\eta_s} + 2 \lambda^{\phi_l \phi_\nu \eta_s} v_\nu) v_{\eta_s}^2 =0,&& \label{eq2}\\
(\mu_{\phi_\nu}^2 + \lambda^{H\phi_\nu} v_H^2 + \lambda^{\phi_l\phi_\nu} v_l^2 +
    2\lambda^{\phi_\nu} v_\nu^2+ \lambda^{\phi_\nu \eta} v_\eta^2   -\lambda^{\phi_\nu\phi_s}v_s^2)  v_\nu
     + \Lambda^{H\phi_l\phi_\nu \eta \phi_s}_{2} v_l^2 \hspace{0.65 cm}&&\crn
    +\, \Lambda^{H\phi_l\phi_\nu \eta \phi_s}_{3} v_s^2 + v_{\eta_s}^2 (\lambda^{\phi_l \phi_\nu \eta_s} v_l^2 + \lambda^{\phi_\nu \eta_s} v_\nu + \lambda^{\phi_\nu \phi_s \eta_s} v_s^2)=0,&& \label{eq3}\\
\mu_\eta^2 + \lambda^{H\eta}v_H^2  + 3\lambda^{\phi_l\eta}v_l^2 +
 \lambda^{\phi_\nu \eta}v_\nu^2 + 2 \lambda^{\eta} v_\eta^2+ 2\lambda^{\phi_s\eta}v_s^2 + 2 \Lambda^{\phi_l \phi_\nu\eta\phi_s} v_\nu + \lambda^{\eta\eta_s} v_{\eta_s}^2 =0,&&\label{eq4}\\
2\mu_s^2  +2 \lambda^{H\phi_s} v_H^2 + \lambda^{\phi_l\phi_s}v_l^2 - \lambda^{\phi_\nu\phi_s}v_\nu^2
+ 2\lambda^{\phi_s\eta} v_\eta^2+ 2\lambda^{\phi_s}v_s^2+ 2 \Lambda^{H\phi_l\phi_\nu\eta \phi_s}_{3} v_\nu \hspace{0.65 cm} &&\crn
+\, 2  (\lambda^{\phi_s\eta_s}+ \lambda^{\phi_\nu \phi_s \eta_s} v_\nu) v_{\eta_s}^2=0,&&\label{eq5}\\
\mu_{\eta_s}^2 + \lambda^{\eta\eta_s} v_\eta^2 +
 2 \lambda^{\eta_s} v_{\eta_s}^2 + \lambda^{H\eta_s} v_H^2 +
 3 \lambda^{\phi_l\eta_s} v_l^2 + \lambda^{\phi_\nu\eta_s} v_\nu^2 +
 2 \lambda^{\phi_s\eta_s} v_s^2 \hspace{0.65 cm}&&\crn
+ 2  (\lambda^{\phi_l\phi_\nu \eta_s} v_l^2 + \lambda^{\phi_\nu \phi_s\eta_s} v_s^2) v_\nu=0, &&\\
\lambda^{H} v_H> 0, \hspace{0.1cm}
 \lambda^{\phi_l} v_l>0,\hspace{0.1cm}
\lambda^{\eta} v_\eta>0,\hspace{0.1cm}
\lambda^{\phi_s} v_s> 0, \hspace{0.1cm}\lambda^{\eta_s} v_{\eta_s}>0,&&\label{ineq1}\\
\mu_\nu^2 + \lambda^{\phi_\nu \eta} v_\eta^2 + \lambda^{H\phi_\nu} v_H^2 + \lambda^{\phi_l\phi_\nu} v_l^2 + 6 \lambda^{\phi_\nu} v_\nu^2 - \lambda^{\phi_\nu \phi_s} v_s^2 + \lambda^{\phi_\nu \eta_s} v_{\eta_s}^2>0,&&\label{ineq2}
\eea
where, the following notations have been introduced:
\bea
&&\Lambda^{H\phi_l \phi_\nu \phi_s}=\lambda^{H\phi_l \phi_s}_1 v_l^2 + \lambda^{H\phi_s\phi_\nu}_1 v_s^2, \hs
 \Lambda^{\phi_l \phi_\nu \eta\phi_s}=\lambda^{\phi_l\phi_\nu \eta}_1 v_l^2 + \lambda^{\phi_\nu \eta\phi_s}_1 v_s^2, \crn
&&\Lambda^{H\phi_l\phi_\nu \eta \phi_s}_1=\lambda^{\phi_l\phi_\nu \eta}_1 v_\eta^2 + \lambda^{H\phi_l\phi_s}_1 v_H^2+\lambda^{\phi_l\phi_\nu \phi_s}v_s^2, \,\,
\Lambda^{H\phi_l\phi_\nu \eta \phi_s}_{2}=\lambda^{\phi_l\phi_\nu\eta}_1 v_\eta^2+ \lambda^{H\phi_l\phi_s}_1 v_H^2, \crn
&&\Lambda^{H\phi_l\phi_\nu \eta \phi_s}_{3}=\lambda^{\phi_\nu \eta\phi_s}_1 v_\eta^2 + \lambda^{H\phi_s\phi_\nu}_1 v_H^2 + \lambda^{\phi_l \phi_\nu \phi_s}v_l^2,  \label{Lamix}\\
%%%%%%%%%%%%%%%%%%%%
&&\lambda^{H\eta} = \lambda^{H\eta}_1 + \lambda^{H\eta}_2,\hs
\lambda^{H\phi_l} = \lambda^{H\phi_l}_1 + \lambda^{H\phi_l}_2,\hs
\lambda^{H\phi_\nu} =\lambda^{H\phi_\nu}_1 + \lambda^{H\phi_\nu}_2, \crn
&&\lambda^{H\phi_s} =\lambda^{H\phi_s}_1 +\lambda^{H\phi_s}_2,\hs
\lambda^{\phi_l\eta} = \lambda^{\phi_l\eta}_1 + \lambda^{\phi_l\eta}_2, \hs
\lambda^{\phi_s\eta} =\lambda^{\phi_s\eta}_1 +\lambda^{\phi_s\eta}_2, \crn
&&\lambda^{H\phi_l} = \lambda^{H\phi_l}_1 + \lambda^{H\phi_l}_2, \hs
\lambda^{\phi_l} = 3 \lambda^{\phi_l}_1 + 4 \lambda^{\phi_l}_4, \crn
&&\lambda^{\phi_l\phi_\nu} = 3 \lambda^{\phi_l\phi_\nu}_1 + \lambda^{\phi_l\phi_\nu}_6 + \lambda^{\phi_l\phi_\nu}_7 +
 \lambda^{\phi_l\phi_\nu}_8 + 2\lambda^{\phi_l\phi_\nu}_9-2\lambda^{\phi_l\phi_\nu}_{10},\crn
&&\lambda^{\phi_l\phi_s} = 6 \lambda^{\phi_l\phi_s}_1 + 4 \lambda^{\phi_l\phi_s}_4
+ 4\lambda^{\phi_l\phi_s}_6 + \lambda^{\phi_l\phi_s}_7+ \lambda^{\phi_l\phi_s}_8+ 6 \lambda^{\phi_l\phi_s}_9 -2 \lambda^{\phi_l\phi_s}_{10},\crn
&&\lambda^{\phi_\nu\phi_s} = - 2\lambda^{\phi_\nu\phi_s}_1 +\lambda^{\phi_\nu\phi_s}_2 + \lambda^{\phi_\nu\phi_s}_3- 2 \lambda^{\phi_\nu\phi_s}_9+2 \lambda^{\phi_\nu\phi_s}_{10}, \crn
&&\lambda^{\phi_\nu \eta} =\lambda^{\phi_\nu \eta}_1 + \lambda^{\phi_\nu \eta}_2, \,
\lambda^{\phi_\nu} =\lambda^{\phi_\nu}_1 +\lambda^{\phi_\nu}_2 + \lambda^{\phi_\nu}_3, \,
\lambda^{\phi_s} = \lambda^{\phi_s}_2 + \lambda^{\phi_s}_3 + 4 (\lambda^{\phi_s}_1+ \lambda^{\phi_s}_4),\crn
&&\lambda^{\phi_l \phi_\nu \phi_s} = 2 \lambda^{\phi_l \phi_\nu \phi_s}_1 + 3 \lambda^{\phi_l \phi_\nu \phi_s}_3 + 2 (\lambda^{\phi_l \phi_\nu \phi_s}_5 + \lambda^{\phi_l \phi_\nu \phi_s}_7)\Blue{,} \crn
%\eea
%\bea
&&\lambda^{H\eta_s}=\lambda^{H\eta_s}_1+\lambda^{H\eta_s}_2, \hs \lambda^{\phi_l \eta_s}=\lambda^{\phi_l\eta_s}_1+\lambda^{\phi_l\eta_s}_2,\hs
\lambda^{\phi_\nu \eta_s}=\lambda^{\phi_\nu \eta_s}_1+\lambda^{\phi_\nu \eta_s}_2,\crn
&&\lambda^{\eta\eta_s}=\lambda^{\eta\eta_s}_1 + \lambda^{\eta\eta_s}_2, \hs \lambda^{\phi_s\eta_s}=\lambda^{\phi_s\eta_s}_1 + \lambda^{\phi_s\eta_s}_2.
\eea
Assuming that $\mu_{H},\,\mu_{\phi_l},\, \mu_{\phi_\nu},\, \mu_{\eta}$ and $\mu_{s}$ are negative and of the same order of magnitude as that of the SM \cite{PDG2020}\footnote{In th SM \cite{PDG2020}, $|\mu|=88.4$ GeV; thus, we use $|\mu_H|\sim |\mu_\phi|\sim|\mu_l|\sim|\mu_\nu|=10^2$\, GeV for their scales.},
\bea
\mu^2_{H}=\mu^2_{\phi_l}= \mu^2_{\phi_\nu}=\mu^2_{\eta}=\mu^2_{\eta_s}=\mu^2_{s} = - 10^4\, \mathrm{GeV}. \label{assume}
\eea
Using (\ref{SMscale})-(\ref{vevsingletscale}), (\ref{assume}) and the following benchmark point
\bea
&&\lambda^{H\phi_l} = \lambda^{H\phi_\nu} =\lambda^{H\eta} = \lambda^{H\phi_s}
 =\lambda^{\phi_l\phi_\nu}= \lambda^{\phi_l\eta} =  \lambda^{\phi_l\phi_s}=\lambda^{\phi_\nu\eta} \crn
 &&\hspace{0.9cm}= \lambda^{\phi_\nu\phi_s} =\lambda^{\phi_s\eta}=\lambda^{H\eta_s} = \lambda^{\phi_l\eta_s} =
  \lambda^{\phi_\nu\eta_s} = \lambda^{\eta\eta_s} = \lambda^{\phi_s\eta_s}= \lambda^{x},\crn
&&\lambda^{\phi_l \phi_\nu\phi_s} =\lambda^{H\phi_l\phi_s}_1 = \lambda^{\phi_l\phi_\nu \eta}_1 = \lambda^{\phi_\nu\eta\phi_s}_1 =\lambda^{H\phi_s\phi_\nu}_1 =\lambda^{\phi_l\phi_\nu \eta_s} = \lambda^{\phi_\nu \phi_s\eta_s}= \lambda^{y}, \label{benchmarkpoint1}
\eea
the quantities $\frac{\partial^2 V_{\mathrm{scalar}}}{\partial \langle \Phi \rangle^2}\, (\Phi =H, \phi_l, \phi_\nu, \eta, \phi_s , \eta_s)$ depend on two parameters $\lambda^{x}$ and $\lambda^{y}$ with $\lambda^{x} \in (-10^{-2}, -10^{-3})$ and $\lambda^{y} \in (-10^{-3}, -10^{-4})$. %which is plotted in Fig. \ref{gilaxlay}.
Fig.\ref{gilaxlay} implies that\footnote{As an example, in the case of $\lambda^{x}=-10^{-3}$ and $\lambda^{y}=-10^{-4}$ we get $\frac{\partial^2 V_{\mathrm{scalar}}}{\partial \langle H\rangle^2}\sim 10^{43},\hs \frac{\partial^2 V_{\mathrm{scalar}}}{\partial \langle \phi_l\rangle^2}\sim \frac{\partial^2 V_{\mathrm{scalar}}}{\partial \langle \phi_s\rangle^2}$ $\sim 10^{34}$, $\frac{\partial^2 V_{\mathrm{scalar}}}{\partial \langle \phi_\nu\rangle^2}\sim 10^{53}$ and $\frac{\partial^2 V_{\mathrm{scalar}}}{\partial \langle \eta \rangle^2} \sim \frac{\partial^2 V_{\mathrm{scalar}}}{\partial \langle \eta_s \rangle^2}  \sim 10^{42}$.}, the inequalities in (\ref{ineq1}) and (\ref{ineq2}) are always satisfied by the solutions of Eqs. (\ref{eq1})--(\ref{eq5}) as given in Appendix \ref{solution}. This proves that the VEV alignments in Eq. (\ref{scalarvev}) is the natural solution of the minimum condition of $V_{\mathrm{scalar}}$ as expected.
%%%%%%%%%%%%%%%%%%%%%
%%%%%%%%%%%%%%%%%
\section{\label{ASmixing}3+1 active - sterile neutrino mixing}
%\subsection{Charged -lepton sector\label{clep}}
With the help of (\ref{scalarvev}) and the Clebsch-Gordan coefficients of $A_4$ group \cite{ishi}, from Eq. (\ref{Lylep0}) we get the charged lepton masses and the corresponding mixing matrices:
% the mass Lagrangian for the charged leptons can be rewritten in the form
%\begin{equation}
%\mathcal{L}_{\mathrm{clep}}^{\mathrm{mass}}=-(\bar{l}_{1L},\bar{l}_{2L},\bar{l}_{3L})M_{l}(l_{1R},l_{2R},l_{3R})^{T}+H.c,
%\end{equation}%
%where
%\bea
%M_{l}&=&\left(
%\begin{array}{ccc}
%2x_{1cl} v_H\frac{v^2_{l}}{\Lambda^2} &x_{2cl}v_H\frac{ v_{l}}{\Lambda}& x_{3cl}v_H \frac{ v^*_{l}}{\Lambda}\\
%2x_{1cl} v_H\frac{v^2_{l}}{\Lambda^2}&\om x_{2cl}v_H\frac{ v_{l}}{\Lambda} &\om^2  x_{3cl}v_H \frac{ v^*_{l}}{\Lambda}\\
%2x_{1cl} v_H\frac{v^2_{l}}{\Lambda^2}&\om^2  x_{2cl}v_H\frac{ v_{l}}{\Lambda} &\om x_{3cl}v_H \frac{ v^*_{l}}{\Lambda}\\
%\end{array}%
%\right), \label{Mlep}
%\eea
%which can be diagonalized as
%\bea U^\dagger_L M_l U_R=\sqrt{3} v_H \mathrm{diag}\left(2x_{1cl} \frac{v^2_{l}}{\Lambda^2}, \, x_{2cl}\frac{ v_{l}}{\Lambda}, \, x_{3cl}\frac{ v^*_{l}}{\Lambda}\right)\equiv
% \mathrm{diag}(m_e, \, m_\mu , \,m_\tau),\label{Mld}\eea
%where
\bea
m_{e} &=&2\sqrt{3} x_{1cl} v_H \left(\frac{v_l}{\Lambda}\right)^2, \hs m_{\mu} =\sqrt{3} x_{2cl} v_H \frac{v_l}{\Lambda}, \hs
 m_{\tau} =\sqrt{3} v_H x_{3cl} \frac{v_l}{\Lambda},  \label{memt}\\
 U^+_L&=&\fr{1}{\sqrt{3}}\left(%
\begin{array}{ccc}
  1 &\,\,\, 1 &\,\,\, 1 \\
  1 &\,\,\, \om^2 &\,\,\, \om \\
  1 &\,\,\, \om &\,\,\, \om^2 \\
\end{array}%
\right),\hs U_R=1, \hs \om=e^{i2\pi/3}. \label{Uclep}\eea
The Eq. (\ref{memt}) tells us that $m_e$ %the electron mass
is suppressed by a factor of $v_H \left(\frac{v_l}{\Lambda}\right)^2 \sim 10^{-4}\, \mathrm{GeV}$ compared to $m_\mu$ %the muon
and $m_\tau$ %the tau masses
 of the order of $\frac{v_H v_l}{\Lambda} \sim 10^{-1}\, \mathrm{GeV}$. Thus, the considered model naturally explain the charged-lepton mass hierarchy.

By comparing $m_{e,\mu,\tau}$ in (\ref{memt}) with the data %experimental values  for the masses of  the charged leptons
given in Ref. \cite{PDG2020}, $m_e\simeq 0.511 \,\mathrm{MeV},  m_\mu \simeq  105.66\,\mathrm{MeV}$ and $m_\tau \simeq  1776.86 \,\mathrm{MeV}$,  along with the help of Eqs. (\ref{SMscale})-(\ref{vevsingletscale}), %putting the VEV of the SM Higgs $v_H=174 \, \mathrm{GeV}$ and assuming that the VEV of $\phi_l$ is at a very high scale $v_l=10^{10}\, \mathrm{GeV}$,
we obtain
\bea
 x_{1cl}=0.848, \hs  x_{2cl}=0.351,\hs  x_{3cl}=5.90.
\eea
%\subsection{Neutrino masses and mixings\label{neutrino}}
%\newpage
Coming back to the neutrino sector. Using the Clebsch-Gordan coefficient property of $A_4$ group \cite{ishi}, from Eq. (\ref{Lylep0}), the Lagrangian for the neutrino sector becomes
\bea
 -\mathcal{L}_{\nu}&=&x_{1\nu} \left(\bar{\psi}_{1L} \nu_{1R} +\bar{\psi}_{2L}\nu_{2R}+\bar{\psi}_{3L}\nu_{3R}\right)\widetilde{H} \crn
 &+& \frac{x_{2\nu}}{\Lambda} \left[(\bar{\psi}_{2L}\nu_{3R}+\bar{\psi}_{3L}\nu_{2R})\widetilde{H}\phi_{1\nu}+(\bar{\psi}_{3L}\nu_{1R}+\bar{\psi}_{1L}\nu_{3R})\widetilde{H}\phi_{2\nu}+
(\bar{\psi}_{1L}\nu_{2R}+\bar{\psi}_{2L}\nu_{1R})\widetilde{H}\phi_{3\nu}\right] \crn
 &+& \frac{x_{3\nu}}{\Lambda}\left[(\bar{\psi}_{2L}\nu_{3R}-\bar{\psi}_{3L}\nu_{2R})\widetilde{H}\phi_{1\nu}
 + (\bar{\psi}_{3L}\nu_{1R}-\bar{\psi}_{1L}\nu_{3R})\widetilde{H}\phi_{2\nu}
 +(\bar{\psi}_{1L}\nu_{2R}-\bar{\psi}_{2L}\nu_{1R})\widetilde{H}\phi_{3\nu}\right]\crn
 &+&\frac{y_\nu}{2}\left(\bar{\nu}^c_{1R} \nu_{1R}+\bar{\nu}^c_{2R}\nu_{2R}+\bar{\nu}^c_{2R} \nu_{3R}\right)\eta
 +  \frac{y_s}{\Lambda}  \left(\bar{\nu}^c_{s}\nu_{1R} \eta_{\Blue{s}}\phi_{1s}+\bar{\nu}^c_{s}\nu_{2R} \eta_{\Blue{s}}\phi_{2s}+\bar{\nu}^c_{s}\nu_{3R} \eta_{\Blue{s}}\phi_{3s}\right)\crn
 &+& \mathrm{H.c}.\label{Lny}\eea
%With the VEV given in Eq. (\ref{scalarvev}), a
After symmetry breaking, we find the $7\times 7$ neutrino mass matrix in the basis $(\nu_{L},\nu^c_{R}, S^c)$ as follows
\bea
M_{\nu}^{7\times7} = \begin{pmatrix}
0     & M_{D} &0 \\
M^T_{D}  & M_{R} &M^T_S \\
0 & M_S   & 0
\end{pmatrix}, \label{Mnu77}
\eea
where $M_{D}, M_{R}$ and $M_S$ %are the Dirac, Majorana and sterile neutrino mass matrices that
take the form:
\bea M_D&=&
v_H\left(%
\begin{array}{ccc}
 x_{1\nu}                                                                               & 0   & \frac{v_\nu}{\Lambda}\left(x_{2\nu} -x_{3\nu} \right)\\
 0                                                                                             & x_{1\nu} & 0 \\
\frac{v_\nu}{\Lambda}\left(x_{2\nu} +x_{3\nu} \right)  & 0   & x_{1\nu} \\
\end{array}%
\right),  \crn
 M_R&=&y_\nu v_\eta \mathbf{I}, \hs\hs M_s= \frac{y_s}{\Lambda}v_{\eta_\Blue{s}} v_s (1 \hspace{0.35 cm} 0\hspace{0.35 cm} 1). \label{MDRS}\eea
 In the case where  $M_R \gg M_S > M_D$, %similar to the typical type-I seesaw model, one can block-diagonalize the full mass matrix $7\times 7$ by using the seesaw formula,
 the effective $4\times4$ light neutrino
mass matrix in the basis $(\nu_L, S^c)$ is constructed by the following expression \cite{Barry2011, Zhang2012, Das2019ea, Krishnan2020}:
\begin{equation}
M_{\nu}= -\left( \begin{array}{cc}
              M_{D} M^{-1}_R M^T_D &  M_D M^{-1}_R M^T_S\\
              M_S (M^{-1}_R)^TM^T_{D} & M_SM^{-1}_{R}M^T_S
                      \end{array} \right).
\label{Mnu44}
\end{equation}
The expression (\ref{MDRS}) shows that, there are five complex parameters thus has ten real parameters in the neutrino sector. %However, not all of them are physical.
Considering the case of real VEV for the singlet scalars $\phi_\nu, \, \phi_s$ and $\eta$, the phase redefinition of the leptonic fields $\psi_L, \nu_R$ and $\nu_s$ allows to rotate away the phase of four neutrino Yukawa couplings $x_{2\nu}, x_{3\nu}, y_{\nu}$ and $y_s$, reducing to six physical parameters. On the other hand, one parameter is absorbed by defining the hermitian matrix. Thus, in the neutrino sector, there are five real parameters, including $\al_1=\mathrm{arg}(x_{1\nu})$, $k_{2, 3}, m_{0}$ and $m_s$ which are defined are defined as follows:
\bea
&&k_2=\frac{v_\nu x_{2\nu}}{\Lambda |x_{1\nu}|},\hs k_3=\frac{v_\nu x_{3\nu}}{\Lambda |x_{1\nu}|}, \hs m_0=\frac{v^2_{H} |x_{1\nu}|^2}{v_\eta y_\nu}, \hs  m_{s}=\frac{2 v^2_{\eta_s} v^2_{s} y_{s}^2}{v_{\eta} \Lambda^2 y_\nu}. \label{k2k3m0ms}
\eea
Combining Eqs.(\ref{MDRS}) and (\ref{Mnu44}) yields:% the $4\times 4$ active-sterile mass matrix in the explicit form as follows
\bea
 &&M^{4\times4}_{\nu} =-\left(
\begin{array}{cccc}
\hspace{-0.1 cm} m_{0}\left(\begin{array}{ccc}
e^{2 i \al_1} + (k_2 - k_3)^2 \hs  & 0 & 2k_2 e^{i \al_1} \\
                       0 & e^{2 i \al_1} & 0\\
2k_2 e^{i \al_1}& 0 &\hs e^{2 i \al_1} + (k_2 + k_3)^2\\
                      \end{array}\right) & \hspace{-0.1 cm} \sqrt{\frac{m_{0} m_{s}}{2}}\left(\begin{array}{c}
                    e^{i \al_1}+(k_2-k_3) \\
                     0\\
                  e^{i \al_1}+(k_2+k_3)\\
                      \end{array}\right)\\
\hspace{-0.75 cm}\sqrt{\frac{m_{0} m_{s}}{2}} \left(\begin{array}{ccc}
                     e^{i \al_1}+(k_2-k_3)\hspace{0.115 cm}& 0\hspace{0.05 cm}&\hs  e^{i \al_1}+(k_2+k_3) \\
                      \end{array}\right) & \quad\quad\quad \hs m_s \\
\end{array}\right),\crn
 \label{M44separate}
\eea
where $\al_1=\mathrm{arg}(x_{1\nu})$, and $k_{2,3 }, m_0$ and $m_s$ are defined in Eq. (\ref{k2k3m0ms}).

Suppose that $M_D < M_s$, we can therefore apply the type-I seesaw mechanism on Eq.(\ref {Mnu44}) to obtain the active neutrino mass matrix as \cite{Barry2011, Zhang2012, Das2019ea, Krishnan2020}
\bea
 M_\nu &=&  M_D M_R^{-1}M_s^T (M_s M_R^{-1} M_s^T)^{-1} M_s M_R^{-1} M_D^T - M_D M_R^{-1} M_D^T \crn
 &=& -\frac{m_0}{2}\left(
\begin{array}{ccc}
\left(e^{i \al_1}-(k_{2}-k_{3})\right)^2 & 0 &k_{3}^2-(e^{i \al_1}-k_{2})^2\\
 0 & 2e^{2i \al_1} & 0 \\
k_{3}^2-(e^{i \al_1}-k_{2})^2 & 0 & \left(e^{i \al_1}-(k_{2}+k_{3})\right)^2 \\
\end{array}
\right).\label{Mnu33}
\eea
Since $k_{2,3}$ and $m_0$ are real, $M_\nu$ is complex. To diagonalise the matrix $M_\nu$, let us define a Hermitian matrix $\mathbf{m}^{2}_{\nu}$, given by
\bea
\mathbf{m}^{2}_{\nu}=& M_\nu M^\+_{\nu}
=\frac{m^2_0}{2}  \left(%
\begin{array}{ccc}
 A_1+A_2&0 & -B_1+i B2  \\
 0          &2 & 0  \\
-B_1-i B_2& 0 & A_1-A_2\\
\end{array}%
\right),\label{Mnu33}
\eea
where
\bea
&&A_1=\left(1 + k_2^2 + k_3^2 - 2 k_2 \cos\al_1\right)^2,\crn
&&A_2 = 2 k_3 (\cos\al_1-k_2) (1 + k_2^2 + k_3^2 - 2 k_2 \cos\al_1),\crn
&&B_1 = (1 + k_2^2 - k_3^2 - 2 k_2 \cos\al_1) (1 + k_2^2 + k_3^2 - 2 k_2 \cos\al_1),\crn
&&B_2 = 2 k_3 (1 + k_2^2 + k_3^2 - 2 k_2 \cos\al_1) \sin\al_1. \label{A12B12}
\eea

The matrix $\mathbf{m}^{2}_{\nu}$ in Eq. (\ref{Mnu33}) is diagonalized by %the rotation matrix
$U_{\nu}$ satisfying
\be
U_{\nu }^+ \mathbf{m}^{2}_{\nu} U_{\nu }=\left\{
\begin{array}{l}
\left(%
\begin{array}{ccc}
0& 0 &0 \\
0 & m^2_0 & 0 \\
0& 0 & k^2_0 m^2_0\\
\end{array}%
\right),\hspace{0.1cm} U_{\nu }=\left(%
\begin{array}{ccc}
\frac{1}{x_1+i x_2} & 0 &\hs -\frac{1}{y_1+i y_2} \\
 0 & 1 & 0 \\
 \frac{1}{x_0}\hs & 0 & \frac{1}{y_0}\\
\end{array}%
\right) \hspace{0.2cm}\mbox{for NH,}\ \  \\
\left(%
\begin{array}{ccc}
k^2_0 m^2_0& 0 &0 \\
0 & m^2_0 & 0 \\
0& 0 & 0\\
\end{array}%
\right),\hspace{0.1cm} U_{\nu }=\left(%
\begin{array}{ccc}
-\frac{1}{y_1+i y_2} & 0 &\hs \frac{1}{x_1+i x_2} \\
 0 & 1 & 0 \\
 \frac{1}{y_0}\hs & 0 & \frac{1}{x_0}\\
\end{array}%
\right) \hspace{0.2cm}\mbox{for IH,}
\end{array}%
\right.  \label{Unu}
\ee
thus the light neutrino masses are
\bea
\left\{
\begin{array}{l}
m^2_1=0, \hspace{0.725 cm} m^2_2 = m^2_0, \hspace{0.1 cm} m^2_3=m^2_0 k^2_0\hspace{0.2cm}\mbox{for  NH},    \\
m^2_1=m^2_0 k^2_0, \hspace{0.05 cm} m^2_2 = m^2_0, \hspace{0.1 cm} m^2_3=0\hspace{0.8cm}\,\mbox{for IH},
\end{array}%
\right. \label{m1m2m3}
\eea
where
\bea
&& k_0=1 + k_2^2 + k_3^2 - 2 k_2 \cos\al_1, \label{k0expres}\\
&& x_0=\sqrt{\frac{2 k_0}{k_0+2k_3 (\cos\al_1-k_2)}},\hspace{0.05cm} y_0=\sqrt{\frac{2k_0}{k_0+2k_3 (k_2-\cos\al_1)}},\\
&& x_1=\frac{(k_0-2k^2_3) x_0}{k_0+2k_3 (k_2 -\cos\al_1)},\hspace{0.45cm} y_1=\frac{(k_0-2k^2_3) y_0}{k_0+2k_3 (\cos\al_1-k_2)},\\
&& x_2=\frac{2k_3 x_0 \sin\al_1}{k_0+2k_3 (k_2 -\cos\al_1)}, \hspace{0.45cm} y_2=\frac{2 k_3 y_0\sin\al_1}{k_0+2k_3 (\cos\al_1-k_2)}. \label{atheta}
\eea
%%%%%%%%%%%%
The corresponding leptonic mixing matrix is
\be
U_{lep}=U_{L}^{\dag} U_{\nu }=\left\{
\begin{array}{l}
\fr{1}{\sqrt{3}}\left( \begin{array}{ccc}
\frac{1}{x_0}+\frac{1}{x_1+i x_2}      &1 & \frac{1}{y_0}-\frac{1}{y_1+i y_2} \\
\frac{\om}{x_0}+\frac{1}{x_1+i x_2}\hs &\om^2 &\hs \frac{\om}{y_0} - \frac{1}{y_1 + i y_2}  \\
\frac{\om^2}{x_0}+\frac{1}{x_1+i x_2}\hs &\om &\hs \frac{\om^2}{y_0} - \frac{1}{y_1 + i y_2}  \\
\end{array}\right) \hspace{0.2cm}\mbox{for \, NH},  \label{Ulep}  \\
\fr{1}{\sqrt{3}}\left( \begin{array}{ccc}
\frac{1}{y_0}-\frac{1}{y_1+i y_2}   &1 & \frac{1}{x_0}+\frac{1}{x_1+i x_2}    \\
 \frac{\om}{y_0} - \frac{1}{y_1 + i y_2}\hs &\om^2 &\hs \frac{\om}{x_0}+\frac{1}{x_1+i x_2}  \\
 \frac{\om^2}{y_0} - \frac{1}{y_1 + i y_2} \hs &\om &\hs \frac{\om^2}{x_0}+\frac{1}{x_1+i x_2} \\
\end{array}\right) \hspace{0.2cm}\mbox{for \, IH},
\end{array}%
\right.
\ee
which being being the $\mathrm{TM}_2$ form. Eq. (\ref{m1m2m3}) implies the neutrino mass ordering should be either $(0, m_0, m_0 k_0)$ or $(m_0 k_0, m_0, 0)$. It is very interesting to note that depending on the sign of $\cos\alpha_1$, $k_0$ can be greater or less than 1 so both NH ($m_1<m_2$) and IH ($m_2<m_1$) are consistent with the experimental data given that $k_2$ and $k_3$ are real parameters. Threrefore, our model predicts both NH and IH of the active neutrino masses which is in consistent with the data taken from Ref. \cite{Salas2020} and different from that of Ref. \cite{Krishnan2020}.

%%%%%%%%%%%%
 The lepton mixing angles, in the three-neutrino scheme, is defined as \cite{PDG2020}
\bea &&s_{13}^2=\left| \mathrm{U}_{e 3}\right|^2=\left\{
\begin{array}{l}
\fr{2}{3} \frac{k_3^2}{k_0}\hspace{0.3cm}\mbox{for \, NH},    \\
\frac{2}{3}\left(1-\frac{k^2_3}{k_0}\right)\hspace{0.25cm}\,\mbox{for \, IH},
\end{array}%
\right.  \label{s13sq}\\
&& s_{12}^2 =\frac{\left| \mathrm{U}_{e 2}\right|^2}{1-\left| \mathrm{U}_{e 3}\right|^2}=
\frac{1}{3c^2_{13}}\hspace{0.25cm}\,\mbox{for \, both \,NH\, and \, IH}, \label{s12sq}\\
&& s_{23}^2=\frac{\left| \mathrm{U}_{\mu 3}\right|^2}{1-\left| \mathrm{U}_{e 3}\right|^2}=
\left\{
\begin{array}{l}
\fr{1}{2}+\frac{\sqrt{3} k_3 \sin \al_1}{3k_0-2k_3^2}\hspace{0.3cm}\mbox{for \, NH},    \\
\fr{1}{2}-\frac{\sqrt{3} k_3 \sin \al_1}{k_0+2k_3^2}\hspace{0.25cm}\,\mbox{for \, IH},
\end{array}%
\right. \label{s23sq}\eea
where $\theta_{ij}$ are neutrino mixing angles and $t_{12}=s_{12}/c_{12}$, $t_{23}=s_{23}/c_{23}$, $c_{ij}=\cos \theta_{ij}$, $s_{ij}=\sin \theta_{ij}$.

The Jarlskog invariant, %constraining the size of CP violation in lepton sector,
determined from Eq. (\ref{Ulep}), takes the form \cite{Jarlskog1, Jarlskog2, Jarlskog3, PDG2020}
\bea
J_{CP}&=&\mathrm{Im} (U_{12} U_{23} U^*_{13} U^*_{22})=\left\{
\begin{array}{l}
\frac{k_3 (\cos\al_1-k_2)}{3\sqrt{3}k_0} \hspace{0.3cm}\mbox{for \, NH},    \\
\frac{k_3 (k_2 - \cos\al_1)}{3\sqrt{3}k_0} \hspace{0.25cm}\,\mbox{for \, IH}.
\end{array}%
\right. \label{Jm}
\eea
From Eqs. (\ref{Unu}) and (\ref{s13sq})-(\ref{Jm}), we can express $m_0, k_0, k_{2, 3}$ and $\sin\delta_{CP}$ in terms of four observables $\Delta m^2_{21}, \Delta m^2_{31}$, $s^2_{23}$ and $s^2_{13}$ and $\alpha_1$ as follows:
\bea
&&m^2_{0}=\left\{
\begin{array}{l}
\Delta m^2_{21}\hspace{1.575cm}\mbox{for NH},    \\
\Delta m^2_{21}-\Delta m^2_{31} \hspace{0.1cm}\,\mbox{for\,  IH},
\end{array}%
\right. \label{m0v}\\
&&k^2_{0}=\left\{
\begin{array}{l}
\frac{\Delta m^2_{31}}{m^2_0}\hspace{0.4cm}\mbox{for \, NH},    \\
\frac{-\Delta m^2_{31}}{m^2_0}\hspace{0.1cm}\,\mbox{for \, IH},
\end{array}%
\right. \label{k0v}\\
&&k_{2}=\cos\alpha_1\mp \sqrt{k_0-k_3^2-\sin^2\alpha_1} \hspace{0.25cm}\mbox{for\ both\ NH\ and IH,} \label{k2}\\
&&k^2_{3}=\left\{
\begin{array}{l}
\frac{3}{2}k_0 s^2_{13}\hspace{1.2cm}\mbox{for \, NH},    \\
\frac{k_0}{2} \left(2-3 s^2_{13}\right)
\hspace{0.05cm}\,\mbox{for \, IH},
\end{array}%
\right. \label{k3}\\
%&&\sin\alpha_1=\left\{
%\begin{array}{l}
%\frac{\sqrt{k_0} c^2_{13} (2 s^2_{23}-1)}{\sqrt{2} s_{13}}\hspace{0.55cm}\mbox{for \, NH},    \\
%\sqrt{\frac{3 k_0}{2}}\frac{c^2_{13} (1-2 s^2_{23})}{\sqrt{2-3 s^2_{13}}}\hspace{0.1cm}\,\mbox{for \, IH},
%\end{array}%
%\right. \label{sinal1}\\
&&\sin\delta_{CP}=%\frac{J_{CP}}{c_{12} c_{13}^2 c_{23} s_{12} s_{13} s_{23}}=
\frac{3J_{CP}}{s_{13} s_{23} c_{23} \sqrt{2-3s^2_{13}}} \hspace{0.25cm}\mbox{for\ both\ NH\ and IH,}\label{sd}
\eea
with $J_{CP}$ is determined in Eq. (\ref{Jm}).

Using the approximation $M_D < M_s$, we obtain the mass of the $4^\text{th}$ mass eigenstate \cite{Barry2011, Zhang2012, Das2019ea, Krishnan2020},
\bea
 m_4= - M_s M_R^{-1}M_s^T=m_s, \label{m4}
\eea
with $m_s$ is defined in Eq. (\ref{k2k3m0ms}). Combining Eqs. (\ref{m1m2m3}) and (\ref{m4}) yields:
\bea
m_{s}=m_4=\left\{
\begin{array}{l}
\sqrt{\Delta m^2_{41}}\hspace{0.4cm}\mbox{for  NH},    \\
\sqrt{\Delta m^2_{41}+m^2_0 k^2_0} \hspace{0.1cm}\,\mbox{for IH}.
\end{array}%
\right. \label{msm4}
\eea

The $4\times 4$ light neutrino mixing matrix is given by \cite{Barry2011, Zhang2012, Das2019ea, Krishnan2020}:
\bea
U =U^\dagger_L U_\nu= \left(\begin{array}{cc}
       U^\dagger_L(1-\frac{1}{2}R R^\dagger) U_\nu & U^\dagger_L R \\
       -R^\dagger U_\nu & 1-\frac{1}{2}R^\dagger R
       \end{array}\right), \label{U44}
\eea
where the three-component column vector $R$ is given by \cite{Barry2011, Zhang2012, Das2019ea, Krishnan2020}
\bea
  R = M_D M_R^{-1}M_s^T (M_s M_R^{-1}M_s^T)^{-1}=\sqrt{\frac{m_0}{2 m_s}}\left(
\begin{array}{c}
 e^{i \al_1}+ (k_2-k_3) \\
 0 \\
e^{i \al_1}+ (k_2+k_3) \\
\end{array}
\right). \label{Rmatrix}
\eea
Combining Eqs. (\ref{Uclep}) and (\ref{Rmatrix}), the strength of the active-sterile mixing is given by
\bea
U^\dagger_L R&=&\frac{1}{\sqrt{6}}\sqrt{\frac{m_0}{m_s}}\left(
\begin{array}{c}
 2 (e^{i \al_1}+ k_2)  \\
 \frac{1}{2}\left[\left(1-i \sqrt{3}\right) \left(k_2+e^{i \al_1}\right)-\left(3+i \sqrt{3}\right) k_3\right] \\
 \frac{1}{2}\left[\left(1+i \sqrt{3}\right) \left(k_2+e^{i \al_1}\right)-\left(3-i \sqrt{3}\right) k_3\right] \\
\end{array}
\right)\equiv \left(
\begin{array}{c}
U_{e4}  \\
U_{\mu 4} \\
U_{\tau 4} \\
\end{array}
\right), \label{Uemutau4}
\eea
which leads to
\bea
&& |U_{14}|^2=\frac{2}{3}\frac{m_{0}}{m_{s}} \left(1+ k_{2}^2+2 k_{2} \cos\alpha_1\right), \label{u14sq}\\
&&|U_{2 4}|^2=\frac{1}{6}\frac{m_{0}}{m_{s}}\left(1+ k_{2}^2+3 k_3^2+2 k_{2} \cos\alpha_1 -2 \sqrt{3} k_{3} \sin\alpha_1 \right), \label{u24sq}\\
&& |U_{3 4}|^2=\frac{1}{6}\frac{m_{0}}{m_{s}}\left(1+ k_{2}^2+3 k_3^2+2 k_{2} \cos\alpha_1 + 2 \sqrt{3} k_{3} \sin\alpha_1\right). \label{u34sq}
\eea
The effective neutrino masses \cite{betdecay1,betdecay2,betdecay3}%which can in principle determine the absolute neutrino mass scale
, determined from Eqs. (\ref{m1m2m3}), (\ref{Ulep}), (\ref{m4}) and (\ref{Uemutau4}), possess the following forms\footnote{Eqs. (\ref{meeexpr}) and (\ref{mbexpr}) are satisfied for both NH and IH.}:
\bea
&&\langle m_{ee}\rangle = \left| \sum^4_{i=1} U_{ei}^2 m_i \right|=\frac{m_0}{3} \sqrt{\frac{\Gamma_{\mathrm{e}}}{k_0+2k_3 (k_2-2\cos\alpha_1)}}, \label{meeexpr}\\
&& m_{\beta }= \sqrt{\sum^4_{i=1} \left|U_{ei}\right|^2 m_i^2}= \frac{m_0}{\sqrt{3}} \sqrt{1+2 k_0 k_3^2 +\frac{2m_s}{m_0}\left(1+k_2^2+2k_2 \cos\alpha_1\right)}, \label{mbexpr}\\
%\eea
%\bea
&&\langle m^{(3\nu)}_{ee}\rangle = \left| \sum^3_{i=1} U_{ei}^2 m_i \right|=\frac{m_0}{3} \sqrt{\frac{\Gamma^{(3)}_{\mathrm{e}}}{k_0+2k_3 (k_2-2\cos\alpha_1)}}, \label{meeexpr33}\\
&& m^{(3\nu)}_{\beta }= \sqrt{\sum^3_{i=1} \left|U_{ei}\right|^2 m_i^2}= \frac{m_0}{\sqrt{3}} \sqrt{1 + 2 k_0 k_3^2 }, \label{mbexpr33}
\eea
with
\bea
\Gamma_{\mathrm{e}}&=&5 + (1 + 8 k_2^2 + 4 k_2^4)k_2^2 +2  [4 k_2^2 (3 + k_2^2)-7] k_2k_3 + [12 (2 + k_2^2) k_2^2+5] k_3^2 \crn
 &+& 8 (2 k_2^2+1) k_2 k_3^3 + 4 (3 k_2^2+2) k_3^4 + 4(2 k_2  + k_3)k_3^5
 + 2 \left[4 (k_2^3  + 3 k_2^2 k_3 + 2 k_3^3) k_2^2\right.\crn
  &-&\left.  7 k_3+ (8 k_3^2-4) k_2^3 +
     (5 + 8 k_3^2 + 4 k_3^4)k_2 - 4 (1 + k_3^2)k_3^3\right] \cos\alpha_1\crn
    &-& 4 \left[(2 k_2^4-1) +  (3 + 2 k_3^2) k_2^2 + 4  (1 + k_3^2) k_2 k_3 - 2 (1 + k_3^2) k_3^2\right] \cos 2\alpha_1 \crn
    &-& 4 \left[(1 + 2 k_2^2) (k_2 + k_3) + 4 k_3^3\right] \cos 3\alpha_1 +
 8 k_3^2 \cos 4 \alpha_1,\\
\Gamma^{(3)}_{\mathrm{e}}&=&%1 + k_3^2 + 8 k_3^4 + 4 k_3^6 +
% 2 k_2 k_3 (1 + 2 k_3^2)^2 + k^2_2(1 + 2 k_3^2)^2 -
% 2 (k_2 + k_3) (1 + 2 k_3^2)^2 \cos\alpha_1 + 4 k_3^2 \cos 2\alpha_1
1 + 4 k_3^2 \left(k^2_3+\cos 2\alpha_1\right) +
(1 + 2 k_3^2)^2 \left[(k_2 + k_3)^2 -2 (k_2 + k_3)\cos\alpha_1\right].
\eea
%where $k_0, k_2, k_3$ and $\alpha_1$ are defined in Eqs. (\ref{k0v})--(\ref{sinal1}).

Although the NH seems favored by some global analysis\cite{Capozzi2017, Salas2018, Esteban2019}, however, it is the fact that the neutrino mass spectrum is currently unknown and it can be normal ($m_1< m_2 < m_3$) or inverted ($m_3< m_1< m_2$) ordering depending on the sign of $\Delta m^2_{31}$ \cite{Salas2020, Kelly2021}. In the next section, we will show that our model can explain the stirile-active neutrino mas and mixing as well as the recent three neutrino oscillation data given in Table \ref{experconstrain} in the minimal extended seesaw framework for both NH and IH.

\section{\label{NR} Numerical analysis}
In this work, eleven independent experimentally measured quantities in the neutrino sector, including $\sin^2 \theta_{12}$, $\sin^2 \theta_{23}$, $\sin^2 \theta_{13}$, $\sin \delta$, $\Delta m^2_{21}$, $\Delta m^2_{31}$, $\langle m_{ee} \rangle$, $\Delta m^2_{41}$, $|U_{14}|^2$, $|U_{24}|^2$ and $|U_{34}|^2$, can be expressed in terms of five model parameters including $k_2$, $k_3$, $m_0$, $m_s$ and $\al_1$. We calculate the model parameters using the experimental data and also make predictions. To calculate the allowed ranges of the model parameters $k_2$, $k_3$, $m_0, m_s$ and $\al_1$ as well as predictive ranges for the experimental parameters $\sin^2 \theta_{12}$, $\sin \delta$, $\langle m_{ee} \rangle$, $|U_{14}|^2$, $|U_{24}|^2$ and $|U_{34}|^2$, we utilize the observables $\Delta m^2_{21}$, $\Delta m^2_{31}, \Delta m^2_{41}$, $\sin^2 \theta_{13}$ and $\sin^2 \theta_{23}$ whose experimental values are shown in Table \ref{experconstrain}.

Firstly, using the $3\sigma$ experimental range of %neutrino mass-squared splittings
$\Delta m^2_{21}$ and $\Delta m^2_{31}$ taken from Tab. \ref{experconstrain}, we get the constraints
\bea
&&m^2_0\in \left\{
\begin{array}{l}
 (69.4, 81.4)\, \mathrm{meV}^2 \hspace{0.25cm}\mbox{for  NH},  \\
(2439, 2611) \, \mathrm{meV}^2\hspace{0.25cm}\mbox{for IH},
\end{array}%
\right. \label{constraintm0}\\
&&k^2_0\in \left\{
\begin{array}{l}
 (30.34, 37.90)\, \mathrm{meV} \hspace{0.25cm}\mbox{for  NH},  \\
(0.967, 0.973) \, \mathrm{meV}\hspace{0.25cm}\mbox{for IH},
\end{array}%
\right. \label{constraintk0}
\eea
In $3\sigma$ range of $s_{13}$ \cite{Salas2020},  $2.000\times 10^{-2}<s^2_{13}<2.405\times 10^{-2}$ for NH and  $2.018\times 10^{-2}<s^2_{13}<2.424\times 10^{-2}$ for IH; thus, Eq. (\ref{s12sq}) implies that
\bea
&&s^2_{12}\in \left\{
\begin{array}{l}
(0.3401, 0.3415) \hspace{0.25cm}\mbox{for  NH},  \\
(0.3402, 0.3416) \hspace{0.25cm}\mbox{for IH},
\end{array}%
\right. \label{s12range}
\eea
i.e.,
\bea
&&\theta_{12} (^\circ)\in \left\{
\begin{array}{l}
(35.68, 35.76) \hspace{0.25cm}\mbox{for  NH},  \\
(35.68, 35.77)\hspace{0.25cm}\mbox{for IH},
\end{array}%
\right. \label{theta12range}
\eea
which all belong to $3\sigma$ range of $s^2_{12}$ \cite{Salas2020} as given in Tab. \ref{experconstrain}.

\begin{figure}[h]
\begin{center}
%\vspace{0.15 cm}
\hspace{-3.0 cm}\includegraphics[width=0.65\textwidth]{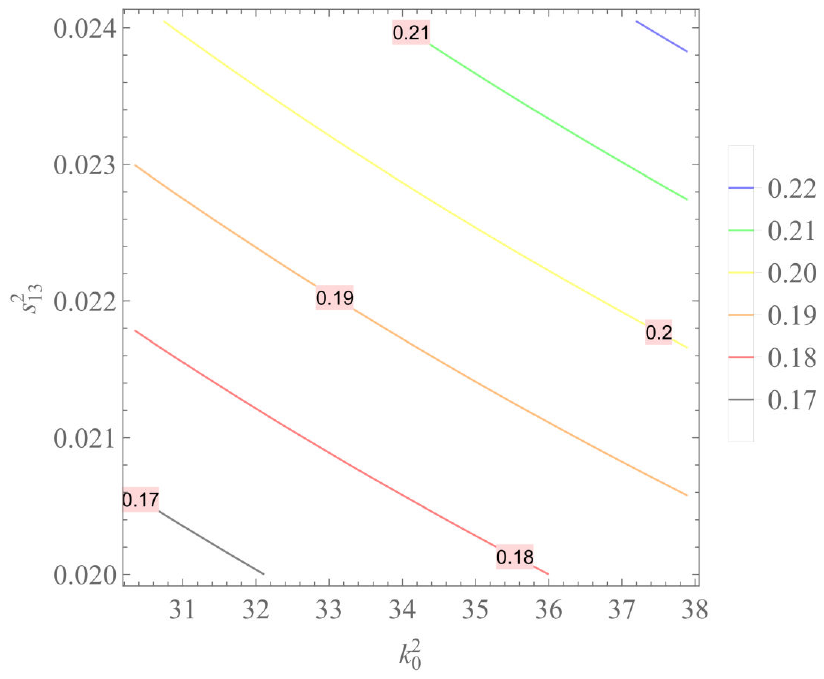}\hspace{-3.35 cm}
\includegraphics[width=0.65\textwidth]{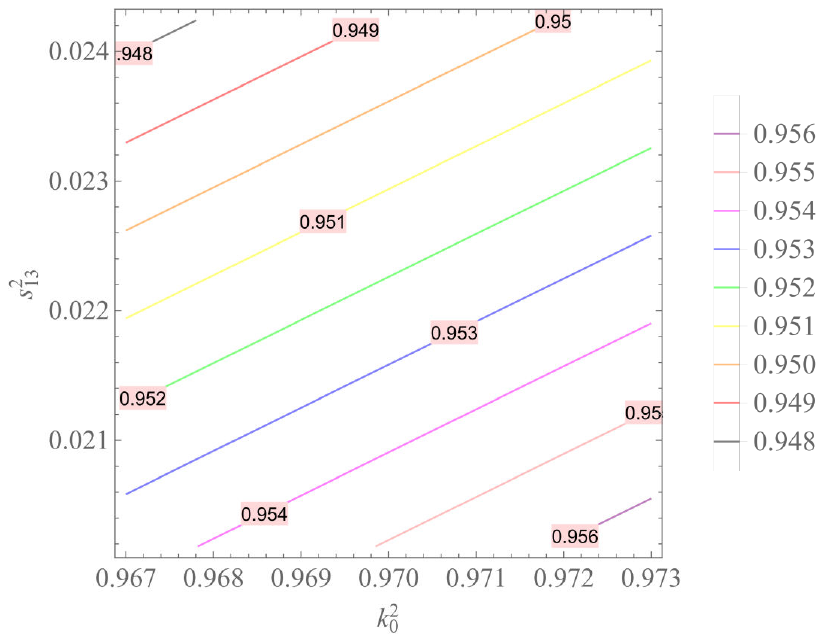}\hspace*{-3.5 cm}
\end{center}
\vspace{-9.5 cm}
\caption{The contour plot of $k^2_3$ as a function of $k^2_0$ and $s^2_{13}$ with $k^2_0 \in (30.34, 37.9)$ and $s^2_{13}\in (2.00, 2.405) 10^{-2}$ for NH (left panel) while $k^2_0 \in (0.967, 0.973)$ and $s^2_{13}\in (2.018, 2.424) 10^{-2}$ for IH (right panel).}
\label{k3F}
\end{figure}

\begin{center}
\begin{figure}[h]
%\begin{center}
\vspace{-0.5 cm}
\hspace{-3.0 cm}
\includegraphics[width=0.65\textwidth]{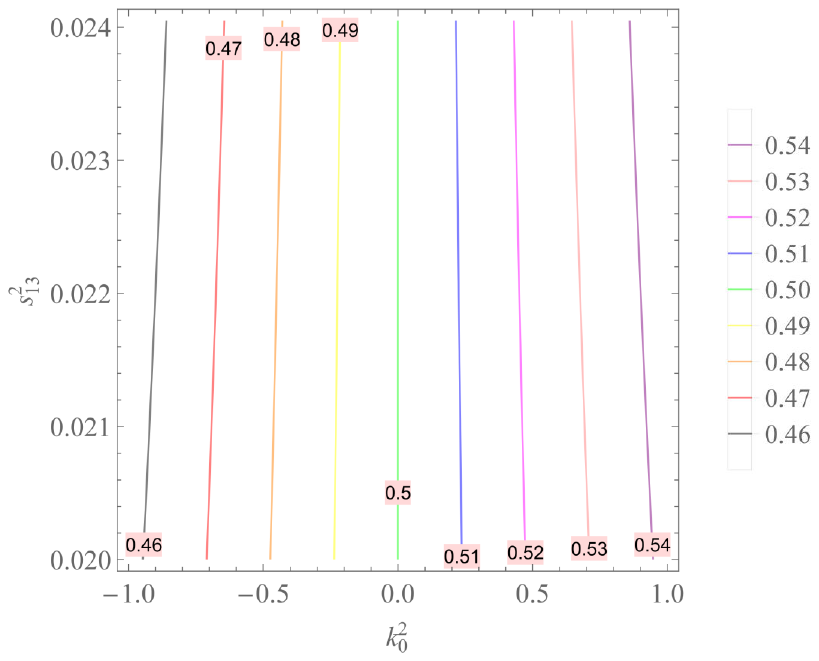}\hspace{-3.35 cm}
\includegraphics[width=0.65\textwidth]{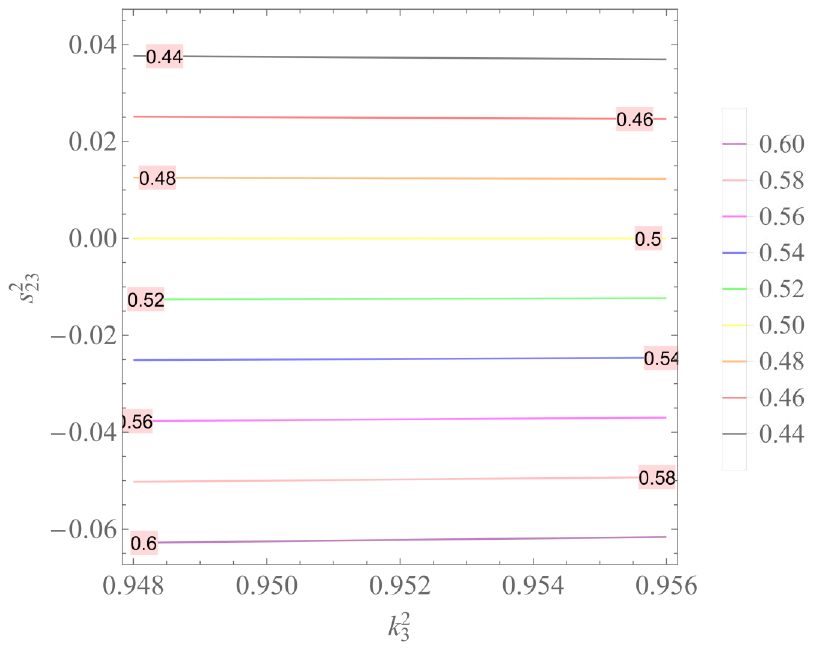}\hspace{-3.5 cm}
%\end{center}
\vspace{-9.0 cm}
\caption{$s^2_{23}$ versus $k^2_3$ and $\sin\alpha_1$ with $k^2_{3}\in (0.17, 0.22)$ and $\sin\alpha_1\in (-1,\, 1)$ for NH (left panel) while $k^2_{3}\in (0.948, 0.956)$ and $\sin\alpha_1\in (-0.065,\, 0.045)$ for IH (right panel).}
\label{s23F}
\end{figure}
\end{center}

\begin{center}
\begin{figure}[h]
%\begin{center}
\vspace{-0.75 cm}
\hspace{-2.0 cm}
\includegraphics[width=0.65\textwidth]{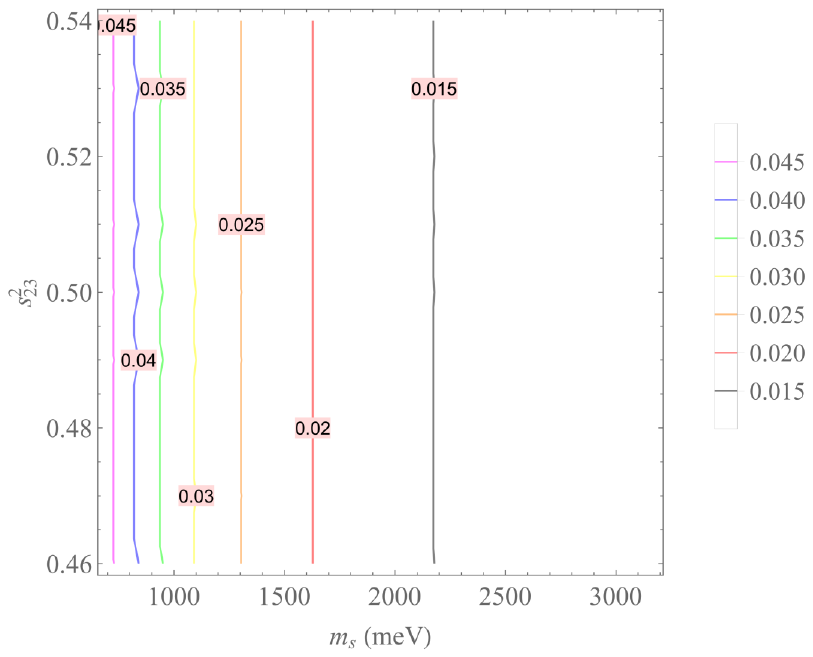}\hspace{-3.35 cm}
\includegraphics[width=0.65\textwidth]{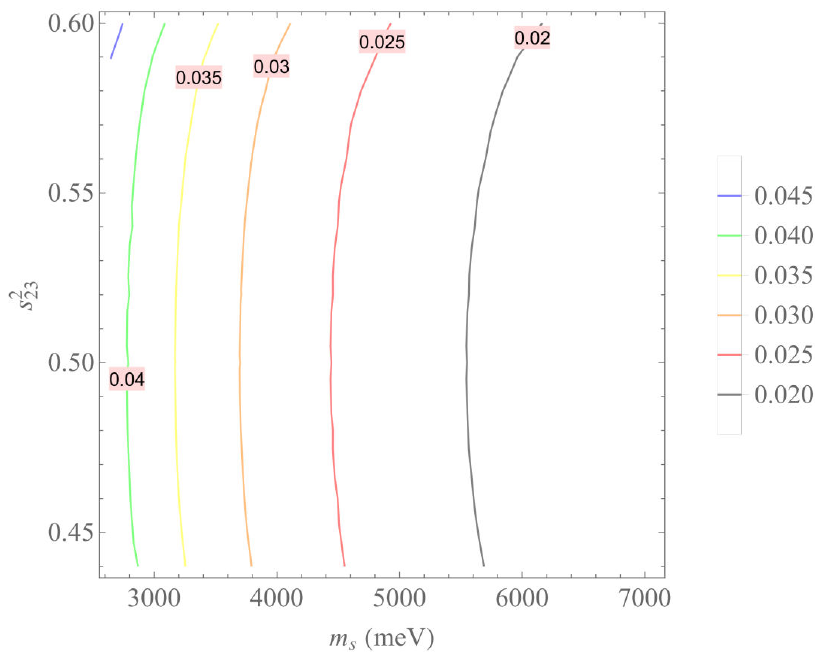}\hspace*{-2.2 cm}
%\end{center}
\vspace{-9.0 cm}
\caption{The contour plot of $|U_{1 4}|^2$ as a function of $m_s$ and $s^2_{23}$ with $m_s\in (707.11,\, 3162.28)\, \mathrm{meV} \, \left[\mathrm{i.e.}, \Delta m^2_{41} \in (0.5, \, 10.0)\, \mathrm{eV}^2\right]$ and $s^2_{23} \in (0.46, 0.54)$ for NH (left panel)
while $m_s \in (2646.22, 7071.24)\, \mathrm{meV}\, \left[\mathrm{i.e.}, \Delta m^2_{41} \in (7.0, \, 50.0)\, \mathrm{eV}^2\right]$ and $s^2_{23}\in (0.44, 0.60)$ for IH (right panel).}
\label{u14sqF}
\end{figure}
\end{center}

\begin{center}
\begin{figure}[h]
%\begin{center}
\vspace{-0.75 cm}
\hspace{-2.0 cm}
\includegraphics[width=0.65\textwidth]{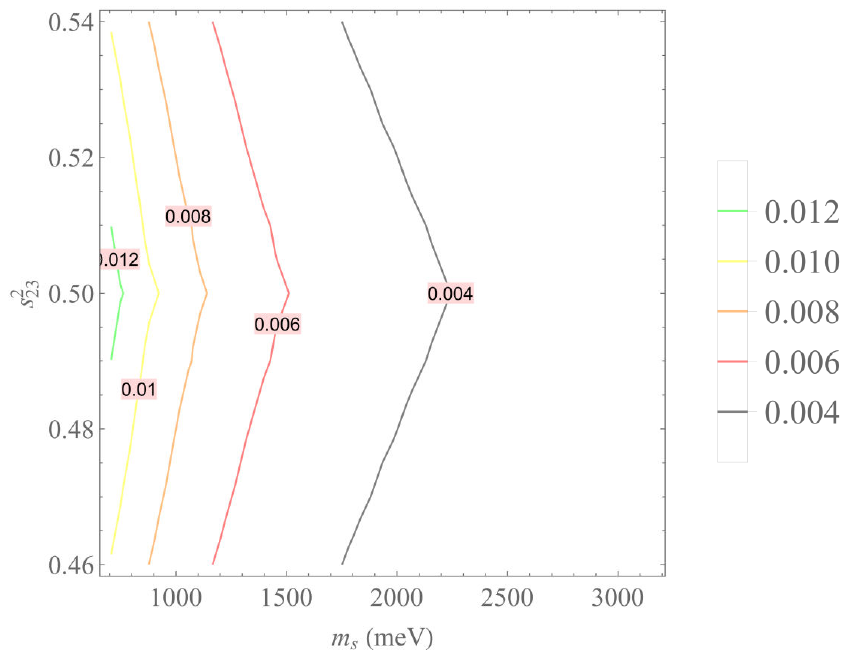}\hspace{-3.25 cm}
\includegraphics[width=0.65\textwidth]{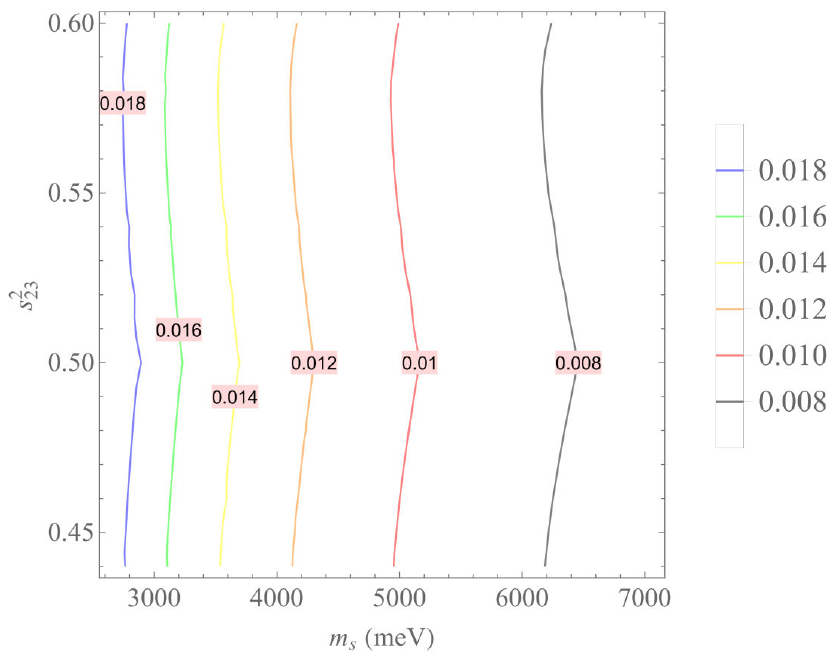}\hspace{-2.2 cm}
%\end{center}
\vspace{-9.0 cm}
\caption{The contour plot of $|U_{2 4}|^2$ as a function of $m_s$ and $s^2_{23}$ with $m_s\in (707.11,\, 3162.28)\, \mathrm{meV} \, \left[\mathrm{i.e.}, \Delta m^2_{41} \in (0.5, \, 10.0)\, \mathrm{eV}^2\right]$ and $s^2_{23} \in (0.46, 0.54)$ for NH (left panel)
while $m_s \in (2646.22, 7071.24)\, \mathrm{meV}\, \left[\mathrm{i.e.}, \Delta m^2_{41} \in (7.0, \, 50.0)\, \mathrm{eV}^2\right]$ and $s^2_{23}\in (0.44, 0.60)$ for IH (right panel).}
\label{u24sqF}
\end{figure}
\end{center}

\begin{center}
\begin{figure}[h]
%\begin{center}
\vspace{-0.45 cm}
\hspace{-2.0 cm}
\includegraphics[width=0.65\textwidth]{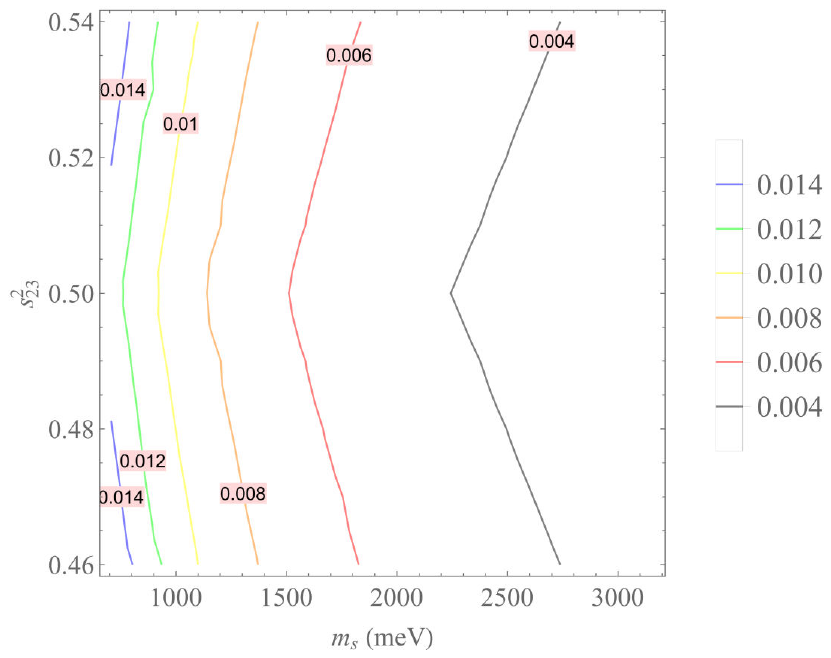}\hspace{-3.35 cm}
\includegraphics[width=0.65\textwidth]{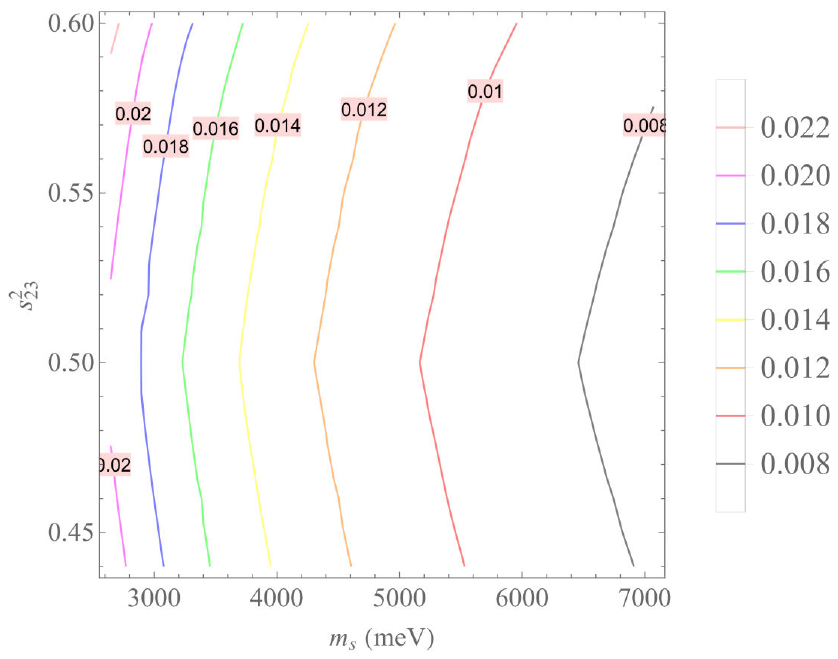}\hspace{-2.2 cm}
%\end{center}
\vspace{-9.0 cm}
\caption{The contour plot of $|U_{3 4}|^2$ as a function of $m_s$ and $s^2_{23}$ with $m_s\in (707.11,\, 3162.28)\, \mathrm{meV} \, \left[\mathrm{i.e.}, \Delta m^2_{41} \in (0.5, \, 10.0)\, \mathrm{eV}^2\right]$ and $s^2_{23} \in (0.46, 0.54)$ for NH (left panel) while $m_s \in (2646.22, 7071.24)\, \mathrm{meV}\, \left[\mathrm{i.e.}, \Delta m^2_{41} \in (7.0, \, 50.0)\, \mathrm{eV}^2\right]$ and $s^2_{23}\in (0.44, 0.60)$ for IH (right panel).}
\label{u34sqF}
\end{figure}
\end{center}

Next, using the constraints of $k_0$ in Eq. (\ref{constraintk0}) and the $3\sigma$  range of reactor mixing angle $s_{13}$, i.e., $s^2_{13}\in (2.00,\, 2.405)10^{-2}$ for NH and $s^2_{13}\in (2.018,\, 2.424)10^{-2}$
for IH, from Eq. (\ref{k3}), we can depict the dependence of $k^2_3$ on $k^2_0$ and $s^2_{13}$ as shown in Fig. \ref{k3F}, which implies that
\bea
&&k^2_{3}\in \left\{
\begin{array}{l}
 (0.17,\, 0.22) \hspace{0.25cm}\mbox{for  NH},  \\
(0.948, 0.956)\hspace{0.25cm}\mbox{for IH}.
\end{array}%
\right. \label{constraintk3}
\eea
%%%%%%%%%%%%%%%%%%%%%
%%%%%%%%%%%%%%%%%
Equation (\ref{s23sq}) shows that $s^2_{23}$ depends on $k_0, k_{3}$ and $\sin \alpha_1$. At the best-fit values of $\Delta m^2_{21}$ and $\Delta m^2_{31}$, we get
\bea
&&m^2_{0}= \left\{
\begin{array}{l}
75.0\, \mathrm{meV}^2 \hspace{0.25cm}\mbox{for  NH},  \\
2525.0\, \mathrm{meV}^2\hspace{0.25cm}\mbox{for IH},
\end{array}%
\right. \hs k^2_{0}= \left\{
\begin{array}{l}
34.0 \hspace{0.25cm}\mbox{for  NH},  \\
0.9703\hspace{0.25cm}\mbox{for IH},
\end{array}%
\right. \label{m0k0values}\\
%\eea
%\bea
&&\left\{
\begin{array}{l}
m_{1}=0\, \mathrm{meV},\hs m_{2}=8.66\, \mathrm{meV},\hs m_{3}=50.50\, \mathrm{meV} \hspace{0.25cm}\mbox{for  NH},  \\
m_{1}=49.50\, \mathrm{meV},\hs m_{2}=50.25\, \mathrm{meV},\hs m_{3}=0\, \mathrm{meV}\hspace{0.25cm}\mbox{for IH},
\end{array}%
\right.  \label{m1m2m3values}
\eea
and we find the ranges of $\sin \alpha_1$ that of $s^2_{23}$ lies in $3 \sigma$ range given in Table \ref{experconstrain} as  depicted in Fig. \ref{s23F} which implies
%with $s^2_{23}\in (0.46, 0.54)$ and $s^2_{13}\in (2.00,\, 2.405)10^{-2}$ for NH while $s^2_{23}\in (0.433, 0.608)$ and $s^2_{13}\in (2.018,\, 2.424)10^{-2}$ for IH, as  depicted in Fig. \ref{s1Fv}, which predicts:
 \bea
&&s^2_{23}\in \left\{
\begin{array}{l}
(0.46, 0.54) \hspace{0.25cm}\mbox{for  NH},  \\
(0.44, 0.60)\hspace{0.25cm}\mbox{for IH}.
\end{array}%
\right. \label{constraints23}
\eea
Further, Eqs. (\ref{Jm}) and (\ref{k2})-(\ref{sd}) impliy that, at the best-fit values of $\Delta m^2_{21}$ and $\Delta m^2_{31}$, $k_2$ and $\sin\delta_{CP}$ depend on two parameters $\theta_{13}$ and $\theta_{23}$ which are plotted in Figs. \ref{k2F1} and \ref{sdFv}, respectively. These figures indicates  that
\bea
&&k_{2}\in \left\{
\begin{array}{l}
(2.00, 3.40) \hspace{0.25cm}\mbox{for  NH},  \\
(0.68, 0.80) \hspace{0.25cm}\mbox{for IH},
\end{array}%
\right. \label{k2F1}\\
&&\sin\delta_{CP}\in \left\{
\begin{array}{l}
(-0.60, -0.20) \hspace{0.25cm}\mbox{for  NH},  \\
(-0.95, -0.60) \hspace{0.25cm}\mbox{for IH},
\end{array}%
\right. \label{sdFv}
\eea
\begin{center}
\begin{figure}[h]
%\begin{center}
\vspace{-0.5 cm}
\hspace{-2.0 cm}
\includegraphics[width=0.65\textwidth]{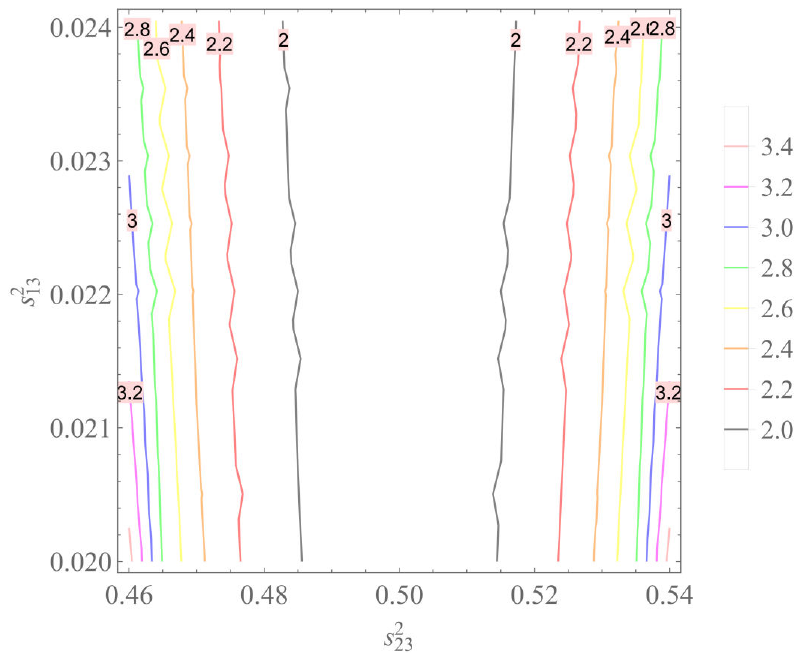}\hspace{-3.45 cm}
\includegraphics[width=0.65\textwidth]{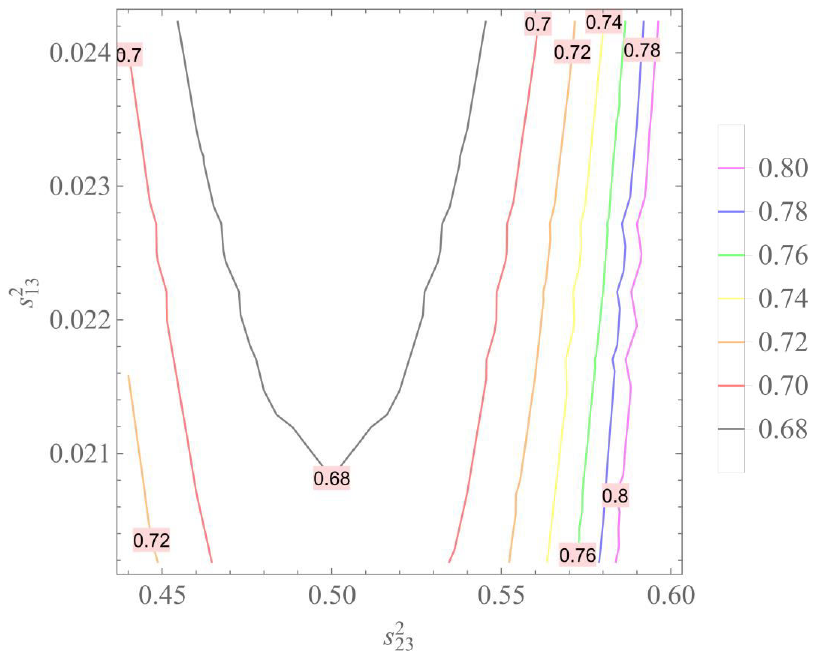}\hspace{-3.0 cm}
%\end{center}
\vspace{-9.25 cm}
\caption{$k_2$ versus $s^2_{13}$ and $s^2_{23}$ with $s^2_{13}\in (2.00, 2.405)\,10^{-2}$ and $s^2_{23}\in (0.46, 0.54)$ for NH (left panel) while $s^2_{13}\in (2.018, 2.424)\,10^{-2}$ and $s^2_{23}\in (0.44, 0.60)$ for IH (right panel).}
\label{k2F1}
\end{figure}
\end{center}
%%%%%%%%%%%%%%%%
\begin{center}
\begin{figure}[h]
%\begin{center}
\vspace{-1.25 cm}
\hspace{-2.5 cm}
\includegraphics[width=0.65\textwidth]{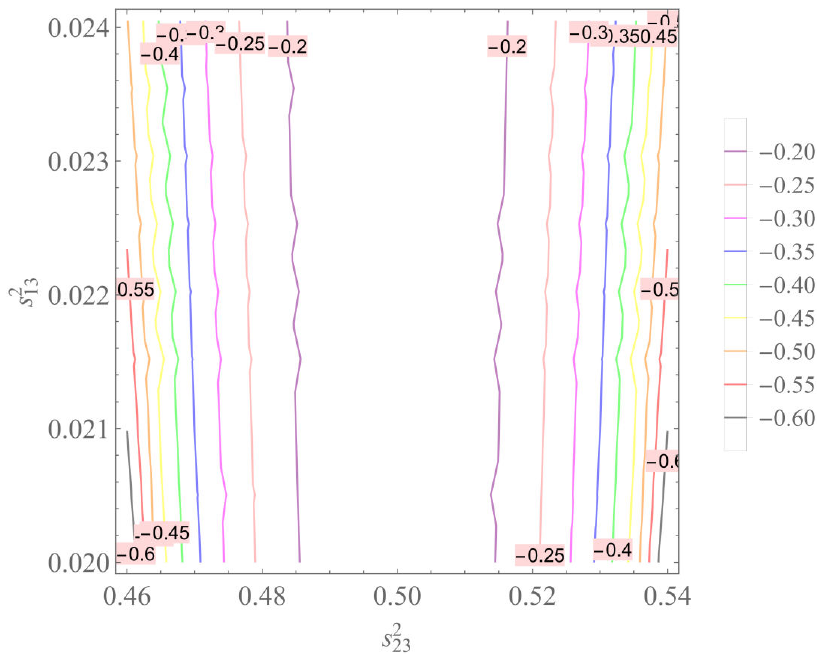}\hspace{-3.35 cm}
\includegraphics[width=0.65\textwidth]{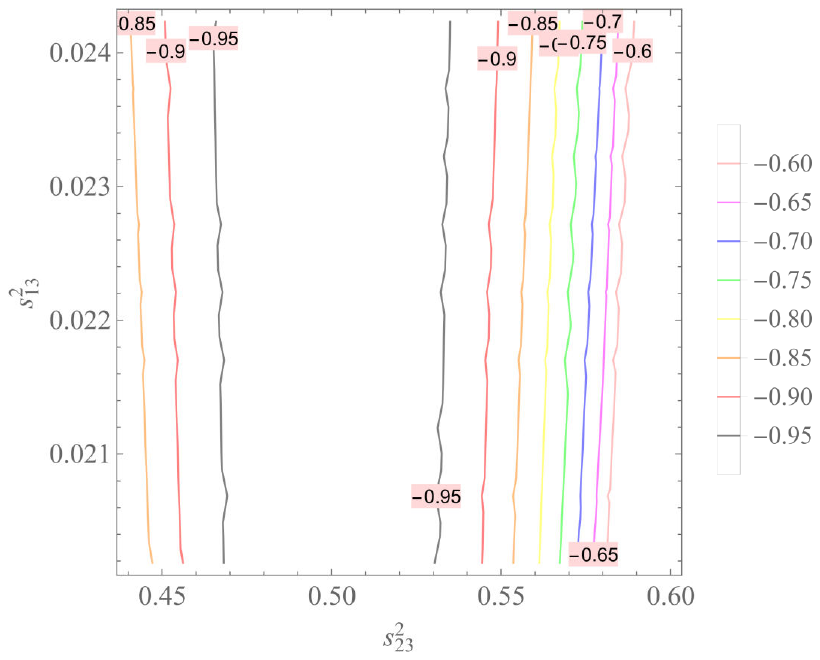}\hspace{-3.0 cm}
%\end{center}
\vspace{-9.0 cm}
\caption{$\sin\delta_{CP}$ versus $s^2_{13}$ and $s^2_{23}$ with $s^2_{13}\in (2.00, 2.405)\,10^{-2}$ and $s^2_{23}\in (0.46, 0.54)$ for NH (left panel) while $s^2_{13}\in (2.018, 2.424)\,10^{-2}$ and $s^2_{23}\in (0.44, 0.60)$ for IH (right panel).}
\label{sdFv}
\end{figure}
\end{center}
\vspace{-1.25 cm}
%%%%%%%%%%%%%%%%
At present, there are various experimental bounds on $\Delta m^2_{41}$ \cite{Aguilar2001, Arevalo2013, Acero2008, Arevalo2010, Mention2011, An2014, Arevalo2018, Gariazzo17, Adamson2019, Adamson2020, Aartsen2020, Beheraa2019, BeheraPRD2020}, for example, $\Delta m^2_{41} \in (0.01, 1.0)\, \mathrm{eV}^2$ \cite{Arevalo2013}, $\Delta m^2_{41}>10^{-2} \, \mathrm{eV}^2$ \cite{Adamson2019}, $\Delta m^2_{41} =0.041\, \mathrm{eV}^2$ \cite{Arevalo2018},  $\Delta m^2_{41}\in (0.1, 1.0)\, \mathrm{eV}^2$ \cite{Arevalo2010},  $\Delta m^2_{41} \in (0.2, 10.0)\, \mathrm{eV}^2$ \cite{Aguilar2001},   $\Delta m^2_{41} \geq 0.1\, \mathrm{eV}^2$ \cite{Acero2008}, $\Delta m^2_{41} =1.0\, \mathrm{eV}^2$ \cite{Beheraa2019, BeheraPRD2020}, $\Delta m^2_{41} > 1.5\, \mathrm{eV}^2$ \cite{Mention2011}, $\Delta m^2_{41} =1.7\, \mathrm{eV}^2$ \cite{Gariazzo17}, $\Delta m^2_{41} <10.0\, \mathrm{eV}^2$ \cite{Adamson2020} $\Delta m^2_{41} =1.45\, \mathrm{eV}^2$ \cite{Aartsen2020}. Expressions (\ref{s12sq})-%(\ref{Jm}), (\ref{m0v})-
(\ref{k3}), (\ref{m4}), (\ref{u14sq})-(\ref{u34sq}) implies that $|U_{14}|^2$, $|U_{24}|^2$ and $|U_{34}|^2$ depend on five experimental parameters $\Delta m^2_{21}$, $\Delta m^2_{31}, \Delta m^2_{41}$, $\sin^2 \theta_{13}$ and $\sin^2 \theta_{23}$. At the best-fit values of $\Delta m^2_{21}, \Delta m^2_{31}$ and $s^2_{13}$, three mixing elements $|U_{14}|^2$, $|U_{24}|^2$ and $|U_{3 4}|^2$ depend on two parameters $m_s$ (or $\Delta m^2_{41}$) and $\theta_{23}$. We found the possible values of $\Delta m^2_{41}$ and $\theta_{23}$ with $\Delta m^2_{41} \in (0.5, \, 10.0)\, \mathrm{eV}^2, s^2_{23} \in (0.46, 0.54)$ for NH and $\Delta m^2_{41} \in (7.0, \, 50.0)\, \mathrm{eV}^2, s^2_{23}\in (0.44, 0.60)$ for IH that satisfy the experimental constraints on $|U_{14}|^2$, $|U_{24}|^2$ and $|U_{3 4}|^2$  are shown in Table \ref{experconstrain} which are plotted in Figs. \ref{u14sqF}, \ref{u24sqF} and \ref{u34sqF}. Thus in this work, $\Delta m^2_{41}$
is chosen
%is assumed
in the range of $\Delta m^2_{41}\in (0.5, 10)\, \mathrm{eV}^2$ for NH while $\Delta m^2_{41}\in (7.0, 50.0)\, \mathrm{eV}^2$ for IH, and Eq. (\ref{msm4}) implies
%\newpage
\bea
m_s\in \left\{
\begin{array}{l}
(707.11,\, 3162.28)\, \mathrm{meV}\hspace{0.25cm}\mbox{for  NH},  \\
(2646.22,\, 7071.24)\, \mathrm{meV}\hspace{0.2cm}\mbox{for IH}.
\end{array}%
\right.
\eea
%%%%%%%%%%%%%%%%
%%%%%%%%%%%%%%%%
%%%%%%%%%%%%%%%%
\vspace{-3.0 cm}
Figures \ref{u14sqF}, \ref{u24sqF} and \ref{u34sqF} show that our model predicts the range of $|U_{e 4}|^2, |U_{\mu 4}|^2$ and $|U_{\tau 4}|^2$ as follows
\bea
&&|U_{1 4}|^2 \in\left\{
\begin{array}{l}
(0.015,\,0.045) \,\ \  \, \,\, \mbox{for \ \  NH,} \\
(0.020, \, 0.045) \,\ \,\,\,\,\,  \mbox{for \ \  IH,}%
\end{array}%
\right.   \label{msranges}\\
&&|U_{2 4}|^2\in\left\{
\begin{array}{l}
(0.004,\,\, 0.012)\ \ \    \mbox{for \ \ \ NH,} \\
(0.008,\, 0.018)\ \ \ \  \mbox{for \ \ \ IH,}%
\end{array}%
\right.   \label{U24ranges} \\
&&|U_{3 4}|^2\in\left\{
\begin{array}{l}
 (0.004,\, 0.014)\ \ \ \   \mbox{for \ \ \ NH,} \\
(0.008,\,\, 0.022)\ \ \ \  \mbox{for \ \ \ IH.}%
\end{array}%
\right.   \label{U34ranges}
\eea

\begin{figure}[h]
\begin{center}
\vspace{-0.35 cm}
\hspace{-3.0 cm}
\includegraphics[width=0.65\textwidth]{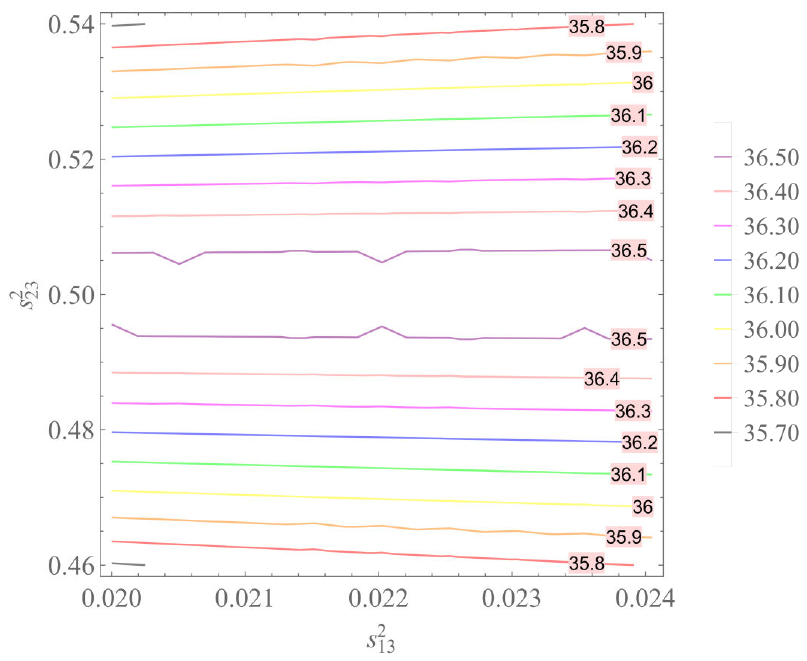}\hspace{-3.5 cm}
\includegraphics[width=0.65 \textwidth]{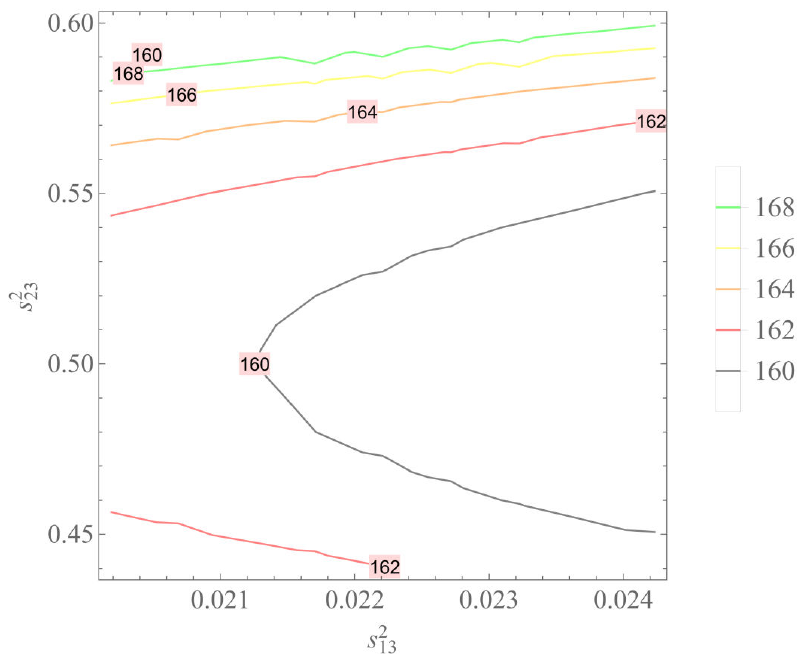}\hspace*{-3.5 cm}
\end{center}
\vspace{-9.25 cm}
\caption{$\langle m_{ee} \rangle$ versus $s^2_{13}$ and $s^2_{23}$ with $s^2_{13} \in (2.00, 2.405) 10^{-2}$ and $s^2_{23}\in (0.46, 0.54)$ for NH (left panel) while $s^2_{13} \in (2.018, 2.424) 10^{-2}$ and $s^2_{23}\in (0.44, 0.60)$ for IH (right panel).}
\label{meeF}
\end{figure}

\begin{figure}[h]
\begin{center}
\vspace{-0.5 cm}
\hspace{-3.0 cm}
\includegraphics[width=0.65\textwidth]{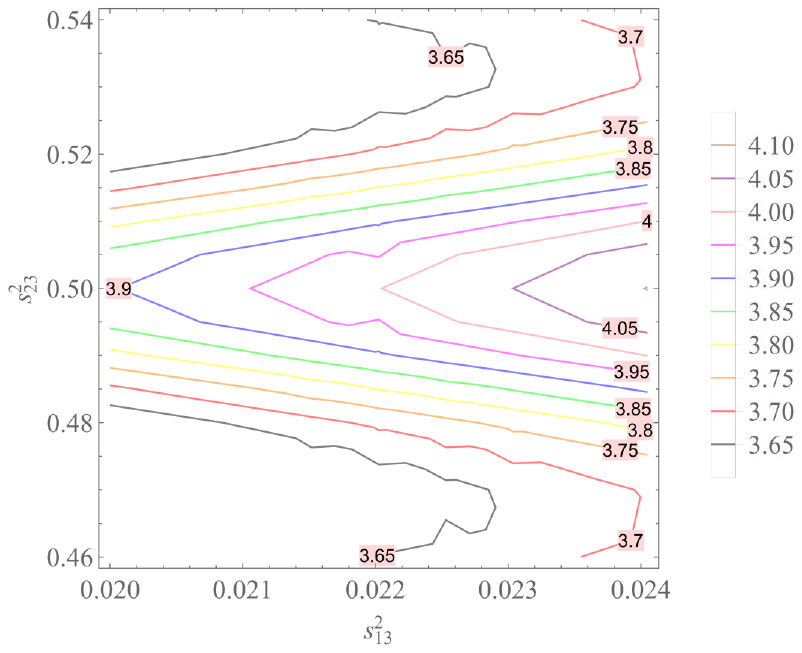}\hspace{-3.5 cm}
\includegraphics[width=0.65 \textwidth]{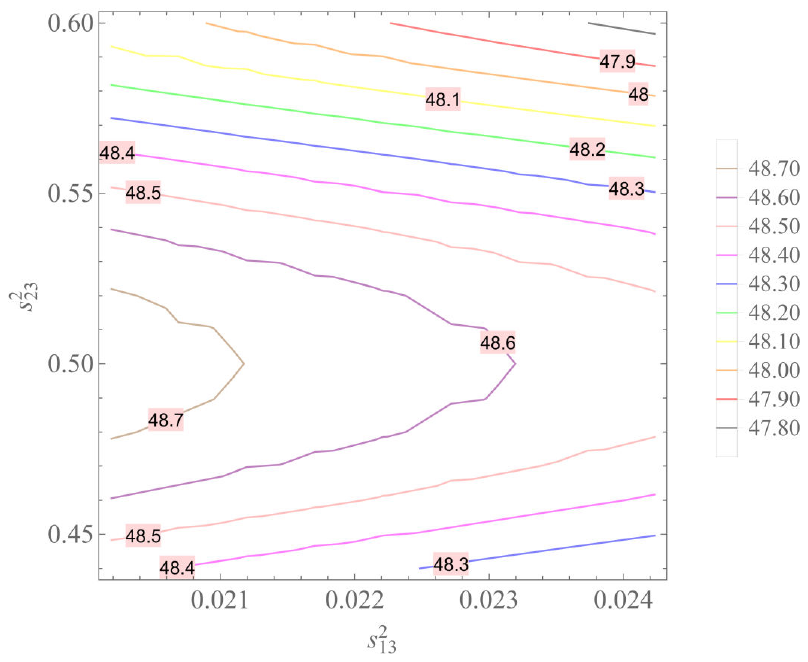}\hspace*{-3.5 cm}
\end{center}
\vspace{-9.25 cm}
\caption{$\langle m^{(3)}_{ee} \rangle$ versus $s^2_{13}$ and $s^2_{23}$ with $s^2_{13} \in (2.00, 2.405) 10^{-2}$ and $s^2_{23}\in (0.46, 0.54)$ for NH (left panel) while $s^2_{13} \in (2.018, 2.424) 10^{-2}$ and $s^2_{23}\in (0.44, 0.60)$ for IH (right panel).}
\label{mee33F}
\end{figure}

\begin{figure}[h]
\begin{center}
\vspace{-0.5 cm}
\hspace{-3.0 cm}
\includegraphics[width=0.65\textwidth]{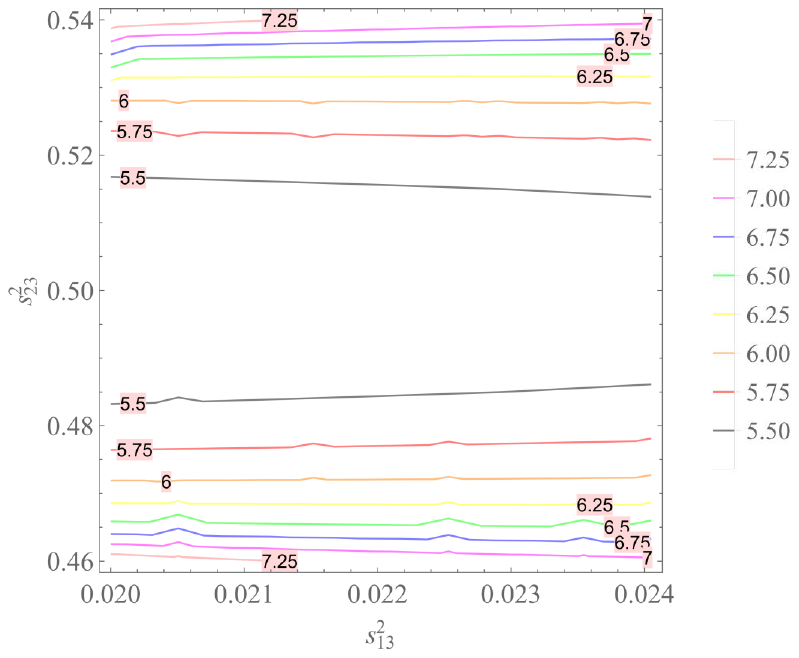}\hspace{-3.5 cm}
\includegraphics[width=0.65\textwidth]{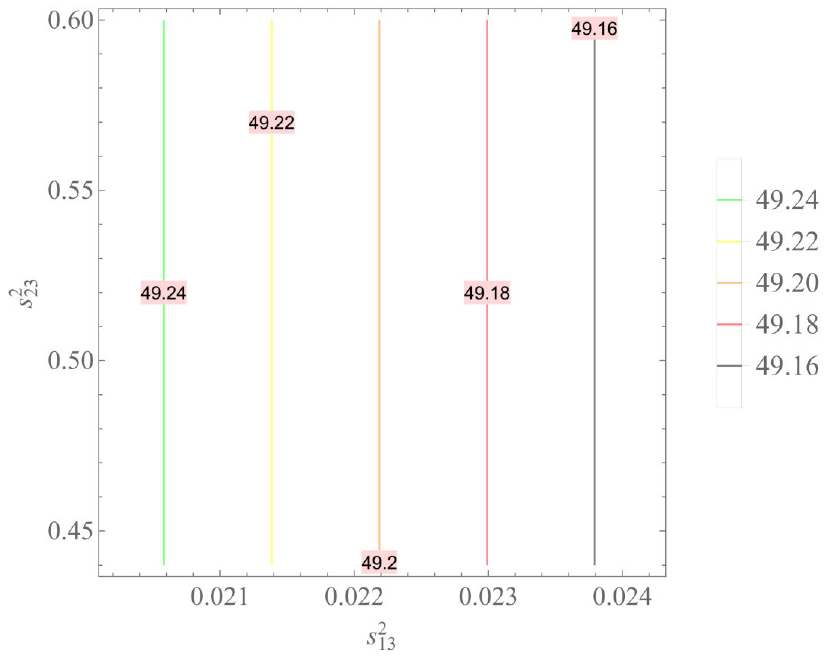}\hspace*{-3.5 cm}
\end{center}
\vspace{-9.25 cm}
\caption{The contour plot of $m^{(3)}_{\beta}$ as a function of $s^2_{13}$ and $s^2_{23}$ with $s^2_{13} \in (2.00, 2.405) 10^{-2}$ and $s^2_{23}\in (0.46, 0.54)$ for NH (left panel) while $s^2_{13} \in (2.018, 2.424) 10^{-2}$ and $s^2_{23}\in (0.44, 0.60)$ for IH (right panel).}
\label{mb33F}
%\vspace{0.5 cm}
\end{figure}

Further, Eqs. (\ref{m0v})--(\ref{k3}) and (\ref{u14sq})--%(\ref{u34sq}), (\ref{meeexpr}) and
(\ref{mbexpr}) imply that $\langle m_{ee}\rangle$, $\langle m^{(3)}_{ee}\rangle$ and $m^{(3)}_\beta$ depend on four parameters $\Delta m^2_{21}, \Delta m^2_{31}$, $\theta_{13}$ and $\theta_{23}$ while $m_\beta$, $|U_{e 4}|^2, |U_{\mu 4}|^2$ and $|U_{\tau 4}|^2$ depend on five parameters $\Delta m^2_{21}, \Delta m^2_{31}$, $\theta_{13}, \theta_{23}$ and $m_s$. At the best-fit values of $\Delta m^2_{21}, \Delta m^2_{31}$
taken from Table \ref{experconstrain}, we get
\bea
&&k^2_{3}= \left\{
\begin{array}{l}
0.1924\hspace{0.25cm}\mbox{for  NH},  \\
0.9522\hspace{0.25cm}\mbox{for IH},
\end{array}%
\right.
\eea
and the effective neutrino masses $\langle m_{ee}\rangle$, $\langle m^{(3)}_{ee}\rangle$ and $m^{(3)}_\beta$ depend on $\theta_{13}$ and $\theta_{23}$ which is depicted in Figs. \ref{meeF}, \ref{mee33F} and \ref{mb33F}.
%%%%%%%%%%%%%%%%%%%%
%%%%%%%%%%%%%%%%%%%%
%%%%
%%%%%%%%%%%%%%%%%%

\begin{figure}[h]
\begin{center}
\vspace{-0.25 cm}
\hspace{-3.0 cm}
\includegraphics[width=0.65\textwidth]{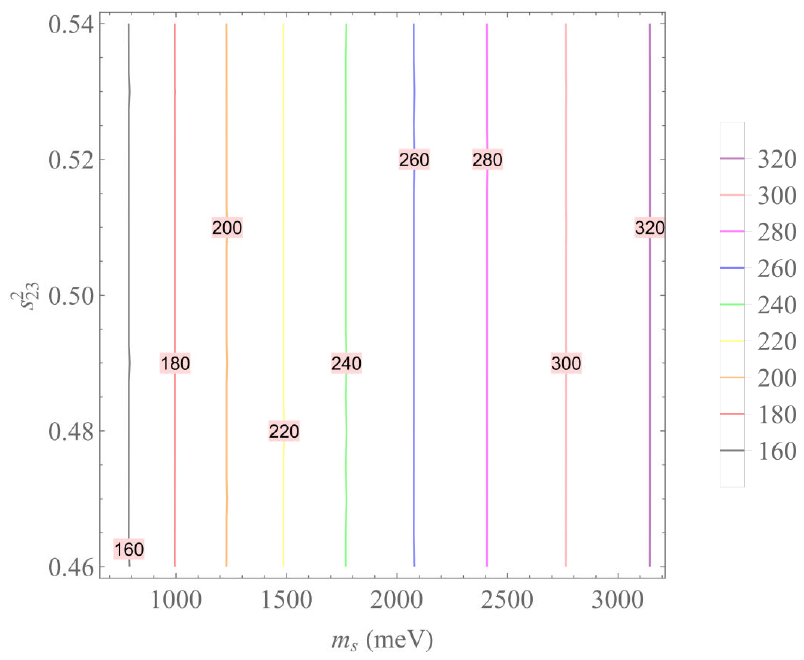}\hspace{-3.5 cm}
\includegraphics[width=0.65\textwidth]{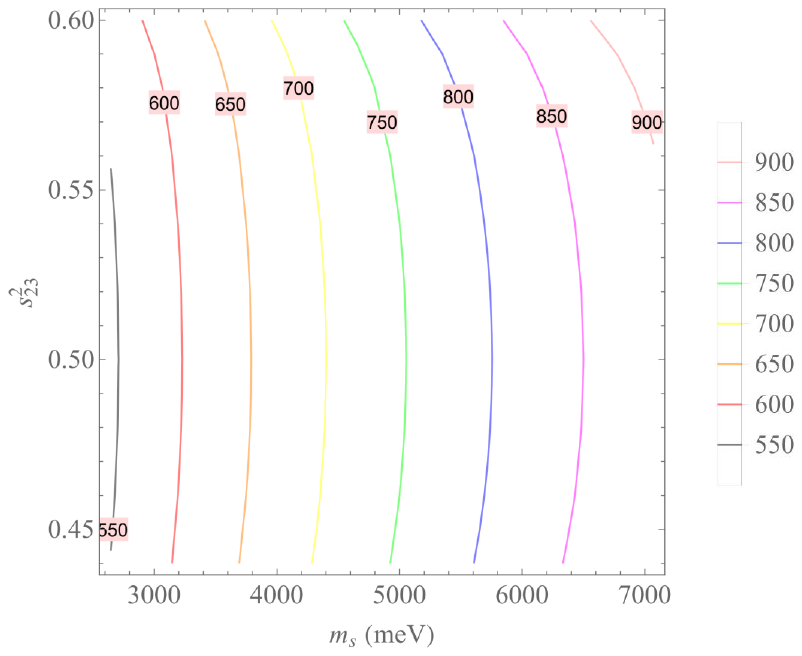}\hspace*{-3.5 cm}
\end{center}
\vspace{-9.25 cm}
\caption{The contour plot of $m_{\beta}$ as a function of $m_s$ and $s^2_{23}$ with $m_s\in (707.11,\, 3162.28)\, \mathrm{meV} \, \left[\mathrm{i.e.}, \Delta m^2_{41} \in (0.5, \, 10.0)\, \mathrm{eV}^2\right]$ and $s^2_{23}\in (0.46, 0.54)$ for NH (left panel)
while $m_s \in (2646.22, 7071.24)\, \mathrm{meV}\, \left[\mathrm{i.e.}, \Delta m^2_{41} \in (7.0, \, 50.0)\, \mathrm{eV}^2\right]$ and $s^2_{23}\in (0.44, 0.60)$ for IH (right panel).}
\label{mbF}
%\vspace{0.5 cm}
\end{figure}

At the best-fit values of $\Delta m^2_{21}, \Delta m^2_{31}$ and $s^2_{13}$, $m_{\beta }$, $|U_{e4}|^2$, $|U_{\mu4}|^2$ and $|U_{\tau 4}|^2$ depend on $m_s$ and $\theta_{23}$ which are plotted in Figs. \ref{mbF},  \ref{u14sqF}, \ref{u24sqF} and \ref{u34sqF}.
%%%%
%%%%%%%
Figures \ref{meeF}--\ref{mbF} show that our model predicts the range of the effective neutrino mass parameters as follows
\bea
&&\langle m_{ee}\rangle \in\left\{
\begin{array}{l}
(35.70,\,\, 36.50) \, \mbox{meV}\hspace{0.25 cm}  \mbox{for \hspace{0.05 cm}  NH,} \\
(160.0,\, 168.0)\, \mbox{meV}\hspace{0.25 cm}  \mbox{for \hspace{0.05 cm} IH,}%
\end{array}%
\right.   \label{meeranges}\\
&&\langle m^{(3)}_{ee}\rangle \in\left\{
\begin{array}{l}
(3.65,\,\, 4.10) \,\, \mbox{meV}\hspace{0.425 cm} \mbox{for \hspace{0.05 cm}NH,} \\
(47.80,\, 48.70)\,\, \mbox{meV}\hspace{0.1 cm} \mbox{for \hspace{0.05 cm}IH,}%
\end{array}%
\right.   \label{mee33ranges} \\%\eea
%\bea
&&m^{(3)}_{\beta}\in\left\{
\begin{array}{l}
(5.50,\,\, 7.25) \,\, \mbox{meV}\hspace{0.425 cm}\mbox{for \hspace{0.05 cm}NH,} \\
(49.16,\, 49.24)\,\, \mbox{meV} \hspace{0.1 cm}\mbox{for \hspace{0.05 cm} IH,}%
\end{array}%
\right.   \label{mb33ranges}\\
&&m_{\beta}\in\left\{
\begin{array}{l}
(160.0,\,\, 320.0) \,\, \mbox{meV} \hspace{0.275 cm}   \mbox{for \hspace{0.05 cm} NH,} \\
(550.0,\, 900.0)\, \mbox{meV}\hspace{0.2 cm}  \mbox{for \hspace{0.05 cm} IH\Revised{.}}%
\end{array}%
\right.   \label{mbranges}
\eea
We see that the resulting effective neutrino mass for three neutrino scheme in Eq. (\ref{mee33ranges}), for both  normal and inverted hierarchies, are below all the upper bounds taken from GERDA \cite{Agostini18}$\langle m_{ee} \rangle < 120 \div 260 \,\mathrm{meV}$, MAJORANA \cite{MAJO} $\langle m_{ee} \rangle < 24 \div 53 \,\mathrm{meV}$, CUORE \cite{CUORE18} $\langle m_{ee} \rangle < 110 \div 500 \,\mathrm{meV}$, KamLAND-Zen \cite{KamLAND16} $\langle m_{ee} \rangle <61\div 165 \,\mathrm{meV}$, GERDA \cite{GERDA19}
 $\langle m_{ee} \rangle < 104 \div 228 \,\mathrm{meV}$ and CUORE \cite{CUORE20} $\langle m_{ee} \rangle < 75 \div 350 \,\mathrm{meV}$. Furthermor, the resulting effective neutrino mass for 3+1 scheme in Eq. (\ref{meeranges}) are below all the upper bounds taken from CUPID-Mo experiment \cite{CUPID2021}, $\langle m_{ee} \rangle < 300 \div 500 \,\mathrm{meV}$.

Finally, from the above analysis, we can summarize the obtained parameters of the model as shown in Table \ref{parameterrangeA4}.
\begin{table}[ht]
\caption{The model prediction compared to the experimental range}
{\begin{tabular}{@{}cc|cc|cccccc@{}} \hline%\toprule
& Parameters & Prediction (NH) &Experimental range (NH)&Prediction (IH)&Experimental range (IH)\\\hline
& \,\,$\sin^2\theta_{12}$ &\, $0.3401\rightarrow 0.3415$&\, $0.271\rightarrow0.369$&\, $0.3402\rightarrow 0.3416$&\, $0.271\rightarrow0.369$&\\
  & \,\,$\sin^2\theta_{23}$ &\, $0.460\rightarrow 0.540$&\, $0.434\rightarrow0.610$&\, $0.44\rightarrow 0.60$&\, $0.433\rightarrow 0.608$&\\
  & \,\,$\delta_{CP} (^\circ)$ &\, $-0.60 \rightarrow -0.20$&\, $-1.0\rightarrow 0.79$&\, $-0.95\rightarrow -0.6$&\, $-1.0 \rightarrow -0.125$&\\
%  & \,\,$m_s \, (\mathrm{eV})$ &\, $\rightarrow $&\, $\rightarrow$&\, $\rightarrow $&\, $\rightarrow$&\\
  &\,\, $|U_{1 4}|^2$ &\, $0.015\rightarrow 0.045$&\, $0.012\rightarrow 0.047$&\, $0.020\rightarrow 0.045$&\, $0.012\rightarrow 0.047$&\\
  &\,\, $|U_{2 4}|^2$ &\, $0.004\rightarrow 0.012$&\, $0.005\rightarrow 0.03$&\, $0.008\rightarrow 0.018$&\, $0.005\rightarrow 0.03$&\\
  &\,\, $|U_{3 4}|^2$ &\, $0.004\rightarrow 0.014$&\, $0 \rightarrow 0.16$&\, $0.008\rightarrow 0.022$&\, $0 \rightarrow 0.16$&\\
 \hline
\end{tabular}} \label{parameterrangeA4}
\end{table}

In the case $s^2_{23}=0.47\, (\theta_{23}=43.30^\circ)$ and $\Delta m^2_{41} =1.45\, \mathrm{eV}^2$ \cite{Aartsen2020} for NH while  $s^2_{23}=0.578\, (\theta_{23}=49.50^\circ)$ and $\Delta m^2_{41} =15.0\, \mathrm{eV}^2 \,(m_s=1873.30\, \mathrm{meV})$ for IH, we get the explicit parameters as listed in Tab. \ref{parameterA4}.
\begin{table}[ht]
\caption{The obtained parameters}
{\begin{tabular}{@{}cccccccccc@{}} \hline%\toprule
& Parameters & The derived values (NH) &The derived values (IH)\\\hline
  & \,\,$k_2$ &\, $1.54 $&\, $0.868$&\\
  & \,\,$\alpha_1 (^\circ)$ &\, $137.50$&\, $352.34$&\\
  & \,\,$\delta_{CP} (^\circ)$ &\, $160.20$&\, $316.70$&\\
  & \,\,$m_s \, (\mathrm{eV})$ &\, $1.204 $&\, $3.873$&\\
  &\,\, $|U_{1 4}|^2$ &\, $0.027$&\, $0.0301$&\\
  &\,\, $|U_{2 4}|^2$ &\, $0.00622$&\, $0.0127$&\\
  &\,\, $|U_{3 4}|^2$ &\, $0.00883$&\, $0.0147$&\\
  &\,\, $\langle m_{ee}\rangle\, (\mathrm{meV})$ &\, $36.00$&\, $164.00$&\\
    &\,\, $\langle m^{(3)}_{ee}\rangle\, (\mathrm{meV})$ &\, $3.61$&\, $48.10$&\\
      &\,\, $m_\beta\, (\mathrm{meV})$ &\, $198.00$&\, $673.00$&\\
        &\,\, $m^{(3)}_\beta\, (\mathrm{meV})$ &\, $6.12$&\, $49.20$&\\
 \hline
\end{tabular}} \label{parameterA4}
\end{table}
%\newpage
\section{\label{conclusion} Conclusions}
The combination of $B-L$ model and discrete symmetry $A_4\times Z_3\times Z_4$ can successfully explain both the active and the sterile mixing observables in 3+1 scheme. In the model, the tiny neutrino mass the mass hierarchy are obtained by the type-I seesaw mechanism and the minimal extended seesaw and the charged lepton mass hierarchy is satisfied by a factor of $v_H \left(\frac{v_l}{\Lambda}\right)^2 \sim 10^{-4}\, \mathrm{GeV}$ of the electron mass compared to the muon and the tau masses of the order of $\frac{v_H v_l}{\Lambda} \sim 10^{-1}\, \mathrm{GeV}$.  The observables are given in terms of five model parameters, $k_{2,3}$, $m_0, m_s$ and $\alpha_1$.
The experimental ranges of the reactor mixing angle ($s^2_{13}$) and the two mass-squared differences ($\Delta m^2_{21},\, \Delta m^2_{31}$) are chosen to obtain the allowed values of $s_{12}$ and $k_3$ for both NH and IH. The possible range values of $\sin\alpha_1$  gives the predictive range of $s_{23}$. $\alpha_1$ and $s_{13}$ are used to predict $k_{2}$ and $\delta_{CP}$. The model parameter $m_s$ (corresponds to the sterile neutrino mass) is determined by the active-sterile mass-squared difference $\Delta m^2_{41}$ for NH and by $\Delta m^2_{41}, \Delta m^2_{31}$ and $\Delta m^2_{21}$ [Eqs. (\ref{m0v}), (\ref{k0v}) and (\ref{msm4})] for IH. The experimental ranges of $s_{13}$ and the extracted values of $s_{23}$ are used to evaluate the effective neutrino masses governing the neutrinoless double-beta decay while the extracted values of $m_s$ and $s_{23}$ are used to evaluate the active-sterile mixing.

The $3+1$ active-sterile neutrino mixings are predicted to be $0.015 \leq|U_{1 4}|^2\leq 0.045$, $0.004 \leq|U_{2 4}|^2\leq 0.012$, $0.004 \leq|U_{3 4}|^2\leq 0.014$ for normal hierarchy and $0.020\leq|U_{1 4}|^2\leq 0.045$, $0.008 \leq|U_{2 4}|^2\leq 0.018$, $0.008\leq|U_{3 4}|^2\leq 0.022$ for inverted hierarchy. Sterile neutrino masses are predicted to be $0.7 \lesssim  m_s \, (\mathrm{eV}) \lesssim 3.16$ for normal hierarchy and $2.6 \lesssim  m_s \, (\mathrm{eV}) \lesssim 7.1$  for inverted hierarchy. For three neutrino scheme the model predicts $0.3401 \leq \sin^2\theta_{12}\leq 0.3415, \, 0.460 \leq \sin^2\theta_{23}\leq 0.540,\, -0.60 \leq \sin\delta_{CP}\leq -0.20$ for normal hierarchy and $0.3402 \leq \sin^2\theta_{12}\leq 0.3416,\, 0.434\leq\sin^2\theta_{23}\leq 0.610,\, -0.95 \leq \sin\delta_{CP}\leq -0.60$ for inverted hierarchy. The effective neutrino masses are predicted to be $35.70 \leq \langle m_{ee}\rangle [\mbox{meV}] \leq 36.50$ in 3+1 scheme and $3.65 \leq \langle m^{(3)}_{ee}\rangle [\mbox{meV}] \leq 4.10$ in three neutrino scheme for NH while $160.0 \leq \langle m_{ee}\rangle [\mbox{meV}] \leq 168.0$ in 3+1 scheme and $47.80 \leq \langle m^{(3)}_{ee}\rangle [\mbox{meV}] \leq 48.70$ in three neutrino scheme for for IH which are all in agreement with the recent experimental bounds.

%\section*{Acknowledgments}
%This research is funded by Vietnam National Foundation for Science
%and Technology Development (NAFOSTED) under grant number 103.01-2017.341.
%\newpage
\appendix
\section{\label{Higgspotential} Higgs potential invariant under $\mathbf{\Gamma}$ symmety}
The total scalar potential invariant under $\mathrm{\Gamma}$ symmetry, up to five-dimension, is given by\footnote{Here, $V(\alpha_1 \rightarrow \alpha_2, \beta_1 \rightarrow \beta_2,\cdots)
\equiv V(\alpha_1, \beta_1,\cdots)\!\!\!\mid_{\{\alpha_1=\alpha_2, \beta_1=\beta_2,\cdots \}}$.}:
\bea V_{\mathrm{scalar}}&=& V(H)+V(\phi_l)+V(\phi_\nu)+V(\eta)+V(\phi_s) \crn
&+&V(H, \phi_l) + V(H, \phi_\nu)+V(H, \eta) +V(H, \phi_s)+V(\phi_l, \phi_\nu)\crn
&+&V(\phi_l, \eta) +V(\phi_l, \phi_s)+V(\phi_\nu, \eta) +V(\phi_\nu, \phi_s)+V(\eta, \phi_s)
+ V_{\mathrm{triple}}, \label{V}\eea
where
\bea
&&V(H)=\mu_{H}^2 (H^\+ H)_{\underline{1}} +\lambda_{H} ({H}^\+H)_{\underline{1}}({H}^\+H)_{\underline{1}}, \label{VH}\\
&&V(\phi_l)=\mu_{\phi_l}^2 (\phi_l^\+ \phi_l)_{\underline{1}}
+\lambda^{\phi_l}_1 (\phi_l^\+\phi_l)_{\underline{1}}(\phi_l^\+\phi_l)_{\underline{1}}
+\lambda^{\phi_l}_2 (\phi_l^\+\phi_l)_{\underline{1}^'}(\phi_l^\+\phi_l)_{\underline{1}^{''}}\crn
&&\hspace{1.1 cm}+\, \lambda^{\phi_l}_3 (\phi_l^\+\phi_l)_{\underline{1}^{''}}(\phi_l^\+\phi_l)_{\underline{1}^{'}}
+\lambda^{\phi_l}_4 (\phi_l^\+\phi_l)_{\underline{3}_s}(\phi_l^\+\phi_l)_{\underline{3}_s}
+\lambda^{\phi_l}_5 (\phi_l^\+\phi_l)_{\underline{3}_a}(\phi_l^\+\phi_l)_{\underline{3}_a}, \label{Vphil}\\
&&V(\phi_\nu)=V(\phi_l\rightarrow \phi_\nu), \, V(\eta)=V(H\rightarrow \eta), \, V(\phi_s)=V(\phi_l\rightarrow \phi_s), \, V(\eta_s)=V(\eta_s\rightarrow \eta),\label{Vphis}\\
&&V(H, \phi_l) =\lambda^{H\phi_l}_1 ({H}^\+H)_{\underline{1}}(\phi_l^\+\phi_l)_{\underline{1}}
+\lambda^{H\phi_l}_2 ({H}^\+\phi_l)_{\underline{3}}(\phi_l^\+H)_{\underline{3}},\hs V(H, \phi_\nu)=V(H, \phi_l\rightarrow \phi_\nu),\label{VHphinu}\\
&&V(H, \eta) =\lambda^{H\eta}_1 ({H}^\+H)_{\underline{1}}(\eta^\+\eta)_{\underline{1}}
+\lambda^{H\eta}_2 ({H}^\+\eta)_{\underline{1}}(\eta^\+H)_{\underline{1}},\hs V(H, \phi_s)=V(H, \phi_l\rightarrow \phi_s),\label{VHphis}\\
&&V(H, \eta_s)=V(H, \eta\rightarrow \eta_s), \hs V(\phi_l, \phi_\nu)=
\lambda^{\phi_l\phi_\nu}_1 (\phi_l^\+\phi_l)_{\underline{1}}(\phi_\nu^\+\phi_\nu)_{\underline{1}}
+\lambda^{\phi_l\phi_\nu}_2 (\phi_l^\+\phi_l)_{\underline{1}^'}(\phi_\nu^\+\phi_\nu)_{\underline{1}^{''}}\crn
&&\hspace{1.3 cm}+\, \lambda^{\phi_l\phi_\nu}_3 (\phi_l^\+\phi_l)_{\underline{1}^{''}}(\phi_\nu^\+\phi_\nu)_{\underline{1}^{'}}+\lambda^{\phi_l\phi_\nu}_4 (\phi_l^\+\phi_l)_{\underline{3}_s}(\phi_\nu^\+\phi_\nu)_{\underline{3}_s}
+\lambda^{\phi_l\phi_\nu}_5 (\phi_l^\+\phi_l)_{\underline{3}_a}(\phi_\nu^\+\phi_\nu)_{\underline{3}_a}\crn
&&\hspace{1.3 cm}+\lambda^{\phi_l\phi_\nu}_6 (\phi_l^\+\phi_\nu)_{\underline{1}}(\phi_\nu^\+\phi_l)_{\underline{1}}
+\lambda^{\phi_l\phi_\nu}_7 (\phi_l^\+\phi_\nu)_{\underline{1}^'}(\phi_\nu^\+\phi_l)_{\underline{1}^{''}}
+\lambda^{\phi_l\phi_\nu}_8 (\phi_l^\+\phi_\nu)_{\underline{1}^{''}}(\phi_\nu^\+\phi_l)_{\underline{1}^{'}}\crn
&&\hspace{1.3 cm}+\,\lambda^{\phi_l\phi_\nu}_9 (\phi_l^\+\phi_\nu)_{\underline{3}_s}(\phi_\nu^\+\phi_l)_{\underline{3}_s}
+\lambda^{\phi_l\phi_\nu}_{10} (\phi_l^\+\phi_\nu)_{\underline{3}_a}(\phi_\nu^\+\phi_l)_{\underline{3}_a}, \hs V(\phi_l, \eta) =V(\phi_l, H\rightarrow\eta), \eea
\bea &&V(\phi_l, \phi_s)=V(\phi_l, \phi_\nu\rightarrow \phi_s), \hs V(\phi_l, \eta_s)=V(\phi_l, \eta\rightarrow \eta_s), \hs V(\phi_\nu, \eta) =V(\phi_l\rightarrow \phi_\nu, \eta), \label{Vphils}\\
&& V(\phi_\nu, \phi_s) =V(\phi_l\rightarrow \phi_\nu, \phi_s), \hs V(\phi_\nu, \eta_s)=V(\phi_\nu, \eta\rightarrow \eta_s), \hs V(\phi_s, \eta) =V(\phi_l\rightarrow \phi_s, \eta), \label{Vetaphis}\\
&&V(\eta, \eta_s)=\lambda^{\eta \eta_s}_1 (\eta^\+\eta)(\eta^\+_s\eta_s) +
  \lambda^{\eta \eta_s}_1(\eta_s^\+\eta)(\eta^\+\eta_s), \hs  V(\phi_s, \eta_s) =V(\phi_s, \eta\rightarrow \eta_s),  \label{etaetas}\\
&& V_{\mathrm{triple}}=\lambda^{H\phi_l\phi_s}_1 ({H}^\+H)_{\underline{1}}(\phi_l^\+\phi_l)_{\underline{3}_s} \phi_\nu
+ \lambda^{H\phi_l\phi_s}_2 ({H}^\+H)_{\underline{1}}(\phi_l^\+\phi_l)_{\underline{3}_a} \phi_\nu \crn
&&\hspace{1.1 cm}+\, \lambda^{H\phi_s\phi_\nu}_1({H}^\+H)_{\underline{1}}(\phi_s^\+\phi_s)_{\underline{3}_s} \phi_\nu + \lambda^{H\phi_s\phi_\nu}_2({H}^\+H)_{\underline{1}}(\phi_s^\+\phi_s)_{\underline{3}_a} \phi_\nu \crn
&&\hspace{1.1 cm}+\, \lambda^{\phi_l\phi_\nu\eta}_1 ({\eta}^\+ \eta)_{\underline{1}}(\phi_l^\+\phi_l)_{\underline{3}_s} \phi_\nu+\, \lambda^{\phi_l\phi_\nu\eta}_2 ({\eta}^\+ \eta)_{\underline{1}}(\phi_l^\+\phi_l)_{\underline{3}_a} \phi_\nu
\crn
&&\hspace{1.1 cm}+\, \lambda^{\phi_l\phi_\nu \phi_s}_1 (\phi_l^\+\phi_l)_{\underline{3}_s} \phi_\nu (\phi_s^\+ \phi_s)_{\underline{1}}+ \lambda^{\phi_l\phi_\nu \phi_s}_2 (\phi_l^\+\phi_l)_{\underline{3}_a} \phi_\nu (\phi_s^\+ \phi_s)_{\underline{1}}\crn
&&\hspace{1.1 cm}+\, \lambda^{\phi_l\phi_\nu \phi_s}_3 (\phi_s^\+\phi_s)_{\underline{3}_s} \phi_\nu (\phi_l^\+ \phi_l)_{\underline{1}}+\, \lambda^{\phi_l\phi_\nu \phi_s}_4 (\phi_s^\+\phi_s)_{\underline{3}_a} \phi_\nu (\phi_l^\+ \phi_l)_{\underline{1}}
\crn
&&\hspace{1.1 cm}+\,\lambda^{\phi_l\phi_\nu \phi_s}_5 (\phi_l^\+\phi_s)_{\underline{3}_s} \phi_\nu (\phi_s^\+ \phi_l)_{\underline{1}}
+\lambda^{\phi_l\phi_\nu \phi_s}_6 (\phi_l^\+\phi_s)_{\underline{3}_a} \phi_\nu (\phi_s^\+ \phi_l)_{\underline{1}}
\crn
&&\hspace{1.1 cm}+\,\lambda^{\phi_l\phi_\nu \phi_s}_7 (\phi_s^\+\phi_l)_{\underline{3}_s} \phi_\nu (\phi_l^\+ \phi_s)_{\underline{1}}
+\lambda^{\phi_l\phi_\nu \phi_s}_8(\phi_s^\+\phi_l)_{\underline{3}_a} \phi_\nu (\phi_l^\+ \phi_s)_{\underline{1}}\crn
&&\hspace{1.1 cm}+\, \lambda^{\phi_\nu \eta \phi_s}_1 (\eta^\+\eta)_1(\phi_s^\+\phi_s)_{\underline{3}_{s}} \phi_\nu
 +\lambda^{\phi_\nu \eta \phi_s}_2 (\eta^\+\eta)_1(\phi_s^\+\phi_s)_{\underline{3}_{a}} \phi_\nu \crn
 &&\hspace{1.1 cm}+\, \lambda^{\phi_l \phi_\nu \eta_s} (\phi^*_l\phi_l)_{3_s} \phi_\nu (\eta^*_s \eta_s)_{1}
 +\lambda^{\phi_\nu \phi_s \eta_s} (\phi^*_s\phi_s)_{3_s} \phi_\nu (\eta^*_s \eta_s)_{1}. %\crn
 %&&\hspace{1.1 cm}+\, \Blue{\lambda^{\phi_\nu \phi_s \eta_s}_2 (\phi^*_s\phi_s)_{3_a} \phi_\nu (\eta^*_s \eta_s)_{1}}.
 \label{Vtrip}
\eea
All the other triple, quartic and quintic terms, up to five-dimension, of three or four or five differential scalar fields are forbidden by one (or some) of the model symmetries, for instance, $({H}^\+H)_{\underline{1}}(\phi_l^\+\phi_\nu)_{\underline{1}} $, $({H}^\+H)_{\underline{1}}(\phi_\nu^\+\phi_\nu)_{\underline{3}_s} \phi_l $ and $({H}^\+H)_{\underline{1}}(\phi_\nu^\+\phi_\nu)_{\underline{3}_s} \phi_l $ are forbidden by $Z_3$ and $Z_4$; $({H}^\+H)_{\underline{1}}(\eta^\+\eta)_{\underline{1}}\phi_l $ is forbidden by $A_4, Z_3$ and $Z_4$; $({H}^\+H)_{\underline{1}}\phi_l \phi_\nu\eta$ is forbidden by $U(1)_{B-L}$ and so on.
\newpage
%%%%%%%%%%%%%%%%%%%%%%%%%%%%%%%%%%%%%
\section{\label{solution} The solution of Eqs. (\ref{eq1})--(\ref{eq5})}
%The system of Eqs. (\ref{eq1})--(\ref{eq5}) always own the following solution
\bea
&&\lambda^{H}=-\frac{\mu_H^2 +3 \lambda^{H\phi_l} v_l^2+\lambda^{H\phi_\nu} v_\nu^2 + \lambda^{H\eta} v_{\eta}^2+ 2 \lambda^{H\phi_s} v_s^2 +2 \Lambda^{H\phi_l \phi_\nu \phi_s} v_\nu +\lambda^{H \eta_s} v_{\eta_s}^2}{2 v_H^2}, \label{sollaH}\\
&&\lambda^{\phi_l}=-\frac{3 \mu_{\phi_l}^2+3 \lambda^{H\phi_l} v_H^2+\lambda^{\phi_l\phi_\nu} v^2_\nu +  \lambda^{\phi_l\phi_s} v_s^2+3 \lambda^{\phi_l\eta} v_\eta^2 +2 \Lambda^{H\phi_l\phi_\nu \phi_s\eta}_1 v_\nu }{6 v_l^2}\crn
&&\hspace{0.65 cm}-\, \frac{ (3\lambda^{\phi_l\eta_s} + 2  \lambda^{\phi_l \phi_\nu \eta_s} v_\nu) v_{\eta_s}^2}{6 v_l^2}, \label{sollaphil}\\
&&\lambda^{\phi_\nu}=-\frac{\mu_\nu^2+\lambda^{H\phi_\nu} v_H^2+\lambda^{\phi_l\phi_\nu} v_l^2+\lambda^{\phi_\nu \eta} v_\eta^2-\lambda^{\phi_\nu \phi_s} v_s^2}{2 v_\nu^2}-\frac{\Lambda^{H\phi_l\phi_\nu\eta\phi_s}_2 v_l^2+ \Lambda^{H\phi_l\phi_\nu\eta \phi_s}_3 v_s^2}{2 v_\nu^3}\crn
&&\hspace{0.7 cm} -\frac{ \left(\Lambda^{\phi_l\phi_\nu\eta_s} v_{l}^2+\Lambda^{\phi_\nu\phi_s\eta_s} v_{s}^2+\Lambda^{\phi_\nu \eta_s} v_{\nu}\right) v_{\eta_s}^2}{2 v_{\nu}^3}, \label{sollaphin}\\
&&\lambda^{\eta}=-\frac{\mu_\eta^2+\lambda^{H\eta} v_H^2+3 \lambda^{\phi_l\eta} v_l^2+\lambda^{\phi_\nu\eta} v_\nu^2+2\lambda^{\phi_s\eta} v_s^2+2 (\lambda^{\phi_l\phi_\nu\eta}_1 v_l^2 + \lambda^{\phi_\nu \eta\phi_s} v_s^2) v_\nu +\lambda^{\eta\eta_s} v_{\eta_s}^2}{2 v_\eta^2}, \hspace{0.2 cm}\\
%\crn &&\Blue{-\frac{\lambda^{\eta\eta_s} v_{\eta_s}^2}{2 v_{\eta}^2}},
\label{sollaeta} \\
&&\lambda^{\phi_s}=-\frac{2 \mu_s^2+2 \lambda^{H\phi_s} v_H^2+\lambda^{\phi_l\phi_s} v_l^2-\lambda^{\phi_\nu\phi_s} v_\nu^2+2 \lambda^{\phi_s\eta} v_\eta^2+2 \Lambda^{H\phi_l \phi_\nu\eta\phi_s}_3 v_\nu}{2 v_s^2}\crn
&&\hspace{0.65 cm}-\, \frac{ (\lambda^{\phi_s\eta_s} + \lambda^{\phi_\nu \phi_s\eta_s} v_\nu) v_{\eta_s}^2}{v_s^2}, \hs \label{sollaphis}\\
&&\lambda^{\eta_s}= -\frac{\mu_{\eta_s}^2+\lambda^{\eta\eta_s} v_{\eta}^2+\lambda^{H\eta_s} v_{H}^2+2 \lambda^{\phi_l \phi_\nu \eta_s} v_{l}^2 v_{\nu}+3 \lambda^{\phi_l\eta_s} v_{l}^2+\lambda^{\phi_\nu\eta_s} v_{\nu}^2}{2 v_{\eta_s}^2}\crn
&&\hspace{0.6 cm}-\frac{(\lambda^{\phi_\nu \phi_s\eta_s} v_{\nu}+\lambda^{\phi_s\eta_s}) v_{s}^2 }{v_{\eta_s}^2}.\eea
%\newpage


\begin{thebibliography}{9}
\bibitem{Salas2020} P. F. de Salas \emph{et. al.}, \emph{2020 Global reassessment of the neutrino oscillation picture}, J. High Energ. Phys. 2021, 71 (2021). https://doi.org/10.1007/JHEP02(2021)071.%, arXiv:2006.11237 [hep-ph].
%\bibitem{Esteban2020} I. Esteban \emph{et. al.}, J. High Energy Phys. 09 (2020) 178.%, arXiv: 2007.14792  [hep-ph].

\bibitem{Aguilar2001} A. Aguilar {\it et al.} [LSND Collaboration], \emph{Evidence for neutrino oscillations from the observation of $\bar{\nu}_e$ appearance in a $\bar{\nu}_\mu$ beam}, Phys.\ Rev.\ D 64 (2001) 112007. https://doi.org/10.1103/PhysRevD.64.112007.%, arXiv: hep-ex/0104049.
\bibitem{Maltoni2003} M. Maltoni, T. Schwetz, M. Tortola and J. Valle, \emph{Constraining neutrino oscillation parameters with current solar and atmospheric data}, Phys. Rev. D 67 (2003) 013011. https://doi.org/10.1103/PhysRevD.67.013011.%[hep-ph/0207227].
\bibitem{Arevalo2013} A. A. Aguilar-Arevalo \emph{et al.} [MiniBooNE Collaboration], \emph{Improved Search for $\bar{\nu}_\mu \rightarrow \bar{\nu}_e$ Oscillations in the MiniBooNE Experiment}, Phys. Rev. Lett. 110 (2013) 161801. https://doi.org/10.1103/PhysRevLett.110.161801.%, arXiv:1303.2588. [hep-ex].

\bibitem{Acero2008} M. A. Acero, C. Giunti, M. Laveder, \emph{Limits on $\nu_e\rightarrow \bar{\nu}_e$ disappearance from Gallium and reactor experiments}, Phys.Rev.D 78 (2008) 073009. https://doi.org/10.1103/PhysRevD.78.073009.%, arXiv: 0711.4222 [hep-ph].
\bibitem{Arevalo2010} A. A. Aguilar-Arevalo \emph{et al.} [MiniBooNE Collaboration], \emph{Event excess in the MiniBooNE search for $\bar{\nu}_\mu\rightarrow \bar{\nu}_e$ oscillations}, Phys. Rev. Lett. 105 (2010) 181801. https://doi.org/10.1103/PhysRevLett.105.181801.% arXiv:1007.1150
\bibitem{Mention2011} G. Mention \emph{et al.}, \emph{The Reactor Antineutrino Anomaly}, Phys. Rev. D 83 (2011) 073006. https://doi.org/10.1103/PhysRevD.83.073006.% arXiv: 1101.2755

\bibitem{An2014} F. P. An \emph{et al.} [Daya Bay Collaboration], \emph{Search for a Light Sterile Neutrino at Daya Bay}, Phys. Rev. Lett. 113 (2014) 141802. https://doi.org/10.1103/PhysRevLett.113.141802.%arXiv: 1407.7259
%\bibitem{Abe2015} K. Abe \emph{et al.} [Super-Kamiokande Collaboration], Phys. Rev. D 91 (2015) 052019.%arXiv:1410.2008
\bibitem{Gariazzo2016} S. Gariazzo \emph{et al.,} \emph{Light sterile neutrinos}, J. Phys. G 43 (2016) 033001. https://doi.org/10.1088/0954-3899/43/3/033001.
\bibitem{Arevalo2018} A. A. Aguilar-Arevalo \emph{et al.} [MiniBooNE Collaboration], \emph{Significant Excess of ElectronLike Events in the MiniBooNE Short-Baseline Neutrino Experiment}, Phys. Rev. Lett. 121 (2018) 221801. https://doi.org/10.1103/PhysRevLett.121.221801.% arXiv:1805.12028 [hep-ex].
\bibitem{Adamson2019} P. Adamson \emph{et al.} [MINOS+ Collaboration], \emph{Search for sterile neutrinos in MINOS and MINOS+ using a two-detector fit}, Phys. Rev. Lett. 122 (2019) 091803. https://doi.org/10.1103/PhysRevLett.122.091803.% arXiv: 1710.06488
\bibitem{Adamson2020} P. Adamson \emph{et al.} [Daya Bay Collaboration, MINOS + Collaboration], \emph{Improved Constraints on Sterile Neutrino Mixing from Disappearance Searches in the MINOS, MINOS+, Daya Bay, and Bugey-3 Experiments}, Phys. Rev. Lett. 125 (2020) 071801. https://doi.org/10.1103/PhysRevLett.125.071801. %	arXiv:2002.00301 [hep-ex]
\bibitem{Aartsen2020} M. G. Aartsen \emph{et al.} [IceCube Collaboration], \emph{An eV-scale sterile neutrino search using eight years of atmospheric muon neutrino data from the IceCube Neutrino Observatory}, Phys. Rev. Lett. 125 (2020) 141801. https://doi.org/10.1103/PhysRevLett.125.141801.%arXiv: 2005.12942
\bibitem{Beheraa2019} S. P. Behera, D. K. Mishrab, L. M. Pant, \emph{Sensitivity to sterile neutrino mixing using reactor antineutrinos}, Eur. Phys. J. C 79 (2019) 86. https://doi.org/10.1140/epjc/s10052-019-6591-0. %	arXiv:1901.04746 [hep-ph]
\bibitem{BeheraPRD2020} S. P. Behera, D. K. Mishra, L. M. Pant, \emph{Active-sterile neutrino mixing constraint using reactor antineutrinos with the ISMRAN set-up}, Phys. Rev. D 102 (2020) 013002. https://doi.org/10.1103/PhysRevD.102.013002.%, arXiv: 2007.00392 [hep-ph].




\bibitem{Kang2013} S. K. Kang, Yeong-Duk Kim, Young-Ju Ko, K. Siyeon, \emph{Four-neutrino analysis of 1.5km-baseline reactor antineutrino oscillations}, Adv. High Energy Phys. 2013 (2013) 138109. https://doi.org/10.1155/2013/138109.%, arXiv:1408.3211 [hep-ph].
\bibitem{Girardi2014} Ivan Girardi \emph{et. al.}, \emph{Constraining Sterile Neutrinos Using Reactor Neutrino Experiments}, J. High Energ. Phys. 2014, 57 (2014). https://doi.org/10.1007/JHEP08(2014)057.%,	arXiv:1405.6540 [hep-ph].
\bibitem{Agudelo2015} D. C. Rivera-Agudelo, A. Pérez-Lorenzana, \emph{Generating $\theta_{13}$ from sterile neutrinos in $\mu-\tau$ symmetric models}, Phys. Rev. D 92 (2015) 073009. https://doi.org/10.1103/PhysRevD.92.073009.%, arXiv:1507.07030 [hep-ph].
\bibitem{Gariazzo17} S. Gariazzo, C. Giunti, M. Laveder, and Y. F. Li, \emph{Updated Global 3+1 Analysis of Short-BaseLine Neutrino Oscillations}, J. High Energ. Phys. 2017, 135 (2017). https://doi.org/10.1007/JHEP06(2017)135.%, arXiv: 1703.00860 [hep-ph].
\bibitem{Coloma2018} Pilar Coloma, David V. Forero, Stephen J. Parke, \emph{DUNE sensitivities to the mixing between sterile and tau neutrinos}, J. High Energ. Phys. 2018, 79 (2018). https://doi.org/10.1007/JHEP07(2018)079.%,, arXiv:1707.05348 [hep-ph].
\bibitem{Liu2018}Jun-Hao Liu, Shun Zhou, \emph{Another look at the impact of an eV-mass sterile neutrino on the effective neutrino mass of neutrinoless double-beta decays}, Int. J. Mod. Phys. A 33 (2018) 1850014. https://doi.org/10.1142/S0217751X18500148.%,, arXiv:1710.10359 [hep-ph].
\bibitem{Gariazzo2018plb} S. Gariazzo, C. Giunti, M. Laveder, Y.F. Li, \emph{Model-Independent $\bar{\nu}_e$ Short-Baseline Oscillations from Reactor Spectral Ratios}, Phys. Lett. B 782 (2018) 13. https://doi.org/10.1016/j.physletb.2018.04.057.%,-21, arXiv:1801.06467 [hep-ph].
\bibitem{Dentler2018ea} Mona Dentler \emph{et al.}, \emph{Updated global analysis of neutrino oscillations in the presence of eV-scale sterile neutrinos}, J. High Energ. Phys. 2018, 10 (2018). https://doi.org/10.1007/JHEP08(2018)010.%,, arXiv:1803.10661 [hep-ph].
\bibitem{Gupta2018ea} Shivani Gupta, Zachary M. Matthews, Pankaj Sharma, Anthony G. Williams, \emph{The Effect of a Light Sterile Neutrino at NO$\nu$A and DUNE}, Phys. Rev. D 98 (2018) 035042. https://doi.org/10.1103/PhysRevD.98.035042.%,, arXiv:1804.03361 [hep-ph].
\bibitem{Thakore2018ea} Tarak Thakore, Moon Moon Devi, Sanjib Kumar Agarwalla, Amol Dighe, \emph{Active-sterile neutrino oscillations at INO-ICAL over a wide mass-squared range}, J. High Energ. Phys. 2018, 22 (2018). https://doi.org/10.1007/JHEP08(2018)022. %,, arXiv:1804.09613 [hep-ph].
\bibitem{Dev2019} S. Dev, DeshRaj, Radha Raman Gautam, Lal Singh, \emph{New mixing schemes for (3+1) neutrinos}, Nucl. Phys. B 941 (2019) 401. https://doi.org/10.1016/j.nuclphysb.2019.02.003.%,, arXiv: 1902.01742 [hep-ph].
\bibitem{Miranda2019} Luis Salvador Miranda, Soebur Razzaque, \emph{Revisiting constraints on 3+1 active-sterile neutrino mixing using IceCube data}, J. High Energ. Phys. 2019, 203 (2019). https://doi.org/10.1007/JHEP03(2019)203.%,, arXiv:1812.00831 [hep-ph].
\bibitem{Giunti2019} C. Giunti, T. Lasserre, \emph{eV-scale Sterile Neutrinos}, Annu. Rev. Nucl. Part. Sci. 69 (2019) 163. https://doi.org/10.1146/annurev-nucl-101918-023755.%,-190, arXiv: 1901.08330 [hep-ph].
\bibitem{Boser2020} Sebastian Böser \emph{et al.}, \emph{Status of Light Sterile Neutrino Searches}, Prog. Part. Nucl. Phys. 111 (2020) 103736. https://doi.org/10.1016/j.ppnp.2019.103736.%,, arXiv:1906.01739 [hep-ex].
\bibitem{Giunti2020cz} C. Giunti, Y.F. Li, Y.Y. Zhang, \emph{KATRIN bound on 3+1 active-sterile neutrino mixing and the reactor antineutrino anomaly}, J. High Energ. Phys. 2020, 61 (2020). https://doi.org/10.1007/JHEP05(2020)061.%, arXiv:1912.12956 [hep-ph].
\bibitem{Behera2020} S. P. Behera, D. K. Mishra, L. M. Pant, \emph{Active-sterile neutrino mixing constraint using reactor antineutrinos with the ISMRAN set-up}, Phys. Rev. D 102 (2020) 013002. https://doi.org/10.1103/PhysRevD.102.013002.%,, arXiv: 2007.00392 [hep-ph].
\bibitem{Diaz2020} A. Diaz \emph{et al.}, \emph{Where Are We With Light Sterile Neutrinos?}, Phys. Rep. 884 (2020) 1. https://doi.org/10.1016/j.physrep.2020.08.005.


\bibitem{Nelson2011} A. E. Nelson, \emph{Effects of CP Violation from Neutral Heavy Fermions on Neutrino Oscillations, and the LSND/MiniBooNE Anomalies}, Phys. Rev. D84 (2011) 053001. https://doi.org/10.1103/PhysRevD.84.053001.
\bibitem{Fan2012} J. Fan and P. Langacker, \emph{Light Sterile Neutrinos and Short Baseline Neutrino Oscillation Anomalies},
J. High Energ. Phys. 2012, 83 (2012). https://doi.org/10.1007/JHEP04(2012)083.
\bibitem{Kuflik2012} E. Kuflik, S. D. McDermott, and K. M. Zurek, \emph{Neutrino Phenomenology in a 3+1+1 Framework}, Phys. Rev. D86 (2012) 033015. https://doi.org/10.1103/PhysRevD.86.033015.
\bibitem{Huang2013} J. Huang and A. E. Nelson, \emph{MeV dark matter in the 3+1+1 model}, Phys. Rev. D88 (2013) 033016. https://doi.org/10.1103/PhysRevD.88.033016.
\bibitem{Giunti2013} C. Giunti, M. Laveder, Y. Li, and H. Long, \emph{A Pragmatic View of Short-Baseline Neutrino Oscillations}, Phys. Rev. D88 (2013) 073008. https://doi.org/10.1103/PhysRevD.88.073008.


\bibitem{Kopp2011} J. Kopp, M. Maltoni, and T. Schwetz, \emph{re there sterile neutrinos at the eV scale?}, Phys. Rev. Lett.107 (2011) 091801. https://doi.org/10.1103/PhysRevLett.107.091801.
\bibitem{Kopp2013} J. Kopp, P. A. N. Machado, M. Maltoni, and T. Schwetz, \emph{Sterile Neutrino Oscillations: The Global Picture},  J. High Energ. Phys. 2013, 50 (2013). https://doi.org/10.1007/JHEP05(2013)050.



\bibitem{Sorel2004} M. Sorel, J. Conrad, and M. Shaevitz, \emph{A combined analysis of short-baseline neutrino experiments in the (3+1) and (3+2) sterile neutrino oscillation hypotheses}, Phys. Rev. D 70 (2004) 073004. https://doi.org/10.1103/PhysRevD.70.073004.
\bibitem{Karagiorgi2007} G. Karagiorgi \emph{et al.}, \emph{Leptonic CP violation studies at MiniBooNE in the (3+2) sterile
neutrino oscillation hypothesis}, Phys. Rev. D75 (2007) 013011. https://doi.org/10.1103/PhysRevD.75.013011. [Erratum-ibid.D 80 (2009) 099902]
\bibitem{Maltoni2007} M. Maltoni and T. Schwetz, \emph{Sterile neutrino oscillations after first MiniBooNE results}, Phys. Rev. D76 (2007) 093005. https://doi.org/10.1088/1742-6596/110/8/082011.
\bibitem{Goswami2007} S. Goswami and W. Rodejohann, \emph{MiniBooNE results and neutrino schemes with 2
sterile neutrinos: Possible mass orderings and observables related to neutrino masses}, JHEP10 (2007) 073. https://doi.org/10.1088/1126-6708/2007/10/073.
\bibitem{Karagiorgi2009} G. Karagiorgi \emph{et al.}, \emph{Viability of $\Delta m^2\sim 1\, \mathrm{eV}^2$ sterile neutrino mixing models in light of MiniBooNE electron neutrino and antineutrino data from the Booster and NuMI beamlines}, Phys. Rev. D80 (2009) 073001. https://doi.org/10.1103/PhysRevD.80.073001.
\bibitem{Giunti2011} C. Giunti and M. Laveder, \emph{3+1 and 3+2 Sterile Neutrino Fits}, Phys. Rev. D84 (2011) 073008. https://doi.org/10.1103/PhysRevD.84.073008.
\bibitem{Donini2012} A. Donini \emph{et al.}, The minimal 3+2 neutrino model versus oscillation anomalies,
J. High Energ. Phys. 2012, 161 (2012). https://doi.org/10.1007/JHEP07(2012)161.
\bibitem{Archidiacono2012} M. Archidiacono, N. Fornengo, C. Giunti, and A. Melchiorri, \emph{Testing 3+1 and 3+2 neutrino mass models with cosmology and short baseline experiments}, Phys. Rev. D86 (2012) 065028. https://doi.org/10.1103/PhysRevD.86.065028.


\bibitem{Akhmedov2010} E. Akhmedov and T. Schwetz, \emph{MiniBooNE and LSND data: non-standard neutrino
interactions in a (3+1) scheme versus (3+2) oscillations}, J. High Energ. Phys. 2010, 115 (2010). https://doi.org/10.1007/JHEP10(2010)115.
\bibitem{Liao2016} J. Liao and D. Marfatia, \emph{Impact of nonstandard interactions on sterile neutrino searches at IceCube}, Phys. Rev. Lett. 117 (2016) 071802. https://doi.org/10.1103/PhysRevLett.117.071802.
\bibitem{Babu2016} K. S. Babu, D. W. McKay, I. Mocioiu, and S. Pakvasa, \emph{Light Sterile Neutrinos, Lepton Number Violating Interactions and the LSND Anomaly}, Phys. Rev. D93 (2016) 113019. https://doi.org/10.1103/PhysRevD.93.113019.
\bibitem{Blennow2017} M. Blennow \emph{et. al.}, \emph{Non-Unitarity, sterile neutrinos, and Non-Standard neutrino Interactions}, J. High Energ. Phys. 2017, 153 (2017). https://doi.org/10.1007/JHEP04(2017)153
\bibitem{Esmaili2019} A. Esmaili and H. Nunokawa, \emph{On the robustness of IceCube’s bound on sterile neutrinos in the presence of non-standard interactions}, Eur. Phys. J. C 79, 70 (2019). https://doi.org/10.1140/epjc/s10052-019-6595-9.




\bibitem{Gninenko2012} S. N. Gninenko, \emph{Sterile neutrino decay as a common origin for LSND/MiniBooNe and T2K excess events}, Phys. Rev. D 85 (2012) 051702. https://doi.org/10.1103/PhysRevD.85.051702.
\bibitem{Gninenko2012plb} S. N. Gninenko, \emph{New limits on radiative sterile neutrino decays from a search for single photons in neutrino interactions}, Phys. Lett. B 710 (2012) 86. https://doi.org/10.1016/j.physletb.2012.02.071.



\bibitem{Krishnan2020} R. Krishnan, A. Mukherjee and S. Goswami, \emph{Realization of the minimal extended seesaw mechanism and the $TM_2$ type neutrino mixing}, J. High Energ. Phys. 2020, 50 (2020). https://doi.org/10.1007/JHEP09(2020)050.
\bibitem{Barry2011} J.~Barry, W.~Rodejohann and H.~Zhang, \emph{Light Sterile Neutrinos: Models and Phenomenology},  J. High Energ. Phys. 2011, 91 (2011). https://doi.org/10.1007/JHEP07(2011)091.%, arXiv: 1105.3911 [hep-ph].
\bibitem{Barry2012} J.~Barry, W.~Rodejohann and H.~Zhang, \emph{Sterile Neutrinos for Warm Dark Matter and the Reactor Anomaly in Flavor Symmetry Models}, \emph{https://doi.org/10.1088/1475-7516/2012/01/052}, JCAP 1201 (2012) 052.%, arXiv:1110.6382 [hep-ph].
\bibitem{Zhang2012} H.~Zhang, \emph{Light Sterile Neutrino in the Minimal Extended Seesaw}, Phys.\ Lett. B 714 (2012) 262. https://doi.org/10.1016/j.physletb.2012.06.074. %, arXiv: 1110.6838 [hep-ph].
\bibitem{Borah2017fqj} D. Borah, \emph{Non-zero $\theta_{13}$ with Unbroken $\mu-\tau$ Symmetry of the Active Neutrino Mass Matrix in the Presence of a Light Sterile Neutrino}, Phys. Rev. D 95 (2017) 035016. https://doi.org/10.1103/PhysRevD.95.035016. %, arXiv:1607.05556 [hep-ph].
\bibitem{Das2019ea} P. Das, A. Mukherjee, M. K. Das, \emph{Active and sterile neutrino phenomenology with $A_4$ based minimal extended seesaw}, Nucl. Phys. B 941 (2019) 755. https://doi.org/10.1016/j.nuclphysb.2019.02.024.%,-799, arXiv:1805.09231 [hep-ph].
\bibitem{Sarma2019} Neelakshi Sarma, Kalpana Bora, Debasish Borah, \emph{Compatibility of $A_4$ Flavour Symmetric Minimal Extended Seesaw with (3+1) Neutrino Data}, Eur. Phys. J. C 79 2 (2019) 129. https://doi.org/10.1140/epjc/s10052-019-6584-z.%,, arXiv:1810.05826 [hep-ph].


\bibitem{Suematsu2001} D. Suematsu, \emph{A light sterile neutrino based on the seesaw mechanism}, Prog.Theor.Phys. 106 (2001) 587. https://doi.org/10.1143/PTP.106.587.%-602, arXiv: hep-ph/0105223.
\bibitem{Mohapatra2001} R. N. Mohapatra, \emph{Connecting bimaximal neutrino mixing to a light sterile neutrino}, Phys.Rev. D 64 (2001) 091301. https://doi.org/10.1103/PhysRevD.64.091301.%, arXiv: hep-ph/0107264.
\bibitem{Babu2004} K.S. Babu, G. Seidl, \emph{Simple Model for (3+2) Neutrino Oscillations}, Phys.Lett. B 591 (2004) 127. https://doi.org/10.1016/j.physletb.2004.03.086.%-136, arXiv: hep-ph/0312285.

\bibitem{Mohapatra2005} R.N. Mohapatra, S. Nasri, Hai-Bo Yu, \emph{Seesaw Right Handed Neutrino as the Sterile Neutrino for LSND}, Phys. Rev. D 72 (2005) 033007. https://doi.org/10.1103/PhysRevD.72.033007.%, arXiv: hep-ph/0505021.



\bibitem{Machado2013S3} A. C. B. Machado, V. Pleitez, \emph{Quasi-Dirac neutrinos in a model with local B-L symmetry}, J. Phys. G: Nucl. Part. Phys. 40 (2013) 035002. https://doi.org/10.1088/0954-3899/40/3/035002. %, arXiv:1105.6064 [hep-ph].
\bibitem{Ghosh2012} M. Ghosh, S. Goswami, S. Gupta, 	\emph{Two-Zero mass matrices and sterile neutrinos}, J. High Energ. Phys. 2013, 103 (2013). https://doi.org/10.1007/JHEP04(2013)103.%, arXiv: 1211.0118 [hep-ph].

\bibitem{Zhang2013}Y. Zhang, \emph{Majorana neutrino mass matrices with three texture zeros and the sterile neutrino}, Phys. Rev. D 87 (2013) 053020. https://doi.org/10.1103/PhysRevD.87.053020. %, arXiv:1301.7302 [hep-ph].

\bibitem{Zhang2013v2} Y.~Zhang, X.~Ji and R.~N.~Mohapatra, \emph{A Naturally Light Sterile neutrino in an Asymmetric Dark Matter Model}, J. High Energ. Phys. 2013, 104 (2013). https://doi.org/10.1007/JHEP10(2013)104.%, arXiv:1307.6178 [hep-ph].
\bibitem{Frank2013} M.~Frank and L.~Selbuz, \emph{Sterile neutrinos in $U(1)'$ with R-parity Violation}, Phys.\ Rev.\ D 88 (2013) 055003. https://doi.org/10.1103/PhysRevD.88.055003.%, arXiv:1308.5243 [hep-ph].
\bibitem{Borah2013} D.~Borah and R.~Adhikari, \emph{Common Radiative Origin of Active and Sterile Neutrino Masses}, Phys.\ Lett.\  B 729 (2014) 143. https://doi.org/10.1016/j.physletb.2014.01.018.%, arXiv:1310.5419 [hep-ph].
\bibitem{Merle2014} A. Merle, S. Morisi, W. Winter, \emph{Common origin of reactor and sterile neutrino mixing}, J. High Energ. Phys.1407 (2014) 039. https://doi.org/10.1007/JHEP07(2014)039. %, arXiv: 1402.6332 [hep-ph].
\bibitem{Nath2016} Newton Nath, Monojit Ghosh, Shivani Gupta, \emph{Understanding the Masses and Mixings of One-Zero Textures in 3+1 Scenario}, Int. J. Mod. Phys. A 31 (2016) 1650132. https://doi.org/10.1142/S0217751X16501323.%, No. 24, arXiv:1512.00635 [hep-ph].
\bibitem{Borah2016xkc} D.~Borah \emph{et. al.,} %M.~Ghosh, S.~Gupta, S.~Prakash and S.~K.~Raut,
\emph{Analysis of four-zero textures in 3+1 framework}, Phys.\ Rev. D 94 (2016) 113001. https://doi.org/10.1103/PhysRevD.94.113001. %, arXiv: 1606.02076 [hep-ph].
\bibitem{Borah2016lrl} D. Borah, \emph{Light Sterile Neutrino and Dark Matter in Left-Right Symmetric Models Without Higgs Bidoublet}, Phys. Rev. D94 (2016) 075024. https://doi.org/10.1103/PhysRevD.94.075024.%, arXiv:1607.00244 [hep-ph].
\bibitem{Dev2017} S. Dev, D. Raj, R. R. Gautam, \emph{Deviations in Tribimaximal Mixing From Sterile Neutrino Sector}, Nucl. Phys. B 911 (2016) 744. https://doi.org/10.1016/j.nuclphysb.2016.08.015.%-753,	arXiv:1607.08051 [hep-ph].
\bibitem{Nath2017} N.~Nath, M.~Ghosh, S.~Goswami and S.~Gupta, \emph{Phenomenological study of extended seesaw model for light sterile neutrino}, J. High Energ. Phys. 1703 (2017) 075. https://doi.org/10.1007/JHEP03(2017)075.%, arXiv:1610.09090 [hep-ph].
\bibitem{Borah2017azf} D.~Borah, M.~Ghosh, S.~Gupta and S.~K.~Raut, \emph{Texture zeros of low-energy Majorana neutrino mass matrix in 3+1 scheme}, Phys.\ Rev.\ D 96 (2017) 055017. https://doi.org/10.1103/PhysRevD.96.055017.%, no. 5, , arXiv:1706.02017 [hep-ph].
\bibitem{Bhat2020} Imtiyaz Ahmad Bhat, Rathin Adhikari, \emph{Dark matter mass from relic abundance, an extra U(1) gauge boson, and active-sterile neutrino mixing}, Phys. Rev. D 101 (2020) 075030. https://doi.org/10.1103/PhysRevD.101.075030.%,, arXiv:1906.10185 [hep-ph].
\bibitem{Pires2020} C. A. de S. Pires, \emph{A cosmologically viable eV sterile neutrino model}, . https://doi.org/10.1016/j.physletb.2019.135135.%,, 	arXiv:1908.09313 [hep-ph].
\bibitem{Kumar2020} Priyanka Kumar, Mahadev Patgiri, \emph{Minimal extended seesaw and group symmetry realization of two-zero textures of neutrino mass matrices}, Nucl.Phys. B957 (2020) 115082. https://doi.org/10.1016/j.nuclphysb.2020.115082. %arXiv:
\bibitem{Pinheiro2020} João Paulo Pinheiro, C. A. de S. Pires, \emph{Vacuum stability and spontaneous violation of the lepton number at low energy scale in a model for light sterile neutrinos}, Phys. Rev. D 102 (2020) 015015.https://doi.org/10.1103/PhysRevD.102.015015. %,, arXiv: 2003.02350 [hep-ph].

\bibitem{VienS3EPJC21} V. V. Vien, \emph{3+1 active–sterile neutrino mixing in B-L model with $S_{3}\times Z_4\times Z_2$ symmetry for normal neutrino mass ordering}, Eur. Phys. J. C 81 (2021) 416. https://doi.org/10.1140/epjc/s10052-021-09214-5.



\bibitem{EBS1} R. E. Marshak and R. N. Mohapatra, \emph{Quark - Lepton Symmetry and B-L as the U(1) Generator of the Electroweak Symmetry Group}, Phys. Lett. 91B (1980) 222. https://doi.org/10.1016/0370-2693(80)90436-0.
\bibitem{EBS2} C. Wetterich, \emph{Neutrino masses and the scale of B-L violation}, Nucl. Phys. B 187 (1981) 343. https://doi.org/10.1016/0550-3213(81)90279-0.



\bibitem{BLscales} W. Abdallah, D. Delepine, S. Khalil, \emph{TeV Scale Leptogenesis in B-L Model with Alternative Cosmologies}, 	Phys.Lett. B 725 (2013) 361. https://doi.org/10.1016/j.physletb.2013.07.047.
\bibitem{Pallis2013} C. Pallis, Q. Shafi, \emph{Update on Minimal Supersymmetric Hybrid Inflation in Light of PLANCK}, Physics Letters B 725 (2013) 327. https://doi.org/10.1016/j.physletb.2013.07.029.
\bibitem{buchmuller2014} W. Buchm$\ddot{\mathrm{u}}$llera, V. Domckeb, K. Kamadac, K. Schmitzd, \emph{Hybrid Inflation in the Complex Plane}, JCAP07 (2014) 054. https://doi.org/10.1088/1475-7516/2014/07/054.
\bibitem{Moursy2021}A. Moursy, \emph{No-scale gauge non-singlet inflation inducing TeV scale inverse seesaw
mechanism},  J. High Energ. Phys. 2021, 208 (2021). https://doi.org/10.1007/JHEP10(2021)208. %, arXiv: 2107.06670 [hep-ph].





\bibitem{cutoffscal21} K. Aoki, T. Q. Loc, T. Noumi, and J. Tokuda, \emph{Is the Standard Model in the Swampland? Consistency Requirements from Gravitational Scattering}, Phys. Rev. Lett. 127 (2021) 091602, https://doi.org/10.1103/PhysRevLett.127.091602. %arXiv: 2104.09682


\bibitem{U1X1} A. Davidson, \emph{$B-L$ as the fourth color within an $SU(2)_L\times U(1)_R\times U(1)$ model}, Phys. Rev. D 20 (1979) 776. https://doi.org/10.1103/PhysRevD.20.776.
\bibitem{U1X2} R. N. Mohapatra and R. E. Marshak, \emph{Local $B-L$ Symmetry of Electroweak Interactions, Majorana Neutrinos, and Neutron Oscillations}, Phys. Rev. Lett. 44 (1980) 1316. https://doi.org/10.1103/PhysRevLett.44.1316.
\bibitem{U1X3} S. Khalil, \emph{Low scale B-L extension of the Standard Model at the LHC}, J. Phys. G 35 (2008) 055001. https://doi.org/10.1088/0954-3899/35/5/055001.%, arXiv: hep-ph/0611205.
\bibitem{U1X4} M . Abbas, S. Khalil, \emph{Neutrino masses, mixing and leptogenesis in TeV scale B-L extension of the standard model}, J. High Energ. Phys.04 (2008) 056. https://doi.org/10.1088/1126-6708/2008/04/056.%, arXiv:0707.0841 [hep-ph].
\bibitem{U1X5} S. Iso, N. Okada and Y. Orikasa, \emph{Classically Conformal B-L extended Standard Model}, Phys. Lett. B 676 (2009) 81. https://doi.org/10.1016/j.physletb.2009.04.046. %, arXiv:0902.4050 [hep-ph].
\bibitem{U1X6} S. Iso, N. Okada and Y. Orikasa, \emph{The minimal B-L model naturally realized at TeV scale}, Phys. Rev. D 80 (2009) 115007. https://doi.org/10.1103/PhysRevD.80.115007.%, arXiv:0909.0128 [hep-ph].
\bibitem{U1X7} N. Sahu and U. A. Yajnik, \emph{Dark matter and leptogenesis in gauged B-L symmetric models embedding $\nu$MSM}, Phys. Lett. B635 (2006) 1116. https://doi.org/10.1016/j.physletb.2006.02.040.%, arXiv:hep-ph/0509285.
\bibitem{U1X8} W. Emam, S. Khalil, \emph{Higgs and $Z'$ Phenomenology in B-L extension of the Standard Model at LHC}, Eur.Phys.J.C 55 (2007) 625. https://doi.org/10.1140/epjc/s10052-007-0411-7.%-633, arXiv:0704.1395 [hep-ph].
\bibitem{U1X9} T. Basak and T. Mondal, \emph{Constraining Minimal $U(1)_{B-L}$ model from Dark Matter Observations}, Phys. Rev. D 89 (2014) 063527. https://doi.org/10.1103/PhysRevD.89.063527.%, arXiv:1308.0023 [hep-ph].
\bibitem{U1X10} W. Rodejohann and C. E. Yaguna, \emph{Scalar dark matter in the B-L model}, JCAP 1512 (2015) 032. https://doi.org/10.1088/1475-7516/2015/12/032.%, arXiv: 1509.04036 [hep-ph].
\bibitem{U1X11} J. Guo, Z. Kang, P. Ko, and Y. Orikasa, \emph{Accidental Dark Matter: Case in the Scale Invariant Local B-L Models}, Phys. Rev. D 91 (2015) 115017. https://doi.org/10.1103/PhysRevD.91.115017. %, arXiv:1502.00508 [hep-ph].
\bibitem{U1X12} A. El-Zant, S. Khalil, and A. Sil, \emph{Warm Dark Matter in B-L Inverse Seesaw}, Phys. Rev. D 91 (2015) 035030. https://doi.org/10.1103/PhysRevD.91.035030.%, arXiv:1308.0836 [hep-ph].
\bibitem{U1X13} S. Khalil, H. Okada, \emph{Dark Matter in B-L Extended MSSM Models}, Phys.Rev.D 79 (2009) 083510. https://doi.org/10.1103/PhysRevD.79.083510.%, arXiv: 0810.4573 [hep-ph].

\bibitem{U1X14} T. Higaki, R. Kitano, R. Sato, \emph{Neutrinoful Universe}, J. High Energ. Phys. 2014, 44 (2014). https://doi.org/10.1007/JHEP07(2014)044.%, arXiv:1405.0013 [hep-ph].
%\bibitem{U1X15} F. F. Deppisch, W. Liu and M. Mitra, J. High Energy Phys. 1808 (2018) 181.%, arXiv:1804.04075 [hep-ph].
\bibitem{U1X16} P. S. B. Dev, R. N. Mohapatra, Y. Zhang, \emph{Leptogenesis constraints on B- L breaking Higgs boson in TeV scale seesaw models}, J. High Energ. Phys. 2018, 122 (2018). https://doi.org/10.1007/JHEP03(2018)122.%, arXiv: 1711.07634 [hep-ph].
\bibitem{U1X17} T. Hasegawa, N. Okada and O. Seto, \emph{Gravitational waves from the minimal gauged $U(1)_{B-L}$ model}, Phys. Rev. D 99 (2019) 095039. https://doi.org/10.1103/PhysRevD.99.095039. %, arXiv:1904.03020 [hep-ph].


\bibitem{ishi} H. Ishimori \emph{et. al.}, \emph{Non-Abelian Discrete Symmetries in Particle Physics}, Prog. Theor. Phys. Suppl. 183 (2010) 1. https://doi.org/10.1143/PTPS.183.1. %, arXiv:1003.3552.

\bibitem{PDG2020} P. A. Zyla \emph{et al.} (Particle Data Group), \emph{Review of Particle Physics}, Prog. Theor. Exp. Phys. 2020, 083C01 (2020). https://doi.org/10.1093/ptep/ptaa104.



\bibitem{Jarlskog1} C. Jarlskog, \emph{Commutator of the Quark Mass Matrices in the Standard Electroweak Model and a Measure of Maximal
CP Nonconservation}, Phys. Rev. Lett. 55 (1985) 1039. https://doi.org/10.1103/PhysRevLett.55.1039.
\bibitem{Jarlskog2} D.-d. Wu, \emph{Rephasing invariants and CP violation}, Phys. Rev. D 33 (1986) 860. https://doi.org/10.1103/PhysRevD.33.860.
\bibitem{Jarlskog3} O.W. Greenberg, \emph{Rephasing invariant formulation of CP violation in the Kobayashi-Maskawa framework}, Phys. Rev. D32 (1985) 1841. https://doi.org/10.1103/PhysRevD.32.1841.




\bibitem{betdecay1} M. Mitra, G. Senjanovic, F. Vissani, \emph{Neutrinoless Double Beta Decay and Heavy Sterile Neutrinos}, Nucl. Phys. B 856 (2012) 26. https://doi.org/10.1016/j.nuclphysb.2011.10.035. % (2012), arXiv: 1108.0004.
\bibitem{betdecay2} W. Rodejohann, \emph{Neutrinoless double beta decay and neutrino physics}, J. Phys. G39 (2012) 124008. https://doi.org/10.1088/0954-3899/39/12/124008. %, arXiv: 1206.2560.
\bibitem{betdecay3} J. D. Vergados, H. Ejiri and F. Simkovic, \emph{Theory of neutrinoless double beta decay},  Rep. Prog. Phys. 75 (2012) 106301. https://doi.org/10.1088/0034-4885/75/10/106301. % (2012), arXiv: 1205.0649.





\bibitem{Capozzi2017} F. Capozzi {\it et. al.}, \emph{Global constraints on absolute neutrino masses and their ordering}, Phys. Rev. D 95 (2017) 096014. https://doi.org/10.1103/PhysRevD.95.096014.
\bibitem{Salas2018} P. de Salas {\it et. al.}, \emph{Status of neutrino oscillations 2018: 3$\sigma$ hint for normal mass ordering and improved CP}, Phys. Lett. B 782 (2018) 633. https://doi.org/10.1016/j.physletb.2018.06.019.
\bibitem{Esteban2019} I. Esteban, M.C. Gonzalez-Garcia and M. Maltoni, \emph{Global analysis of three-flavour neutrino oscillations: synergies and tensions in the determination of $\theta_{23}, \delta_{CP}$, and the mass ordering}, J. High Energ. Phys. 2019 (2019) 55. https://doi.org/10.1007/JHEP01(2019)106.
\bibitem{Kelly2021} Kevin J. Kelly {\it et. al.}, \emph{Neutrino mass ordering in light of recent data}, Phys. Rev. D 103 (2021) 013004.  https://doi.org/10.1103/PhysRevD.103.013004.



\bibitem{Agostini18} M. Agostini \emph{et al.} (GERDA Collaboration), \emph{Improved limit on neutrinoless double beta decay of $^{76}Ge$ from GERDA Phase II}, Phys. Rev. Lett. 120 (2018) 132503. https://doi.org/10.1103/PhysRevLett.120.132503.% (2018), arXiv:1803.11100 [nucl-ex].
\bibitem{MAJO} C. E. Aalseth \emph{et al.} (Majorana Collaboration), \emph{Search for Zero-Neutrino Double Beta Decay in $^{76}Ge$ with the Majorana Demonstrator}, Phys. Rev. Lett. 120 (2018) 132502. https://doi.org/10.1103/PhysRevLett.120.132502.% (2018), arXiv: 1710.11608 [nucl-ex].
\bibitem{CUORE18} C. Alduino \emph{et al.}(CUORE collaboration), \emph{First Results from CUORE: A Search for Lepton Number Violation via $0\nu \beta\beta$ Decay of $^{130}Te$}, Phys.Rev.Lett.120 (2018) 132501. https://doi.org/10.1103/PhysRevLett.120.132501.%(2018), arXiv:1710.07988 [nucl-ex].


\bibitem{KamLAND16} A. Gando \emph{et al.} (KamLAND-Zen Collaboration),  Phys. Rev. Lett. 117 (2016) 082503.

\bibitem{GERDA19} M. Agostini \emph{et al.} (GERDA Collaboration), \emph{Probing Majorana neutrinos with double-$\beta$ decay}, Science 365 (2019) 1445. DOI:10.1126/science.aav8613. %, arXiv:1909.02726.
\bibitem{CUORE20} D. Adams \emph{et al.} (CUORE collaboration), \emph{Improved Limit on Neutrinoless Double-Beta Decay in $^{130}Te$ with CUORE}, Phys. Rev. Lett. 124 (2020) 122501. https://doi.org/10.1103/PhysRevLett.124.122501. %, arXiv:1912.10966.
\bibitem{CUPID2021} E. Armengaud et al. (CUPID-Mo Collaboration), \emph{New Limit for Neutrinoless Double-Beta Decay of
$^{100}Mo$ from the CUPID-Mo Experiment}, Phys. Rev. Lett. 126 (2021) 181802. https://doi.org/10.1103/PhysRevLett.126.181802. % – Published 3 May 2021




%%%%%%%%%%%%%%%%%%%%%%%%%%%%%%%%%%
\end{thebibliography}
\end{document}